\documentclass[twocolumn,hyperpdf,
amsmath,amssymb,
aps,prd,10pt,
superscriptaddress,nofootinbib,noeprint,preprintnumbers,floatfix
]{revtex4-1}
                           
\pdfoutput=1 

\usepackage{graphicx, color, xcolor}
\graphicspath{{./figures/}}

\usepackage[letterspace=-10]{microtype} 

\usepackage{mathtools, bm, braket}
\usepackage{multirow, tabularx, arydshln}
\usepackage{float}
\usepackage{rotating}
\usepackage{booktabs}

\usepackage[utf8]{inputenc} 			
\usepackage{hyperref}					

\definecolor{jlab_red}{RGB}{192,39,45}
\definecolor{jlab_orange}{RGB}{249,102,0}
\definecolor{jlab_blue}{RGB}{47,122,121}
\definecolor{jlab_green}{RGB}{65,125,10}
\definecolor{jlab_purple}{RGB}{132,0,193} 
\definecolor{jlab_dark_blue}{RGB}{0,85,156}

\newcommand{\minitab}[2][c]{\begin{tabular}{#1}#2\end{tabular}}

\newcommand{\SlJ}{\prescript{2S+1\!}{}{\ell}_J}
\newcommand{\threelJ}{\prescript{3\!}{}{\ell}_J}

\newcommand{\threeSone}{\prescript{3\!}{}{S}_1} 
\newcommand{\threePzero}{\prescript{3\!}{}{P}_0} 
 
\newcommand{\threePtwo}{\prescript{3\!}{}{P}_2} 
\newcommand{\threeDone}{\prescript{3\!}{}{D}_1}
\newcommand{\threeDtwo}{\prescript{3\!}{}{D}_2}
\newcommand{\threeDthree}{\prescript{3\!}{}{D}_3} 
\newcommand{\threeFtwo}{\prescript{3\!}{}{F}_2}

\newcommand{\threeGthree}{\prescript{3}{}{G}_3}

\newcommand{\piomegaS}{\pi\omega \big\{\! \threeSone \!\big\}}
\newcommand{\piomegaD}{\pi\omega \big\{\! \threeDone \!\big\}}
\newcommand{\piphiS}{\pi\phi \big\{\! \threeSone \!\big\}}
\newcommand{\piomegaSsub}{\pi\omega\{\! \threeSone \!\}}
\newcommand{\piomegaDsub}{\pi\omega\{\! \threeDone \!\}}
\newcommand{\piphisub}{\pi\phi\{\! \threeSone \!\}}
\newcommand{\piphiSsub}{\pi\phi\{\! \threeSone \!\}}

\hypersetup{%
pdftitle = {The $b_1$ resonance in coupled $\pi \omega$, $\pi\phi$ scattering from lattice QCD},
pdfsubject = {QCD},
pdfkeywords = {QCD, Hadron, Physics, Lattice, Meson, Scattering},
pdfauthor = {Hadron Spectrum Collaboration},
colorlinks = {true},
filecolor = {black},
linkcolor = {jlab_blue},
menucolor = {black},
citecolor = {jlab_green},
urlcolor = {jlab_green},
}{}

\newcommand{\cm}{\ensuremath{\mathsf{cm}}}

\makeatletter
\def\adl@drawiv#1#2#3{%
        \hskip.5\tabcolsep
        \xleaders#3{#2.5\@tempdimb #1{1}#2.5\@tempdimb}%
                #2\z@ plus1fil minus1fil\relax
        \hskip.5\tabcolsep}
\newcommand{\cdashlinelr}[1]{%
  \noalign{\vskip\aboverulesep
           \global\let\@dashdrawstore\adl@draw
           \global\let\adl@draw\adl@drawiv}
  \cdashline{#1}
  \noalign{\global\let\adl@draw\@dashdrawstore
           \vskip\belowrulesep}}
\makeatother

\begin{document}

\title{The $b_1$ resonance in coupled $\pi \omega$, $\pi\phi$ scattering from lattice QCD}

\author{Antoni~J.~Woss}
\email{a.j.woss@damtp.cam.ac.uk}
\affiliation{DAMTP, University of Cambridge, Centre for Mathematical Sciences, Wilberforce Road, Cambridge CB3 0WA, UK}
\author{Christopher~E.~Thomas}
\email{c.e.thomas@damtp.cam.ac.uk}
\affiliation{DAMTP, University of Cambridge, Centre for Mathematical Sciences, Wilberforce Road, Cambridge CB3 0WA, UK}
\author{Jozef~J.~Dudek}
\email{dudek@jlab.org}
\affiliation{\lsstyle Thomas Jefferson National Accelerator Facility, 12000 Jefferson Avenue, Newport News, VA 23606, USA}
\affiliation{Department of Physics, College of William and Mary, Williamsburg, VA 23187, USA}
\author{Robert~G.~Edwards}
\email{edwards@jlab.org}
\affiliation{\lsstyle Thomas Jefferson National Accelerator Facility, 12000 Jefferson Avenue, Newport News, VA 23606, USA}
\author{David~J.~Wilson}
\email{djwilson@maths.tcd.ie}
\affiliation{School of Mathematics, Trinity College, Dublin~2, Ireland}

\collaboration{for the Hadron Spectrum Collaboration}
\date{April 8, 2019}
\preprint{DAMTP-2019-12}
\preprint{JLAB-THY-19-2910}


\begin{abstract}

We present the first lattice QCD calculation of coupled $\pi\omega$ and $\pi\phi$ scattering, incorporating coupled $S$ and $D$-wave $\pi\omega$ in $J^P=1^+$. Finite-volume spectra in three volumes are determined via a variational analysis of matrices of two-point correlation functions, computed using large bases of operators resembling single-meson, two-meson and three-meson structures, with the light-quark mass corresponding to a pion mass of $m_\pi \approx 391$ MeV. Utilizing the relationship between the discrete spectrum of finite-volume energies and infinite-volume scattering amplitudes, we find a narrow axial-vector resonance ($J^{PC}=1^{+-}$), the analogue of the $b_1$ meson, with mass $m_{R}\approx1380$ MeV and width $\Gamma_{R}\approx 91$ MeV. The resonance is found to couple dominantly to $S$-wave $\pi\omega$, with a much-suppressed coupling to $D$-wave $\pi\omega$, and a negligible coupling to $\pi\phi$ consistent with the `OZI rule'. No resonant behavior is observed in $\pi\phi$, indicating the absence of a putative low-mass $Z_s$ analogue of the $Z_c$ claimed in $\pi J/\psi$. In order to minimally present the contents of a unitary three-channel scattering matrix, we introduce an $n$-channel generalization of the traditional two-channel Stapp parameterization.

\end{abstract}

\maketitle


\section{Introduction \label{Sec:Introduction}}


Contemporary studies of hadron spectroscopy seek to relate the spectrum of hadron resonances, including their decay properties, to the fundamental theory of quarks and gluons, quantum chromodynamics. The most successful theoretical technique to achieve this has proven to be lattice QCD which considers the theory on a discretized space-time grid of finite size, allowing numerical calculation of correlation functions through averaging over Monte-Carlo generated field configurations. The discrete spectrum in a finite volume corresponding to a particular choice of quantum numbers can be extracted from a matrix of correlation functions, constructed using a basis of operators which resemble the hadronic system being studied. The fact that lattice QCD studies the theory in a finite volume can be turned to our advantage -- an approach introduced by L\"uscher relates the discrete spectrum in a finite volume to hadron-hadron scattering amplitudes. Initially this was only for elastic scattering of spinless particles with the system overall at rest with respect to the lattice~\cite{Luscher:1985dn,Luscher:1986pf,Luscher:1990ck,Luscher:1990ux}, but subsequent extensions generalize the formalism to describe coupled-channels, particles with intrinsic spin, and moving frames~\cite{Briceno:2014oea,Briceno:2012yi,Christ:2005gi,Guo:2012hv,Kim:2005gf,He:2005ey,Rummukainen:1995vs,Gockeler:2012yj}.

This approach has been applied to a number of cases in which several coupled pseudoscalar-pseudoscalar channels are present, for example $\pi \eta, K\overline{K}$ in which the scalar $a_0$ appears as a resonance~\cite{Dudek:2016cru}, or $\pi \pi , K\overline{K}, \eta \eta$ where scalar $f_0$ and tensor $f_2$ resonances appear~\cite{Briceno:2017qmb}. Pseudoscalar-pseudoscalar scattering with relative orbital angular momentum defines the `natural parity' sequence, $J^P=0^+, 1^-, 2^+, \ldots$, where $J$ is the angular momentum and $P$ is the parity. To observe resonances with two-body decays in the `unnatural parity' sequence, $J^P = 0^-, 1^+, 2^- ,\ldots$, we must consider the scattering of mesons with non-zero spin. An experimentally-observed example~\cite{PhysRevD.98.030001} is the $b_1(1235)$ resonance which is dominantly seen through its decay to the $\pi \omega$ final state, where the $\omega$ is the lightest isoscalar vector meson which has a very small decay width to three pions.

Once we move into the pseudoscalar-vector scattering sector, there can often be more than one partial-wave construction having a particular $J^P$. For example, in the $1^+$ case relevant for the $b_1$, we can have the $\pi$ and $\omega$ in a relative $S$--wave or a relative $D$--wave -- indeed, by studying the angular distribution in the decay of the $b_1$, experiments have estimated the amplitudes of these two partial-waves~\cite{Nozar:2002br}.

The finite-volume formalism to handle pseudoscalar-vector scattering is in place~\cite{Briceno:2014oea}, and has been tested previously in a channel which did not feature a resonance, namely $\pi \rho$ scattering in isospin-2 with quark masses sufficiently heavy such that the $\rho$ resonance becomes a bound-state, kinematically stable against decay to two pions~\cite{Woss:2018irj}. That first calculation determined the $S$-- and $D$--wave $J^P=1^+$ amplitudes and their dynamical mixing, finding relatively weak effects as expected in this exotic isospin channel.

In this paper we will report on a study of the $J^P=1^+$ $I^G=1^+$ channel, where $I$ is the isospin and $G$ is $G$-parity, in which we expect to see a $b_1$ resonance decaying to $\pi \omega$. We make use of $N_f=2+1$ lattice configurations generated with a light-quark mass such that the pion has a mass around 391 MeV. With this light-quark mass, the $\omega$ meson is found to have a mass around 881 MeV~\cite{Dudek:2011tt,Dudek:2013yja}, and hence is \emph{stable} against decay to three pions.

To study the $b_1$ we have computed matrices of correlation functions in three lattice volumes, in several moving frames (i.e.\ where systems have overall non-zero momentum with respect to the lattice). To robustly determine the finite-volume spectrum, a wide range of operators resembling both single-hadron and multi-hadron structures were included in the basis.  
These correlation functions provide information which constrains the energy dependence of the $I^G = 1^+$ $J^P=1^+$ scattering matrix whose channels are $\pi \omega$ in $S$ and $D$-wave, and, in addition, $\pi \phi$ which is kinematically open in the considered energy region\footnote{the $\phi$ is stable against decay to $K\overline{K}$ and $\pi\pi\pi$ at the light-quark mass considered}.

A previous lattice QCD study~\cite{Lang:2014tia} of the $b_1$ limited itself to the rest-frame in one rather small volume. By considering only two degenerate flavors of light quarks and no strange quarks, any physics associated with the $\pi \phi$ channel was disallowed. A very small operator basis was used, such that only one usable energy level was obtained and this had a statistical uncertainty at the percent level. Enforcing elastic $S$--wave scattering only, ignoring any effect from the $D$--wave, and fixing the decay coupling of an assumed $b_1$ resonance at a value equal to that extracted from experimental measurements, a crude estimate of the $b_1$ mass was made in the case that the pion mass is 266 MeV. An earlier study~\cite{McNeile:2006bz} used a different approach in which the light-quark mass was tuned such that the $b_1$ decay to $\pi \omega$ is exactly at kinematic threshold. From the time-dependence of a single correlation function, an estimate of the decay coupling was inferred.

In this calculation, we determine a large number of finite-volume energy levels in multiple volumes and moving frames. We use up to 36 of these levels, each typically having statistical uncertainty at the tenth of a percent level, to constrain the coupled-channel scattering matrix. 

As well as the $\pi \omega$ and $\pi \phi$ channels, we pay attention to the fact that three-body channels, $\pi \pi \eta$ and $\pi K \overline{K}$, which have relatively low thresholds even for $m_\pi \approx 391$ MeV, can in principal play a role. Experimentally, three-body decays of resonances are found to be dominated by two-body \emph{isobar} resonances. For example, in a $\pi \pi \eta$ final state at relatively small total energy, the Dalitz plot will be expected to have the bulk of the events in narrow horizontal and vertical bands around $m_{\pi\pi} \sim m_\rho$ and $m_{\pi \eta} \sim m_{a_0}$.\footnote{There will also be a diagonal `reflection' of the $a_0$ band.}

We will explore the role of these three-body channels by including operators in our bases whose construction resembles a meson coupled to a two-body resonance, in a way which respects the symmetries of the finite cubic lattice. No finite-volume formalism capable of rigorously incorporating three-body scattering channels is yet sufficiently mature to be applied in the current case, but there has been significant recent developments~\cite{Briceno:2017tce,Briceno:2018aml,Briceno:2018mlh,Hammer:2017kms,Mai:2017bge,Mai:2018djl,Hansen:2019nir,Blanton:2019igq}. Our explorations will yield evidence that suggests that the three-body channels have a negligible effect in this particular case of a low-lying $b_1$ resonance.

To convert the finite-volume spectra calculated in lattice QCD into scattering amplitudes, we consider parameterizations of the energy dependence of the scattering $t$-matrix and the parameters are found which best describe the finite-volume spectra. This approach allows us to explore the resonance content of each $J^P$ in a rigorous way by searching for the presence of \emph{pole singularities} in $t(s)$ at complex values of $s=E^2$. Poles lying relatively close to the real energy axis typically have the real and imaginary parts of their pole position interpreted in terms of the mass and width of the resonance, and from the residue of $t(s)$ at the pole we can determine the relative couplings of the resonance to its decay channels.

A relatively light $b_1$ resonance is expected based upon an earlier set of calculations, performed on the same lattice configurations used in this paper, in which the operator basis was restricted to a set of fermion bilinears~\cite{Dudek:2010wm,Dudek:2011tt,Dudek:2013yja}. The resulting spectrum, which we expect to be incomplete owing to the lack of multi-meson operators, nevertheless featured a $J^{PC}=1^{+-}$ state near 1400 MeV, which had strong overlap with, in particular, those operators which resemble the $q\bar{q}$ spin-singlet, $P$-wave structure expected for the $b_1$ in the quark model. Such a calculation can do no more than indicate to us the likely presence of a narrow resonance -- in the current calculation we will rigorously determine its presence and properties.

It has been suggested~\cite{Ablikim:2018ofc} that the $\pi \phi$ channel, coupled to $\pi \omega$, may feature a $Z_s$ resonance analogous to the $Z_c$ enhancement that has been claimed in the $\pi J/\psi$ final state~\cite{Liu:2013dau,Ablikim:2013mio}. We will find no evidence of a $Z_s$ resonance in this work.

The remainder of this paper is structured as follows. In Section~\ref{Sec:Spectral_Determination} we briefly review the calculation of finite-volume spectra from correlation functions and describe our single-, two- and three-meson operator constructions. The lattice setup used and relevant hadron masses and thresholds are presented in Section~\ref{Sec:lattice_setup}, and in Section~\ref{Sec:operator_bases} we discuss the partial waves which are present and our choice of operator bases. The finite-volume spectra are presented and commented on in Section~\ref{Sec:finite_volume_spectra}. In Section~\ref{Sec:Scattering_Analysis} we discuss the techniques used to relate these spectra to scattering amplitudes and apply them to determine $\piomegaS$, $\piomegaD$ and $\piphiS$ amplitudes, and in Section~\ref{Sec:Pole_Analysis} we examine the pole singularities of these amplitudes. Systematic tests of our analysis are given in Section~\ref{Sec:Systematic_Analysis} where we examine the effects of additional partial-waves, including those that mix due to the reduced symmetry of the finite-volume, and additional channels that resemble $\pi\pi\eta$ and $\pi K\overline{K}$. An interpretation of the results is provided in Section~\ref{Sec:Interpretation} and we conclude with a summary in Section~\ref{Sec:Summary}.

\section{Spectral Determination and Operator Construction \label{Sec:Spectral_Determination}}


Working in a cubic volume of size $L\times L \times L$ with spatially periodic boundary condition discretises momenta, restricting to values $\vec{P}=(2\pi/L)(n_x,n_y,n_z)$ where $n_i \in \mathbb{Z}$. For particles at rest with respect to the lattice, the infinite-volume $\text{O}(3)$ spatial symmetry is broken to that of the double cover of the octahedral group with parity, $\text{O}^D_h$, and total angular momentum $J$ and parity $P$ labelling the irreducible representations (\emph{irreps}) of $\text{O}(3)$ are replaced by $\Lambda^P$, the irreps of $\text{O}^D_h$. In this work we only encounter integer spin and therefore irreps of the single cover $\text{O}_h$. For particles ``in-flight'', i.e.\ moving with respect to the lattice, parity is no longer a good quantum number and the irreps $\Lambda$ are those of the \emph{little group} of symmetries, $\text{LG}(\vec{P})$, as discussed in Ref.~\cite{Moore:2005dw}. We write lattice irreps $\vec{P}\Lambda$ with shorthand $\vec{P}=[n_xn_yn_z]$, omitting units of $(2\pi/L)$ for brevity.

In order to robustly determine the discrete finite-volume energy eigenstates in each irrep, $\vec{P}\Lambda$, we first compute a large matrix of two-point correlation functions, ${C(t)_{ij}=\braket{0|\mathcal{O}^{}_i(t+t_{\text{src}}) \mathcal{O}_j^\dagger(t_{\text{src}})|0}}$, by employing a diverse basis of operators $\mathcal{O}_i$. These operators are constructed with the desired flavour structure and \emph{subduced} into the irrep $\vec{P}\Lambda$~\cite{Dudek:2010wm,Thomas:2011rh}. A variationally optimal determination of the spectrum~\cite{Michael:1985ne,Luscher:1990ck} follows from solving the generalized eigenvalue problem for each irrep,
\begin{equation}\label{Eq:VM}
C(t)\, v^{\mathfrak{n}} = \lambda_{\mathfrak{n}}(t) \, C(t_0)\, v^{\mathfrak{n}} \, .
\end{equation}
The energy levels $E_\mathfrak{n}$ are determined by fitting \emph{principal correlators} $\lambda_\mathfrak{n}(t)$ to the form,
\begin{equation}\label{Eq:PC}
\lambda_\mathfrak{n}(t)=(1-A_\mathfrak{n})\, e^{-E_\mathfrak{n}(t-t_0)} + A_\mathfrak{n} \, e^{-E'_\mathfrak{n} (t-t_0)} \, ,
\end{equation}
where the second term soaks up any residual excited state contamination. The eigenvector $v^\mathfrak{n}$ can be used to construct a variationally optimised operator, $\Omega_\mathfrak{n}^\dagger = \sum_i v^\mathfrak{n}_i \, \mathcal{O}^\dagger_i$, efficient at interpolating the $\mathfrak{n}^{\text{th}}$ eigenstate in the spectrum. We refer the reader to Refs.~\cite{Dudek:2010wm,Dudek:2007wv} for further details of our implementation and techniques for selecting a reasonable value of $t_0$. 

Previous calculations~\cite{Woss:2018irj,Wilson:2015dqa,Cheung:2017tnt} have demonstrated the importance of having sufficiently `complete' operator bases in order to reliably determine the complete spectra in a given energy region. The region we study includes the opening of several multi-hadron thresholds: $\pi\omega$, $\pi\phi$, $\pi\pi\eta$ and $\pi K \overline{K}$, and we find that this necessitates the inclusion of two-meson-like and three-meson-like operators in our basis, as well as single-meson operators of fermion-bilinear form which we expect to have good overlap with any bound state or relatively-narrow resonance present. Four-meson thresholds lie beyond the energy region we consider, and previous calculations suggest that local tetraquark-like operators have little effect on the spectra~\cite{Cheung:2017tnt,Padmanath:2015era}, so neither of these types of operators are included in the basis. The construction of interpolating operators resembling single-meson, two-meson and three-meson structures is discussed in the subsections which follow.

\subsection{Single-meson operators \label{Sec:single_ops}}


The construction of `single-meson-like' operators follows the procedure detailed in Refs.~\cite{Thomas:2011rh,Dudek:2010wm}. To summarise, fermion bilinears $\bar{\psi}\Gamma\overleftrightarrow{D}...\overleftrightarrow{D}\psi$ are constructed with definite $J^P$ and $z$-component of angular momentum $M$ by appropriately coupling products of gauge-covariant derivatives $\overleftrightarrow{D}$ and Dirac $\gamma$-matrices $\Gamma$. These are then projected onto definite momentum $\vec{P}$ and appropriate linear combinations yield continuum single-meson operators $\mathcal{O}^{\dagger J M}_{\mathbb{M}}(\vec{P},t)$ of definite flavor, labelled by $\mathbb{M}$. Schematically, 
\[
\mathcal{O}^{\dagger J M}_{\mathbb{M}}(\vec{P},t)=\sum_{\vec{x}}e^{i\vec{P}\cdot\vec{x}}\big[\bar{\psi}\Gamma\overleftrightarrow{D}...\overleftrightarrow{D} \psi\big]^{JM}(\vec{x},t) \, ,
\]
where for $\vec{P}\neq \vec{0}$ we use \emph{helicity operators}, labelled by helicity $\lambda$ rather than $M$, as discussed in Ref.~\cite{Thomas:2011rh}. Single-meson operators, transforming irreducibly under the symmetry of the lattice grid and boundary, $\mathcal{O}^{\dagger \Lambda \mu}_{\mathbb{M}}(\vec{P})$, are obtained by subducing,
\[
\mathcal{O}^{\dagger \Lambda \mu}_{\mathbb{M}}(\vec{P})=\sum_M \mathcal{S}^{J M}_{\Lambda \mu} \, \mathcal{O}^{\dagger J M}_{\mathbb{M}}(\vec{P}),
\]
where $\mathcal{S}^{J M}_{\Lambda \mu}$ are \emph{subduction coefficients} tabulated in Refs.~\cite{Dudek:2010wm,Thomas:2011rh}.

A large basis of operators can be constructed by combining $\gamma$-matrices with various numbers of derivatives -- here we use up to three derivatives for operators with zero momentum and up to two otherwise. Single-meson operators are written as $\bar{\psi}\bm{\Gamma}\psi$ for the remainder of this article.

Optimised operators for the stable $\omega$ ($\Omega^\dagger_\omega$) and $\phi$ ($\Omega^\dagger_\phi$) in each relevant irrep follow from variational analysis of a matrix of correlation functions constructed using a basis of quark bilinears with both hidden-light ($\bar{u} \boldsymbol{\Gamma} u + \bar{d} \boldsymbol{\Gamma} d$) and hidden-strange ($\bar{s} \boldsymbol{\Gamma} s$) flavor structure. The required `annihilation' diagrams are computed but, as shown in Figures 4 and 5 of Ref.~\cite{Dudek:2013yja}, they prove to be small in the vector channel in line with the experimentally-motivated `OZI rule'. In each irrep, the $\omega$ appears as the ground state, dominated by overlap with $\bar{u} \boldsymbol{\Gamma} u + \bar{d} \boldsymbol{\Gamma} d$, and the $\phi$ as the first excited state, dominated by $\bar{s} \boldsymbol{\Gamma} s$.

The same flavor basis is used to determine the optimum $\eta$ operator ($\Omega^\dagger_\eta$) in each irrep, but here significant mixing between light and strange is observed through the annihilation diagrams (see Figures 2 and 3 in Ref.~\cite{Dudek:2013yja}), indicating, as is well known, that the OZI rule does not apply in the pseudoscalar channel.

The need to account for `in-hadron annihilation' when considering isoscalar mesons will reappear when the optimized operators are used in two-meson and three-meson constructions as discussed below.

\subsection{Two-meson operators \label{Sec:two-meson_ops}}


Our approach to constructing operators which resemble a two-meson structure has been discussed in detail in Ref.~\cite{Dudek:2012gj} and, in particular, pseudoscalar-pseudoscalar operators have been implemented in many calculations~\cite{Dudek:2012xn,Dudek:2014qha,Dudek:2016cru,Briceno:2016mjc,Briceno:2017qmb,Moir:2016srx,Wilson:2014cna,Wilson:2015dqa}, and vector-pseudoscalar operators are used in Refs.~\cite{Woss:2018irj,Cheung:2017tnt}.

We construct two-meson operators with definite flavor and momentum in irrep $\Lambda$ (row $\mu$) by taking appropriate linear combinations of the products of optimised single-meson operators $\Omega^\dagger_{\mathbb{M}}$, each independently constructed to transform irreducibly in some lattice irrep.  Schematically,
\begin{align}\label{Eq:two-meson}
\mathcal{O}_{\mathbb{M}_1\mathbb{M}_2}^{\dagger \Lambda \mu}(\vec{p}_{12}) = \sum_{\substack{\vec{p}_1,\vec{p}_2 \\ \mu_1,\mu_2}} & \mathcal{C}([\vec{p}_{12}]\Lambda,\mu;[\vec{p}_1]\Lambda_1,\mu_1;[\vec{p}_2]\Lambda_2,\mu_2) \nonumber \\ 
&\times \, \Omega^{\dagger \Lambda_1\mu_1}_{\mathbb{M}_1}(\vec{p}_1)  \, \Omega^{\dagger \Lambda_2\mu_2}_{\mathbb{M}_2}(\vec{p}_2),
\end{align}
where the sum is over the rows $\mu_i$ of the irreps $\Lambda_i$ and the sets of momenta $\{\vec{p}_i\}^*$, containing all momenta related to $\vec{p}_i$ by an allowed lattice rotation with the total momentum $\vec{p}_{12} = \vec{p}_1 + \vec{p}_2$ fixed -- see Eq.~3.3 of Ref.~\cite{Woss:2018irj}. For $|\vec{p}_i|^2 < 9(2\pi/L)^2$, the set $\{\vec{p}_i\}^*$ is equivalently labelled by the magnitude of the momentum $|\vec{p}_i|$.
The sum is weighted by lattice Clebsch-Gordon coefficients, $\mathcal{C}([\vec{p}_{12}]\Lambda,\mu;[\vec{p}_1]\Lambda_1,\mu_1;[\vec{p}_2]\Lambda_2,\mu_2)$~\cite{Dudek:2012gj}.

For energies below three-meson thresholds, previous calculations suggest that a sufficient set of operators for a reliable calculation of the spectra consists of single-meson and two-meson operators. Two-meson operators $\mathcal{O}_{\mathbb{M}_1\mathbb{M}_2}^{\dagger \Lambda \mu}(\vec{p}_{12})$ are efficient at interpolating the finite-volume energy levels near to the associated non-interacting energies,
\[
E^{(2)}_{\text{n.i.}}=\sqrt{m_1^2+|\vec{p}_1|^2}+\sqrt{m_2^2+|\vec{p}_2|^2} \, ,
\]
and truncating the two-meson operator bases when the corresponding non-interacting energies are beyond the energy region of interest has been demonstrated to be sufficient for a robust determination of the spectra~\cite{Woss:2018irj,Wilson:2015dqa,Dudek:2012gj,Dudek:2012xn,Dudek:2014qha,Dudek:2016cru,Briceno:2016mjc,Briceno:2017qmb,Cheung:2017tnt,Moir:2016srx,Wilson:2014cna,Wilson:2015dqa}. Two-meson operators are written $\mathbb{M}_{1\,[\vec{p}_1]}\mathbb{M}_{2\,[\vec{p}_2]}$ in all tables and figures for the remainder of this work.

The fact that a vector meson in flight is subduced into multiple irreps means that there can be multiple $\mathbb{M} \mathbb{M}$ constructions for a single non-interacting energy. For example, $\pi_{001} \omega_{001}$ subduced into the $[000]\, T_1^+$ irrep (which contains $J^P=1^+$) can be constructed independently from $\pi(A_2) \otimes \omega(A_1)$ or from $\pi(A_2) \otimes \omega(E_2)$. Cases such as these where the \emph{multiplicity} of operators is greater than one are discussed in detail in Ref.~\cite{Woss:2018irj}, and we indicate them with a notation $\{n\}$.

Correlation functions with $\mathbb{MM}$ operators at the source and/or sink feature Wick contractions in which quarks annihilate either within an isoscalar meson or between two mesons. Considering a basis with overall $I=1$ as relevant here, with $\mathbb{M} = \bar{u} \boldsymbol{\Gamma} d$ and $\mathbb{MM} = \{ \pi \omega, \pi \phi \}$, we need to evaluate diagrams whose structure is similar to those shown in Figure 1 of~\cite{Wilson:2014cna}.

\subsection{Three-meson operators \label{Sec:three-meson_ops}}


Three-meson operators\footnote{and operators with a structure resembling more than three mesons} can be constructed by iteratively applying the two-meson operator construction outlined above. Schematically,
\begin{align}\label{Eq:three-meson-MMM}
\mathcal{O}_{\mathbb{M}_1\mathbb{M}_2\mathbb{M}_3}^{\dagger \Lambda \mu}(\vec{p}_{123}) 
= \!\sum_{\substack{\vec{p}_{12},\vec{p}_3 \\ \mu_{12},\mu_3 } }&
\mathcal{C}([\vec{p}_{123}]\Lambda,\mu;[\vec{p}_{12}]\Lambda_{12},\mu_{12};[\vec{p}_3]\Lambda_3,\mu_3) \nonumber \\
&\times  \mathcal{O}^{\dagger \Lambda_{12}\mu_{12}}_{\mathbb{M}_1  \mathbb{M}_2}(\vec{p}_{12})\,
\Omega^{\dagger \Lambda_3 \mu_3}_{\mathbb{M}_3}(\vec{p}_3)
\end{align}
where $\mathcal{O}^{\dagger \Lambda\mu}_{\mathbb{M}_1  \mathbb{M}_2}$ is a two-meson operator constructed from a product of optimised single-meson operators as in Section~\ref{Sec:two-meson_ops}. Note that it does not matter with which optimised single-mesons we formed the intermediate two-meson operator, i.e.~$\mathcal{O}_{\mathbb{M}_1\mathbb{M}_2}^{\dagger \Lambda\mu}(\vec{p}_{12})$, $\mathcal{O}_{\mathbb{M}_2\mathbb{M}_3}^{\dagger \Lambda\mu}(\vec{p}_{23})$ or $\mathcal{O}_{\mathbb{M}_1\mathbb{M}_3}^{\dagger \Lambda\mu}(\vec{p}_{13})$, as the tensor product is associative. An argument for determining a sufficient set of three-meson operators, analogous to that presented previously, would suggest calculating the corresponding non-interacting energies
\[
E^{(3)}_{\text{n.i.}}=\sqrt{m_1^2+|\vec{p}_1|^2}+\sqrt{m_2^2+|\vec{p}_2|^2}+\sqrt{m_3^2+|\vec{p}_3|^2}
\]
and enforcing a similar truncation on the basis. While this approach has the advantage of being straightforward, it pays no attention to the fact that we expect certain two-meson pairs to feature resonating behavior, the finite-volume analogue of the Dalitz-plot enhancements mentioned in the introduction.

Consider the example of~$\pi\pi\pi$ in isospin-2. Following the construction above, we would be attempting to describe energy eigenstates of the~$\pi\pi$ isospin-1 subsystem using {$\mathcal{O}^{\dagger \Lambda\mu}_{\mathbb{M}_1\mathbb{M}_2}$} constructed using only `$\pi\pi$'-like operators. To reliably determine the isovector $\pi\pi$ spectra, i.e.~the $\rho$ spectra, an operator basis including both $\bar{\psi}\mathbf{\Gamma}\psi$ and $\pi\pi$-like operators is needed as shown in Figure 1 of Ref.~\cite{Dudek:2012xn}. An alternative approach, based upon this observation and used in Ref.\ \cite{Cheung:2017tnt}, utilizes an \emph{optimised two-meson operator} which will be a linear combination of $\bar{\psi}\mathbf{\Gamma}\psi$ and $\pi\pi$-like operators. We denote such an optimised operator {$\Omega^\dagger_{\mathbb{R}}$}, where $\mathbb{R}$ indicates the meson with the corresponding quantum numbers, i.e.~$\Omega^\dagger_{\rho}$ for the example above.\footnote{Lattice irreps contain more than one spin but for convenience we choose the label $\mathbb{R}$ corresponding to the lightest such meson, e.g.~in $[000]T_1^-$ we choose $\rho$.} In general, multiple optimised operators may be relevant -- {$\Omega^\dagger_{\mathbb{R}^\mathfrak{n}}$} denotes the optimal interpolating operator for the $\mathfrak{n}^\text{th}$ excited state in the relevant meson-meson subsystem.

Combining these operators with an optimized single-meson operator yields an alternative set of three-meson operators, given schematically by 
\begin{align}\label{Eq:three-meson-MR}
\mathcal{O}_{\mathbb{R}_{12}\mathbb{M}_3}^{\dagger \Lambda \mu}(\vec{p}_{123}) 
= \!\!\sum_{\substack{\vec{p}_{12},\vec{p}_3 \\ \mu_{12},\mu_3 } }
&\mathcal{C}([\vec{p}_{123}]\Lambda,\mu;[\vec{p}_{12}]\Lambda_{12},\mu_{12};[\vec{p}_3]\Lambda_3,\mu_3) \nonumber \\
&\times\, \Omega^{\dagger \Lambda_{12}\mu_{12}}_{\mathbb{R}_{12}}(\vec{p}_{12})   
\,\Omega^{\dagger \Lambda_3\mu_3}_{\mathbb{M}_3}(\vec{p}_3) \, .
\end{align}
By design, we anticipate that these three-meson operators will efficiently interpolate finite-volume levels in the region of an energy value
\begin{equation}\label{Eq:ni-MR}
E^{(2+1)}_{\text{n.i.}}=E^{\Lambda_{12}}_{\mathbb{R}^\mathfrak{n}_{12}}(\vec{p}_{12})+\sqrt{m_3^2+|\vec{p}_3|^2},
\end{equation}
where {$E^{\Lambda_{12}}_{\mathbb{R}^\mathfrak{n}_{12}}(\vec{p}_{12})$} are finite-volume energies calculated in the two-meson subsystem in irrep $[\vec{p}_{12}]\Lambda_{12}$, i.e.~they will efficiently capture interaction in the two-meson subsystem assuming weak residual interaction with the third meson. Calculating $E^{(2+1)}_{\text{n.i.}}$ energies, for all possible combinations of two-meson subsystems that together with the third meson give the desired quantum numbers, and truncating at a desired energy, provides a procedure for selecting which of these three-meson operators to include in the basis.
\begin{center}
\rule{0.25\textwidth}{.1pt}
\end{center}
%
%
To illustrate the construction presented above, consider the example of a three-meson operator resembling $\pi\pi\eta$ in the irrep $[000]\,T_1^+$ with $I^G=1^+$. We begin with the construction shown in Eq.~\ref{Eq:three-meson-MMM}. For $\vec{p}_1=\vec{p}_2=\vec{p}_3 = \vec{0}$, there is only one possible irrep,
\[
\overbrace{\underbrace{[000]A_1^-}_{\pi}}^{(I^G=1^-)} \otimes \overbrace{\underbrace{[000]A_1^-}_{\pi}}^{(I^G=1^-)} \otimes  \overbrace{\underbrace{[000]A_1^-}_{\eta}}^{(I^G=0^+)} \rightarrow [000]A_1^- \notag \, ,
\]
so no non-interacting $\pi \pi \eta$ level, or corresponding operator, appears in $[000]\,T_1^+$ at threshold. If the pions are both given one unit of momentum, $\vec{p}_1=\vec{p}_2=[001]$ and $\vec{p}_3 = \vec{0}$ (recalling that directions of momenta $\vec{p}_i$ are summed over as detailed in Section~\ref{Sec:two-meson_ops}), the product
\[
\overbrace{\underbrace{[001]A_2}_{\pi}}^{(I^G=1^-)} \otimes \overbrace{\underbrace{[001]A_2}_{\pi}}^{(I^G=1^-)} \otimes  \overbrace{\underbrace{[000]A_1^-}_{\eta}}^{(I^G=0^+)} \rightarrow \overbrace{[000]T_1^+}^{(I^G=1^+)} \oplus \, ... \notag
\]
appears \emph{once} in $[000]T_1^+$ with $I^G=1^+$. Following the construction outlined in Eq.~\ref{Eq:three-meson-MMM} yields one operator of the form $\mathcal{O}^\dagger_{\pi\pi\eta}$ with corresponding non-interacting energy,
\[
E^{(3)}_{\text{n.i.}}=2\sqrt{\,m_\pi^2+ \left(\tfrac{2\pi}{L}\right)^2}+m_\eta.
\]
%
\begin{center}
\rule{0.25\textwidth}{.1pt}
\end{center}

Now we consider bound-states and resonances in the $\pi\pi$ and $\pi\eta$ two-meson subsystems and construct operators according to Eq.~\ref{Eq:three-meson-MR}. Unlike in the previous construction, the order in which we combine the single-meson operators does matter as the intermediate $\Omega^\dagger_{\mathbb{R}}$ depends on the flavor structure of the two-meson subsystem.
As before, for $\vec{p}_1=\vec{p}_2=\vec{p}_3 = \vec{0}$ there is no $[000]\,T_1^+$, while for $\vec{p}_1=\vec{p}_2=[001]$ and $\vec{p}_3 = \vec{0}$, there are two possible distinct two-meson subsystems.

First, for the $\pi\pi$ subsystem, there are three possible flavor combinations, $I^G=0^+,1^+,2^+$, and three possible irreps with momentum $\vec{p}_{12}=\vec{0}$, namely $[000]\, A_1^+$, $[000]\, T_1^-$ and $[000]\, E^+$. When combined with the $\eta$, only the $\pi\pi$ subsystem with $I^G=1^+$ transforming in $[000]\, T_1^-$ gives the desired overall flavor and irrep. This $\pi\pi$ subsystem contains quantum numbers corresponding to the $\rho$ and the construction is, schematically,
\begin{align}\label{Eq:rhoeta}
\Big(\overbrace{\underbrace{[001]A_2}_{\pi}}^{(I^G=1^-)} \otimes \overbrace{\underbrace{[001]A_2}_{\pi}}^{(I^G=1^-)}\Big) &\otimes \, \overbrace{\underbrace{[000]A_1^-}_{\eta}}^{(I^G=0^+)} \rightarrow \overbrace{[000]T_1^+}^{(I^G=1^+)} \notag \\
\overbrace{\underbrace{[000]T_1^-}_{\rho}}^{(I^G=1^+)} &\otimes \, \overbrace{\underbrace{[000]A_1^-}_{\eta}}^{(I^G=0^+)} \rightarrow \overbrace{[000]T_1^+}^{(I^G=1^+)}.
\end{align}
Calculating the $E^{(2+1)}_{\text{n.i.}}$ energies amounts to determining the $\rho$-like energy eigenstates in $[000]\, T_1^-$ with $I^G=1^+$ and adding these to the $\eta$ energy according to Eq.~\ref{Eq:ni-MR},
\[
E^{(2+1)}_{\text{n.i.}}=E^{T_1^-}_{\rho^\mathfrak{n}}([000]) + m_\eta\, ,
\]
where we recall that $\rho^\mathfrak{n}$ denotes the $\mathfrak{n}^\text{th}$ energy eigenstate within the irrep. In many cases, including here, only the lowest energy two-meson state ($\mathfrak{n}=0$) yields an operator below the energy cut-off.

The second possible construction considers the $\pi\eta$ subsystem where there is only one flavor combination, $I^G=1^-$, and one possible irrep, $[001]\,A_1$. These quantum numbers correspond to the $a_0$ meson. Schematically,
\begin{align}
\Big(\overbrace{\underbrace{[001]A_2}_{\pi}}^{(I^G=1^-)} \otimes \overbrace{\underbrace{[000]A_1^-}_{\eta}}^{(I^G=0^+)}\Big)\otimes \, \overbrace{\underbrace{[001]A_2}_{\pi}}^{(I^G=1^-)}  \, &\rightarrow \overbrace{[000]T_1^+}^{(I^G=1^+)} \notag \\
\overbrace{\underbrace{[001]A_1}_{a_0}}^{(I^G=1^-)}  \otimes
\overbrace{\underbrace{[001]A_2}_{\pi}}^{(I^G=1^-)} &\rightarrow \overbrace{[000]T_1^+}^{(I^G=1^+)},
\end{align}
and, as before, we determine the $E^{(2+1)}_{\text{n.i.}}$ energies by calculating the $a_0$-like energy eigenstates in $[001]A_1$ with $I^G=1^-$ and add these to the $\pi$ energy according to Eq.~\ref{Eq:ni-MR},
\[
E^{(2+1)}_{\text{n.i.}}=E^{A_1}_{{a_0}^\mathfrak{n}}([001]) + \sqrt{m_\pi^2 + \big(\tfrac{2\pi}{L}\big)^2}.
\]
For each $E^{(2+1)}_{\text{n.i.}}$ below some energy cut-off we can construct operators of the form $\mathcal{O}_{a_0 \pi}^{\dagger}$ via Eq.~\ref{Eq:three-meson-MR}. The $E^{A_1}_{{a_0}^\mathfrak{n}}([001])$ energies are an example of a case where it may be prudent to consider multiple states ($\mathfrak{n} \geq 0$) in the two-body sector. Figure 4 of Ref.~\cite{Dudek:2016cru} shows the $[001]A_1$ spectra corresponding to the $E^{A_1}_{{a_0}^\mathfrak{n}}([001])$ energies -- there are many nearby low-lying energy levels on each volume. Following the construction given in Eq.~\ref{Eq:three-meson-MR} leads to multiple operators of the form $\mathcal{O}_{a_0 \pi}^{\dagger}$ corresponding to similar $E^{(2+1)}_{\text{n.i.}}$.

\begin{center}
\rule{0.25\textwidth}{.1pt}
\end{center}

\begin{figure}[h]
\includegraphics[width=0.32\textwidth]{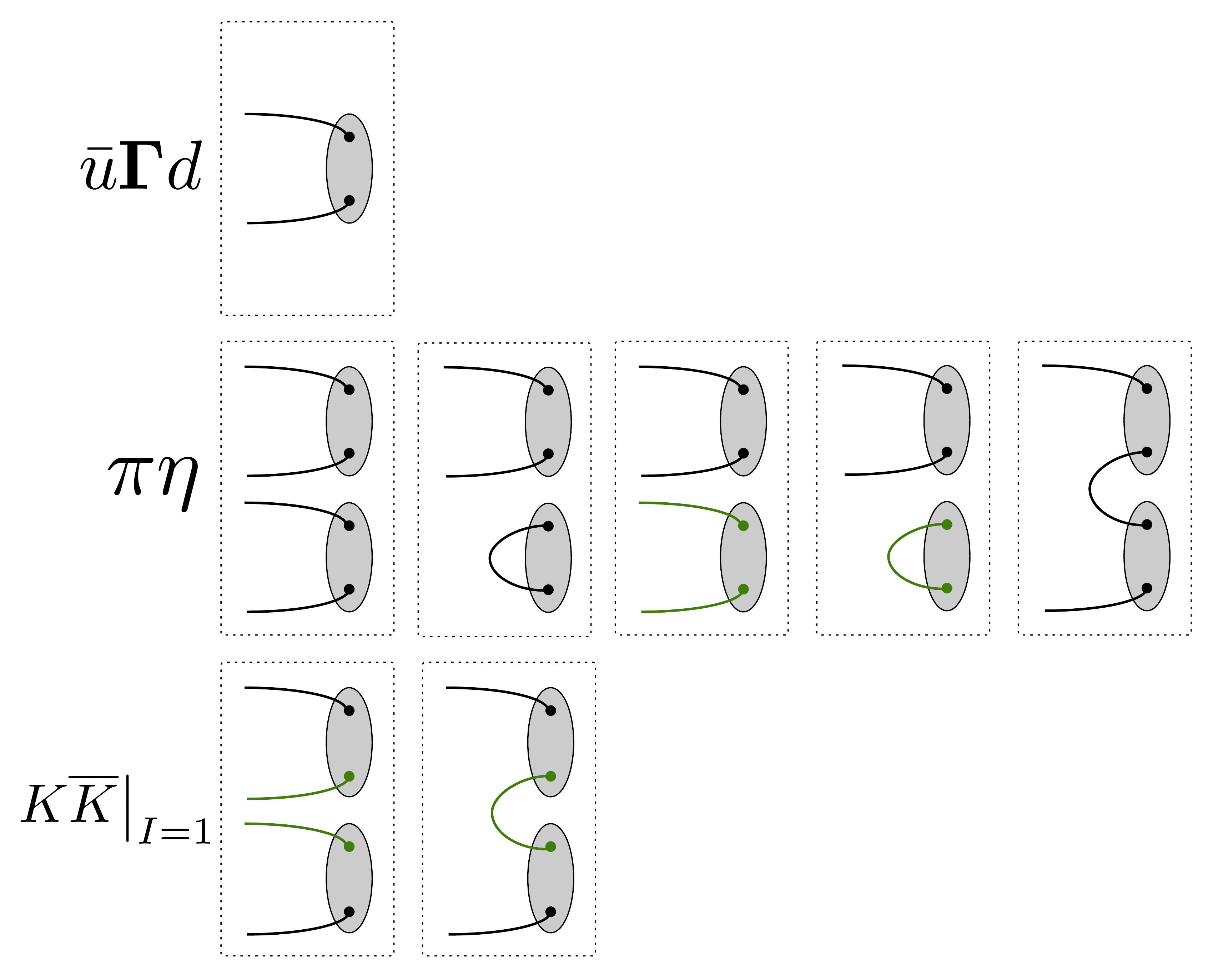}
\caption{Quark propagation lines (black are light quarks, green are strange quarks) from operator constructions featuring in an optimized $a_0$-like operator.}
\label{Fig:a0-wicks}
\end{figure}

\begin{figure}[h]
\includegraphics[width=0.32\textwidth]{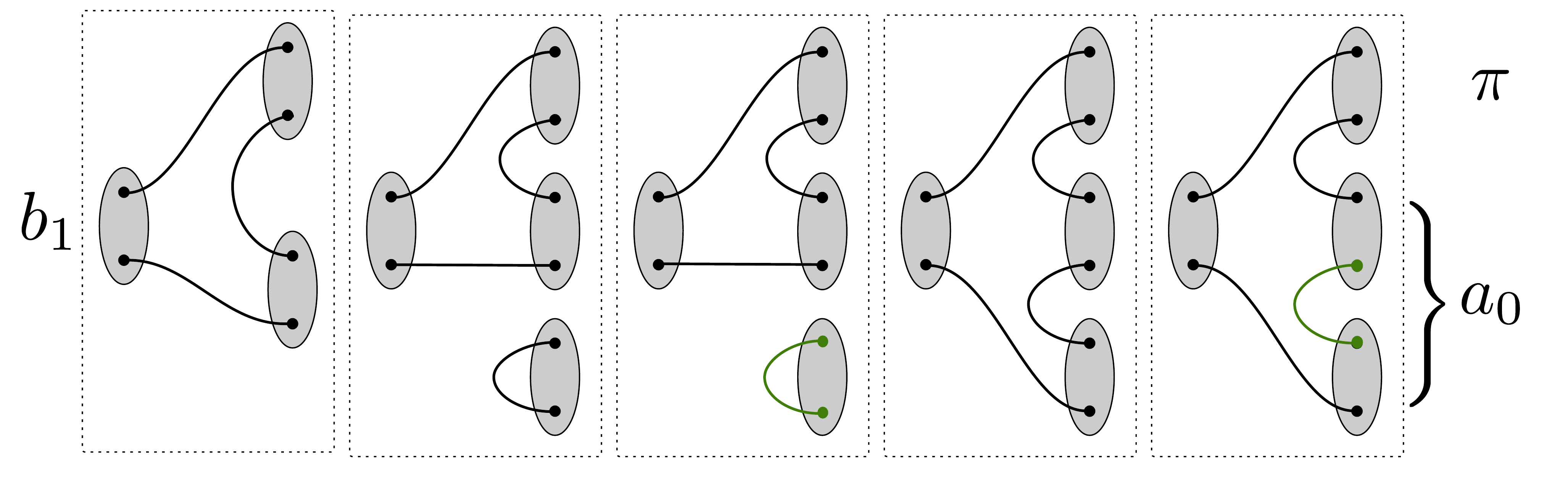}
\caption{Wick contraction topologies for $b_1 - a_0 \pi$. Left meson resembles the $b_1$, upper right meson the $\pi$ and the remaining one or two mesons the $a_0$ (only a subset of the topologies in Figure~\ref{Fig:a0-wicks} are relevant here).}
\label{Fig:M-MR_wicks}
\end{figure}

The use of $\mathbb{RM}$ operators to efficiently interpolate finite-volume states above three-meson thresholds requires the calculation of a large number of diagrams. As an example, consider the case of an $a_0 \pi$ operator at the sink, where the optimized $a_0$ operators are linear superpositions of $\bar{u} \boldsymbol{\Gamma} d$, $\pi \eta$ and $K\overline{K}$ constructions (see Table~\ref{Tab:ops_a0_proj}). This leads to the diagram components shown in Figure~\ref{Fig:a0-wicks}, which need to be connected to the quark lines from the $\pi$ and the source operator to form complete Wick contractions. It follows that even in the simple case of $b_1 - a_0 \pi$ correlators we would have diagrams with the structures shown in Figure~\ref{Fig:M-MR_wicks}.

\section{Lattice Setup \label{Sec:lattice_setup}}


Correlation functions were computed on anisotropic lattices of spatial volumes $(L/a_s)^3 = 16^3$, $20^3$ and $24^3$ each having temporal extent $T/a_t = 128$, where the temporal lattice spacing, $a_t$, is finer than the spatial lattice spacing, $a_s \sim 0.12$ fm, with an anisotropy $\xi=a_s/a_t \sim 3.5$. Gauge fields were generated from a tree-level Symanzik-improved gauge action and a Clover fermion action with $N_f=2+1$ flavors of dynamical quarks where the strange quark is tuned to approximately its physical mass and the degenerate light quarks are such that $m_\pi \sim 391$ MeV~\cite{Edwards:2008ja,Lin:2008pr}. We utilize the \emph{distillation} framework~\cite{Peardon:2009gh} to compute correlation functions as successfully demonstrated in many previous works. All relevant Wick contractions were calculated within this framework without requiring additional propagator inversions beyond the basic set of $t_\mathrm{src} - t$ and $t-t$ `perambulators' for light and strange quarks which were computed for use in previously reported calculations. The very large number of diagrams incurs only a combinatoric cost associated with the contraction of the perambulators with the operator constructions.

Correlation functions were computed using the number of distillation vectors, gauge configurations and time-sources shown in Table~\ref{Tab:Ns}. Typically, we calculated all the elements of the matrix of correlation functions, including the transposes, $C_{ij}$ and $C_{ji}$, which are related by hermiticity. In a few cases where there are a particularly large number of diagrams contributing, we made use of hermiticity to infer $C_{ji}$ from the computed $C_{ij}$.

\begin{figure}[tb]
	\centering
	\includegraphics[width=0.5\textwidth]{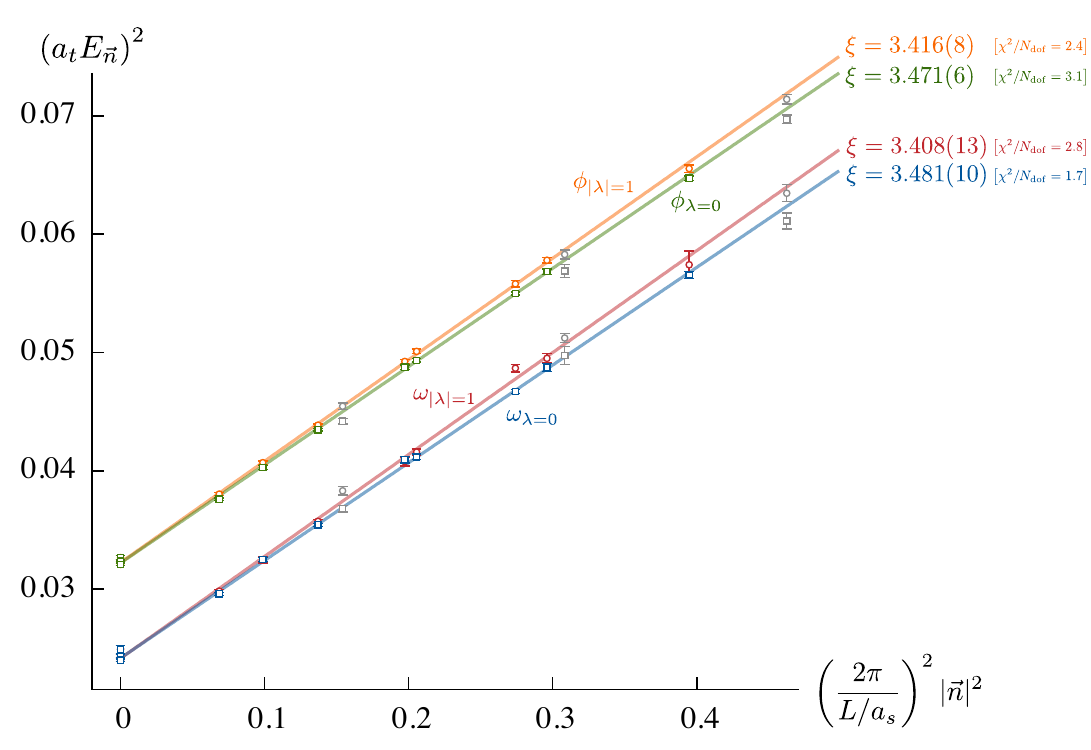}
	\caption{Momentum dependence of $\omega$ and $\phi$ energies and fits to Eq.~\ref{Eq:ani}.  Blue and red lines correspond to the $\omega$ meson with $|\lambda|=0$ and $1$ respectively.  Similarly, green and orange lines correspond to the $\phi$ meson with $|\lambda|=0$ and $1$. Points are shown with statistical uncertainties and grey points show the $(L/a_s)=16$ in-flight energies which are not included in the fit.}
	\label{Fig:dispersion}
\end{figure}

Masses of relevant stable hadrons are shown in Table~\ref{Tab:masses_thresholds}, where $\pi$, $K$, $\eta^{(\prime)}$ and $\sigma$ masses are taken from Refs.~\cite{Dudek:2012gj}, \cite{Wilson:2014cna}, \cite{Dudek:2016cru} and \cite{Briceno:2017qmb} respectively. Using energy levels on three lattice volumes, we determine the masses and anisotropies of the $\omega$ and $\phi$ mesons from fits to the dependence of the energy of a stable hadron of momentum, $\vec{p}=(2\pi/L)\vec{n}$,
\begin{equation}\label{Eq:ani}
(a_t E_{\vec{n}})^2 = (a_t m)^2 + \frac{1}{\xi^2}\bigg(\frac{2\pi}{L/a_s}\bigg)^2 |\vec{n}|^2,
\end{equation}
up to discretisation effects, as shown in Figure~\ref{Fig:dispersion}. We observe the same characteristic splitting between the ${|\lambda|=0,1}$ components as was found for the stable $\rho$ meson in Ref.~\cite{Woss:2018irj} at larger quark mass, and we attribute this splitting to discretisation effects given that the finite-volume effects here are small. The values of $a_t m_\omega$, $a_t m_\phi$ and $\xi$ we use are obtained by taking the largest variations within one standard deviation of the means across the different helicities. This yields the masses given in Table~\ref{Tab:masses_thresholds} and an anisotropy $\xi=3.443(48)$ which is consistent with the anisotropies previously determined for $\pi$, $K$ and $\eta$~\cite{Dudek:2012gj,Wilson:2014cna,Dudek:2016cru}.
\begin{table}
	\centering
		\begin{tabular}{c c c c}
			$(L/a_s)^3 \times (T/a_t)$ & $N_{\text{vecs}}$ & $N_{\text{cfgs}}$ & $N_{\text{tsrcs}}$ \\
			\midrule
			$16^3 \times 128$ & 64 & 479 & 8 -- 16\\
			$20^3 \times 128$ & 128 & 452 -- 603 & 4\\
			$24^3 \times 128$ & 160 & 553 & 4\\
		\end{tabular}
	\caption{Number of distillation vectors $N_{\text{vecs}}$, gauge configurations $N_{\text{cfgs}}$, and time-sources  $N_{\text{tsrcs}}$ used in the computation of correlation functions.}
	\label{Tab:Ns}
\end{table}
\begin{table}
		\centering
		\begin{tabular}{@{\extracolsep{1pt}}rl}
			meson $(J^P)$ & \multicolumn{1}{c}{$a_tm$}  \\
			\midrule
			$\pi(0^-)$ & $0.06906(13)$  \\
			$K(0^-)$ & $0.09698(9)$  \\
			$\eta(0^-)$ & $0.10364(19)$  \\
			$\sigma(0^+)$ & $0.1316(9)$ \\
			$\omega(1^-)$ & $0.15541(29)$  \\
			$\eta'(0^-)$ & $0.1641(10)$  \\
			$\phi(1^-)$ & $0.17949(21)$  \\
		\end{tabular}
		\quad\quad\quad
		\begin{tabular}{@{\extracolsep{1pt}}rc}
			threshold & \multicolumn{1}{c}{$a_tE_{thr}$}  \\
		\midrule
		$\pi\omega$ & $0.22447(32)$  \\
		$\pi\pi\eta$ & $0.24176(26)$  \\
		$\pi\phi$ & $0.24855(25)$  \\
		$\pi K \overline{K}$ & $0.26302(18)$  \\
		$\pi\pi\sigma$ & $0.26972(92)$  \\
		$\pi\pi\pi\pi$ & $0.27624(26)$  \\
		$\pi\pi\eta'$ & $0.30222(102)$  \\
		\end{tabular}
\caption{Left: The masses of relevant stable hadrons with uncertainties.
Right: Relevant threshold energies with uncertainties.}
	\label{Tab:masses_thresholds}
\end{table}

\section{Partial waves and operator bases\label{Sec:operator_bases}}


In this study we are principally interested in irreps that contain $J^P=1^{+}$. For irreps at rest, $J^P=1^{+}$ subduces only into $T_1^+$. However, for in-flight irreps, different helicity components of $J^P=1^{+}$ are subduced across multiple irreps as shown in Table II of Ref.~\cite{Dudek:2012gj} -- for example, $\lambda=0$ and $\pm1$ subduce into $A_2$ and $E_2$ respectively for overall momentum $\vec{P}=[001]$. Furthermore, at non-zero momentum parity is no longer a good quantum number and so many irreps contain both $J^+$ and $J^-$, e.g.~$1^+$ and $1^-$. 

We will restrict our attention to $\vec{P}A_2$ in-flight irreps -- these contain subductions of the $\lambda=0$ part of $J^P=1^{+}$ but, because reflection parity $\tilde{\eta}=P(-1)^J$ is a good quantum number for $\lambda=0$, they do \emph{not} contain $J^P=1^{-}$. In contrast, $[001]E_2$ contains $J^P=1^{-}$ as well as \smash{$J^P=1^{+}$} -- the latter gives comparatively lower-lying $J^P=1^{-}$ levels, as seen in Ref.~\cite{Dudek:2012xn}, and so will lead to a dense spectrum of mixed $J^P=1^{+}$ and $1^-$ energy eigenstates.  Considering only $\vec{P}A_2$ allows us to avoid the complication of disentangling the $J^P=1^{+}$ and $1^-$ scattering amplitudes.

The partial-wave content of a pseudoscalar-vector system for irreps $[000]\,T_1^+$ and $\vec{P}\,A_2$ with $|\vec{P}|^2\leq 4 (2\pi/L)^2$ is given in Table~\ref{Tab:PW}.  There we make use of the $\SlJ$ notation for meson-meson scattering, where $2S+1=3$ reflects the unique spin-coupling in pseudoscalar-vector scattering, and $\ell$ is the relative orbital angular momentum. The use of the $\ell-S$ basis, over say the helicity basis, is for convenience; in particular, the threshold behavior of a partial-wave of definite $\ell$ is known.

Table~\ref{Tab:PW} includes cases where two $\SlJ$ constructions appear with the same $J^P$ -- in these cases the scattering matrix is $2 \times 2$ in the case of single meson-meson channel, e.g. for $J^P = 1^+$ scattering of $\pi \omega$, the $t$-matrix is
\begin{equation}
 \mathbf{t} = \begin{pmatrix} 
 t( \piomegaS | \piomegaS\!) &  t( \piomegaS | \piomegaD\!)  \\[1.0ex]
 t( \piomegaS | \piomegaD\!) &  t( \piomegaD | \piomegaD\!), 
 \end{pmatrix}
\end{equation}
where the symmetric nature of the matrix follows from time-reversal invariance.

\begin{table}[tb] 
	{\renewcommand{\arraystretch}{1.35}
		\begin{tabular}{l @{\hskip 2.0ex} l @{\hskip 2.0ex} l @{\hskip 2.0ex} l} 
			\multicolumn{1}{l}{$[000]\,T_1^+$}&  \multicolumn{1}{l}{$[00n]\,A_2$} &  \multicolumn{1}{l}{$[0nn]\,A_2$} &   \multicolumn{1}{l}{$[nnn]\,A_2$} \\  
			\midrule
			& $0^- \, \left( \threePzero \right)$ & $0^- \, \left( \threePzero \right)$ & $0^- \, \left( \threePzero \right)$ \\[1.5ex]
			$1^+ \left( \begin{matrix} \threeSone \\  \threeDone \end{matrix}   \right)$ & $1^+ \left( \begin{matrix} \threeSone \\  \threeDone \end{matrix}   \right)$& $1^+ \left( \begin{matrix} \threeSone \\  \threeDone\end{matrix}   \right)$ & $1^+ \left( \begin{matrix} \threeSone \\  \threeDone \end{matrix}   \right)$ \\[3.5ex]
			& & $2^+ \, \left( \threeDtwo \right)$ & \\[1.5ex]
			& $2^- \left( \begin{matrix} \threePtwo \\  {\color{gray}\threeFtwo} \end{matrix}   \right)$ & $2^- \left( \begin{matrix} \threePtwo \\  {\color{gray}\threeFtwo} \end{matrix}   \right)_{\!\![2]}$ & $2^- \left( \begin{matrix} \threePtwo \\  {\color{gray}\threeFtwo} \end{matrix}   \right)$ \\[4.5ex]
			$3^+ \, \left( \begin{matrix} \threeDthree \\  {\color{gray}\threeGthree} \end{matrix}   \right)$ & $3^+ \, \left( \begin{matrix} \threeDthree \\  {\color{gray}\threeGthree} \end{matrix}   \right)$ & $3^+ \, \left( \begin{matrix} \threeDthree \\  {\color{gray}\threeGthree} \end{matrix}   \right)_{\!\![2]}$ & $3^+ \, \left( \begin{matrix} \threeDthree \\  {\color{gray}\threeGthree} \end{matrix}   \right)_{\!\![2]}$ \\[0.5ex]
		\end{tabular}
	}
	\caption{Partial-wave $J^P(\threelJ)$ content for pseudoscalar-vector scattering in irreps \smash{$\vec{P}\Lambda$} containing $J^P=1^+$, transcribed from Ref.~\cite{Woss:2018irj}. A subscript $[N]$ indicates that this $J^P$ has $N$ embeddings in that irrep.}
	\label{Tab:PW}
\end{table}

The relevant thresholds for the isovector sector with positive $G$-parity are shown in Table~\ref{Tab:masses_thresholds}. In the construction of correlation matrices we utilize two-meson operators resembling $\pi\omega$ and $\pi\phi$ and three-meson operators resembling $\pi\pi\eta$ and {$\pi K \overline{K}$}. All three-meson operators are of the form {$\mathcal{O}_{\mathbb{RM}}^{\dagger}$} corresponding to $\rho\eta$ and $a_0\pi$ for $\pi\pi\eta$--like operators and {$a_0\pi$}, {$K^*\overline{K}$} for {$\pi K \overline{K}$}--like operators.\footnote{The optimised operators $\Omega^\dagger_{\mathbb{R}}$ for $\rho$, $a_0$ and $K^*$ used in $\mathbb{RM}$ operator constructions are determined independently in each relevant irrep using variational analysis with the operator bases that are presented in Appendix~\ref{App:Tables}.} {$K^*\overline{K}$} operators are constructed with definite $G$-parity analogous to the {$K\overline{K}$} operators in Ref.~\cite{Wilson:2015dqa}. For the $\pi\pi\sigma$--threshold, three-meson operators resembling $\rho\sigma$ and $a_1 \pi$ were considered for inclusion. These appear in a relative $P$-wave in the $[000]T_1^+$ and $\vec{P}A_2$ irreps at values of $E^{(2+1)}_{\text{n.i.}}$ that lie far above $\pi\pi\pi\pi$--threshold. Similarly, relevant $\pi\pi\sigma$ non-interacting energies, $E^{(3)}_{\text{n.i.}}$, are far beyond $\pi\pi\pi\pi$--threshold. Although the construction of operators resembling four-mesons could be done analogously to the three-meson operator construction described above, we do not include these in our basis and choose to restrict to energies below the $\pi\pi\pi\pi$--threshold. 

The operator basis used for the $[000]\,T_1^+$ irrep on each lattice volume is presented in Table~\ref{Tab:ops_b1_rest}. Included are all two-meson and three-meson operators corresponding to $E^{(2)}_{\text{n.i.}}$ and $E^{(2+1)}_{\text{n.i.}}$ below $\pi\pi\pi\pi$--threshold.\footnote{There are no $E^{(3)}_{\text{n.i.}}$ below $4 m_\pi$ in $[000]\,T_1^+$.}  The operator lists for $\vec{P}\,A_2$ irreps with $\vec{P} \neq \vec{0}$ are presented in Appendix~\ref{App:Tables} -- we include, as well as all low-lying two-meson operators, also the lowest three-meson ($\mathbb{RM}$) operator in each irrep, with the intention of robustly determining the spectra up to the lowest {$E^{(2+1)}_{\text{n.i.}}$} or {$E^{(3)}_{\text{n.i.}}$} energy. As well as providing many more energy levels with which to constrain the scattering matrix, moving frames are required to determine the sign of the off-diagonal element, {$t\big(\piomegaS|\piomegaD\big)$}, as previously explored for $\pi \rho$ scattering in Ref.~\cite{Woss:2018irj}.

\begin{table}[tb]
\small	
{\renewcommand{\arraystretch}{1.2}
	\begin{tabular}{c @{\hskip 3.0ex} c @{\hskip 3.0ex} c @{\hskip 3.0ex} c}
	\midrule
	$L/a_s$ & 16 & 20 & 24 \\
	\midrule
	& ${22}\times \bar{\psi}\bm{\Gamma}\psi$ & ${22}\times \bar{\psi}\bm{\Gamma}\psi$ & ${22}\times \bar{\psi}\bm{\Gamma}\psi$ \\[0.5ex]
	& $\pi_{[000]}\omega_{[000]}$ & $\pi_{[000]}\omega_{[000]}$ & $\pi_{[000]}\omega_{[000]}$ \\[0.5ex]
	& $\pi_{[000]}\phi_{[000]}$ & $\pi_{[000]}\phi_{[000]}$ &  $\pi_{[000]}\phi_{[000]}$ \\[0.5ex]
	& $\rho_{[000]}\eta_{[000]}$ & $\rho_{[000]}\eta_{[000]}$ & $\rho_{[000]}\eta_{[000]}$ \\[0.5ex]
	& $K^*_{[000]}\overline{K}_{[000]}$ & $K^*_{[000]}\overline{K}_{[000]}$ & $K^*_{[000]}\overline{K}_{[000]}$ \\[0.5ex]
	& & & $\{2\}\pi_{[001]}\omega_{[001]}$  \\[0.5ex]		
	\end{tabular}
}
\caption{$[000]\,T_1^+$ operator basis for each lattice volume, with operators ordered by increasing $E_\text{n.i.}$. The maximum number of single-meson operators, $N$, is denoted by $N \times \bar{\psi}\bm{\Gamma}\psi$; various subsets of these were considered to obtain robust fits. The number in braces, $\{N_\text{mult}\}$, denotes the \emph{multiplicity} of linearly independent two-meson operators if this is larger than one.}
\label{Tab:ops_b1_rest}
\end{table}

In order to estimate the strength of partial-waves with $J\geq 2$ that appear alongside our desired $J^P=1^+$ in $[000]\,T_1^+$ and $\vec{P}\,A_2$, on the largest volume we also computed spectra in irreps $[000]\,E^-$,~$[000]\,T_2^+$,~$[001]\,B_1$ and~$[001]\,B_2$, whose partial-wave content is presented in Table~\ref{Tab:high_pw}. As well as the pseudoscalar-vector partial waves presented in the table, the $[001]B_1$ and~$[001]B_2$ irreps also contain a pseudoscalar-pseudoscalar $J^P=3^-$ ($\prescript{1\!}{}{F}_3$) partial-wave. The operator bases used for these irreps are presented in Appendix~\ref{App:Tables}. 

\begin{table}[tb] 
	{\renewcommand{\arraystretch}{1.35}
		\begin{tabular}{l @{\hskip 2.5ex} l @{\hskip 2.5ex} l @{\hskip 2.5ex} l} 
			\multicolumn{1}{l}{$[000]\,T_2^+$}&  \multicolumn{1}{l}{$[000]\,E^-$}&  \multicolumn{1}{l}{$[001]\,B_1$}&   \multicolumn{1}{l}{$[001]\,B_2$}\\  
			\midrule
			$2^+ \, \left( \threeDtwo \right)$  & & $2^+ \, \left( \threeDtwo \right)$ & $2^+ \, \left(\threeDtwo \right)$ \\[1.5ex]
			& $2^- \left( \begin{matrix} \threePtwo \\  {\color{gray}\threeFtwo} \end{matrix}   \right)$ & $2^- \left( \begin{matrix} \threePtwo \\  {\color{gray}\threeFtwo} \end{matrix}   \right)$ & $2^- \left( \begin{matrix} \threePtwo \\  {\color{gray}\threeFtwo} \end{matrix}   \right)$ \\[3.5ex]
			$3^+ \, \left( \begin{matrix} \threeDthree \\  {\color{gray}\threeGthree} \end{matrix}   \right)$ & & $3^+ \, \left( \begin{matrix} \threeDthree \\  {\color{gray}\threeGthree} \end{matrix}   \right)$ & $3^+ \, \left( \begin{matrix} \threeDthree\\  {\color{gray}\threeGthree} \end{matrix}   \right)$ \\[0.5ex]
		\end{tabular}
	}
	\caption{Partial-wave $J^P(\threelJ)$ content for pseudoscalar-vector scattering in irreps with lowest $J=2$.} 
	\label{Tab:high_pw}
\end{table}

In summary, because we are considering the $G$-parity positive isovector sector, the neutral channels have charge-conjugation $C=-$. The contributing $J^{PC}$ includes our target $1^{+-}$ where we expect a low-lying $b_1$ resonance, which in the quark model would be a $q\bar{q}$ spin-singlet in a $P$-wave. $2^{--}$ and $3^{--}$ are expected to resonate at a somewhat higher energy, corresponding to $\rho_2$, $\rho_3$ resonances which would be spin-triplet $D$-waves in the quark model. Still higher we might have a $3^{+-}$ resonance, $b_3$, as a spin-singlet $F$-wave $q\bar{q}$. $0^{--}$ and $2^{+-}$ are exotic -- they do not appear in the $q\bar{q}$ quark model and previous lattice calculations~\cite{Dudek:2010wm} suggest that they may resonate in the form of \emph{hybrid mesons} at much higher energy. Because they do not resonate, and feature at least a $P$-wave threshold suppression, it follows that we expect all partial waves except $J^P=1^+$ to be small at low energies, and indeed we will find this to be the case below.

\section{Finite-Volume Spectra \label{Sec:finite_volume_spectra}}


\begin{figure*}[htb]
	\centering
		\includegraphics[trim={0cm 0 0cm 0},clip,width=1\textwidth]{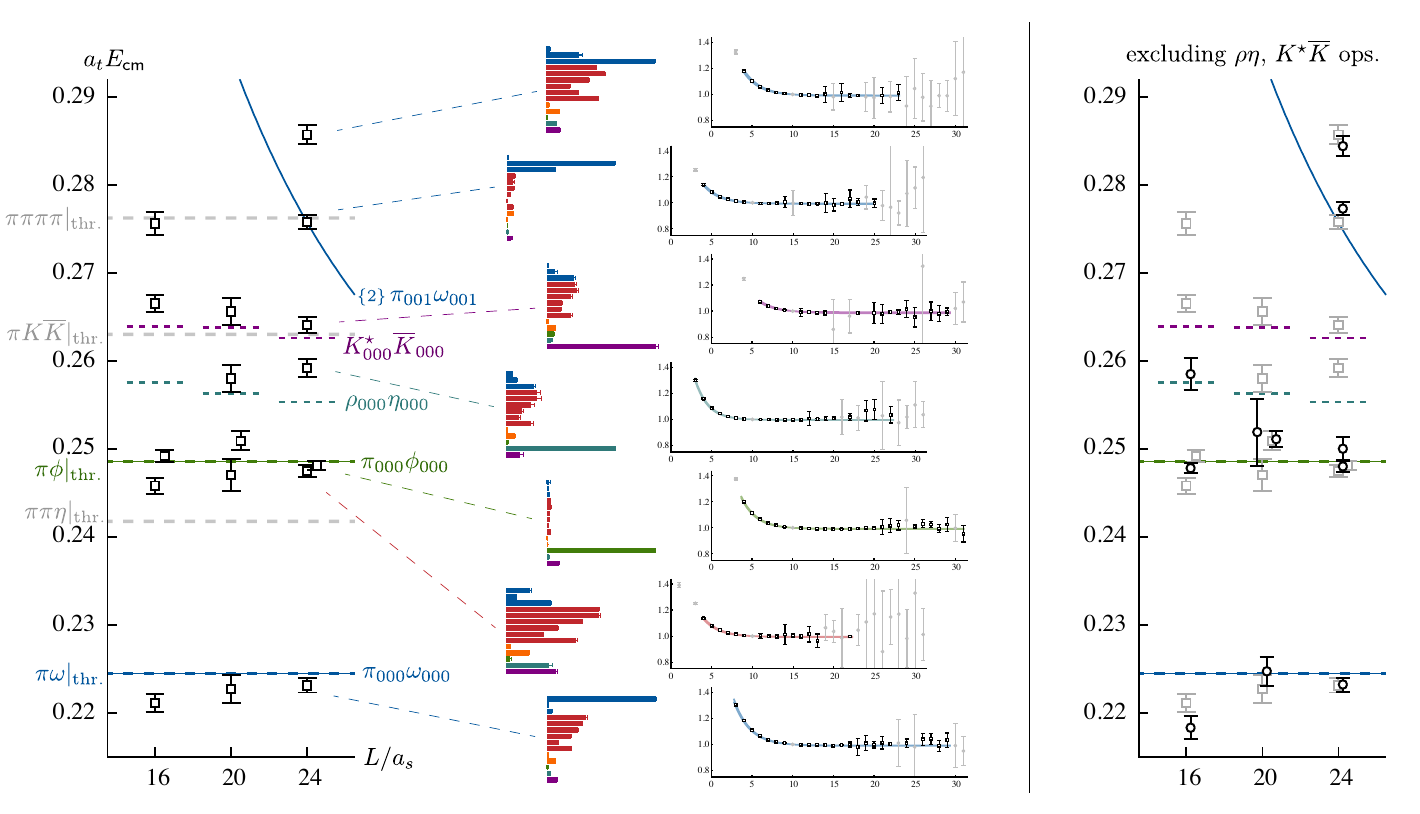}
	    \caption{\textbf{Left:} Finite-volume spectrum in the $[000]T_1^+$ irrep on three lattice volumes. Black points give the energy levels, including statistical uncertainties, from a variational analysis using the operator bases in Table~\ref{Tab:ops_b1_rest}. Solid curves are two-meson non-interacting energies, $a_tE^{(2)}_{\text{n.i.}}$, short dashed horizontal lines are $a_tE^{(2+1)}_{\text{n.i.}}$, and long dashed horizontal lines show the two--, three--, and four--meson thresholds. Multiplicities (if greater than one) are shown as $\{ n \}$. For each energy level on the largest volume, we show the principal correlators, plotted as $\lambda_\mathfrak{n}(t,t_0)\, e^{E_\mathfrak{n}(t-t_0)}$ for $t_0=10 \, a_t$, and histograms showing the operator-state overlap factors, $Z_i^\mathfrak{n}=\braket{ \mathfrak{n}|\mathcal{O}_i^\dagger(0)|0}$, for the $\mathbb{MM}=\pi\omega$~(dark blue), $\pi \phi$~(green) and $\mathbb{RM}=\rho\eta$~(blue-green), $K^*\overline{K}$~(purple) operators along with a sample set of single-meson operators subduced from $J^P=1^+$ (red) and $J^P=3^+$ (orange). The overlaps are normalized such that the largest value for any given operator across all energy levels is equal to one. \textbf{Right:} The spectrum extracted when $\rho\eta$ and $K^*\overline{K}$ operators are excluded from the basis (black) compared with the complete spectrum (gray).
	    }
		\label{Fig:spec-000-T1p}
\end{figure*}

The spectra determined from variational analysis of $[000]\, T_1^+$ correlation matrices on three volumes using the operator bases in Table~\ref{Tab:ops_b1_rest} are presented in Figure~\ref{Fig:spec-000-T1p}. For the largest lattice volume ($L/a_s=24$), the principal correlators and operator-state overlaps, ${Z_i^\mathfrak{n}=\braket{\mathfrak{n}|\mathcal{O}_i^\dagger(0)|0} }$, are also provided for illustration. The typical magnitude of statistical uncertainty on the energy levels, even relatively high in the spectrum, is at the level of a few tenths of a percent. It should be clear from the operator-state overlaps that our operator basis is rather efficiently \mbox{`latching on'} to the finite-volume eigenstates. In some cases an eigenstate has a dominant overlap with only one operator, suggesting that the state closely resembles that particular operator structure.

Consider first the number of energy levels expected below $a_t E_\mathsf{cm} \approx 0.27$ on each volume. In the absence of residual meson-meson interactions we would expect \emph{four} on each lattice volume: one at each of the two $E^{(2)}_{\text{n.i.}}$ corresponding to $\pi_{000}\omega_{000}$ and $\pi_{000}\phi_{000}$, shown as solid horizontal lines in the figure, and one at each of the $E^{(2+1)}_{\text{n.i.}}$ corresponding to $\rho_{000}\eta_{000}$ and $K^*_{000}\overline{K}_{000}$, shown as short dotted horizontal lines. Counting the number of energy levels actually extracted, we find \emph{five}, with an `additional' level appearing near $\pi\phi$ threshold. This may suggest the existence of a narrow resonance, as seen in calculations of the $\rho$ resonance~\cite{Dudek:2012xn,Wilson:2015dqa}, with a mass close to $\pi\phi$ threshold.\footnote{We will later find that the proximity of the resonance to $\pi \phi$ threshold is a coincidence -- this is hinted at by the operator overlaps in Figure~\ref{Fig:spec-000-T1p} as discussed below.} On the largest volume, the effect of there being two ways to construct $\pi_{001}\omega_{001}$ can be seen: two energy levels are found, one very close to the non-interacting energy and one somewhat higher in energy.

In Figure~\ref{Fig:spec-000-T1p} we also present an investigation of the importance of including $\mathbb{RM}$ operators in the basis. The rightmost panel shows the spectrum extracted when $\rho\eta$ and $K^*\overline{K}$ operators are excluded, compared to the spectrum extracted with the full basis -- with the smaller basis we see that typically the levels close to the $\rho\eta$ and $K^*\overline{K}$ `non-interacting' energies are no longer found. The spectrum at lower energies shows only modest discrepancies, except on the smallest lattice volume ($L/a_s=16$) where we might indeed expect the finite-volume effects associated with $\rho\eta$ and $K^*\overline{K}$ to be largest. Finding \mbox{`incorrect'} spectra due to `incomplete' operator bases has been demonstrated in previous works. One example can be seen in Figure 1 of Ref.~\cite{Dudek:2012xn} where including both $\bar{\psi}\bm{\Gamma}\psi$ and $\pi\pi$ operators is shown to be essential in order to robustly determine the $\rho$ spectrum. Figure~\ref{Fig:spec-000-T1p} demonstrates an analogue of this for the case of three-meson operators.

Some qualitative observations about the spectrum can be gleaned from the operator-state overlap factors shown in Figure~\ref{Fig:spec-000-T1p}. The energy level just below $\pi\omega$ threshold on all volumes has significant overlap onto both $\pi_{000}\omega_{000}$ and $\bar{\psi}\bm{\Gamma}\psi$ operators, as one might expect if a $q\bar{q}$-like resonance lies nearby. For the two levels in close proximity to $\pi\phi$ threshold, one appears dominated by $\bar{\psi}\bm{\Gamma}\psi$ operators with some overlap onto $\pi\omega$, $\rho\eta$ and $K^*\overline{K}$ operators, while the other is completely dominated by $\pi_{000}\phi_{000}$. Furthermore, we observe that all other levels have very small overlaps with the $\pi_{000}\phi_{000}$ operator, reflecting the fact that the matrix of correlation functions is approximately block diagonal with respect to $\pi_{000}\phi_{000}$. This suggests that $\pi\phi$ is essentially `decoupled', as might be expected from the `OZI rule' which postulates that $q\bar{q}$ pairs in isoscalar mesons prefer not to annihilate. The states close to the $\rho\eta$ and $K^*\overline{K}$ `non-interacting' energies are observed to have large overlap with $\rho\eta$ and $K^*\overline{K}$ operators respectively. The highest two states shown, near to the $\pi_{001}\omega_{001}$ two-fold degenerate non-interacting energy, differ somewhat in their overlaps. The level shifted up has overlap with both the $\pi_{001}\omega_{001}$ and $\bar{\psi}\bm{\Gamma}\psi$ operators, while the other, which lies on the non-interacting energy, has significant overlap only with the $\pi_{001}\omega_{001}$ operators.

\begin{figure*}[htb]
	\centering
    \includegraphics[trim={0cm 0 0cm 0},clip,width=1\textwidth]{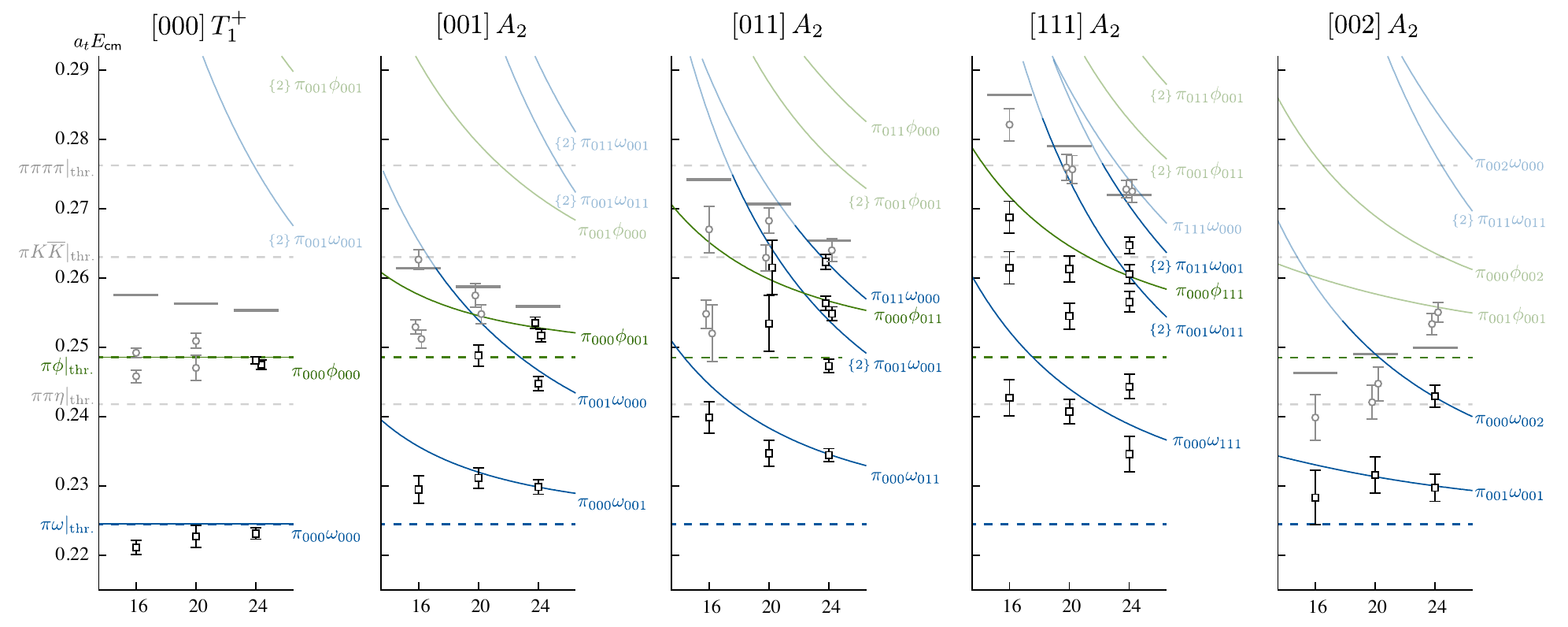}
	\caption{Finite-volume energy levels in the {\cm}-frame for $[000]\,T_1^+$ and $\vec{P}\, A_2$ below the lowest {$E^{(2+1)}_{\text{n.i.}}$} or {$E^{(3)}_{\text{n.i.}}$}.  Black points are used in the scattering analysis in Section~\ref{Sec:Scattering_Analysis} while gray points are excluded from the main analysis as discussed in the text. Solid curves are two-meson non-interacting energies, $a_tE^{(2)}_{\text{n.i.}}$, short solid gray horizontal lines show the lowest {$E^{(2+1)}_{\text{n.i.}}$} or {$E^{(3)}_{\text{n.i.}}$}, and long dashed horizontal lines show the two--, three--, and four--meson thresholds. Multiplicities (if greater than one) are shown as $\{ n \}$. The horizontal axes are in units of $L/a_s$.}
	\label{Fig:spec-A2}
\end{figure*}

In Figure~\ref{Fig:spec-A2}, we present the {\cm}-frame finite-volume spectrum for irreps $[000]\,T_1^+$ and $\vec{P}\,A_2$ on the three volumes, with only those levels found below the lowest $E^{(2+1)}_{\text{n.i.}}$ or $E^{(3)}_{\text{n.i.}}$ shown.\footnote{Errorbars on the energy levels include estimates of systematic uncertainly coming from varying $t_0$ and fitting time ranges, and reasonable variations of the operator basis. Also included is the effect of the uncertainty on the anisotropy which appears when we boost back from the `lab' energy to the {\cm} frame.} Points in grey are levels that prove to be sensitive to the presence of $\rho\eta$, $K^*\overline{K}$ and $a_0 \pi$ operators in the basis, or which are very close to the energy cut-off, and these levels are excluded from the main scattering analysis in Section~\ref{Sec:Scattering_Analysis}. Although we take a conservative approach and exclude these levels, we will find in Section~\ref{Sec:Scattering_Analysis} that they are mainly well described by the scattering amplitudes, and we re-examine these levels in Section~\ref{Sec:Systematic_Analysis}.

For irreps $\vec{P}\, A_2$ with $\vec{P} \neq \vec{0}$, the density of energy levels is much higher than in irreps at rest -- more momentum combinations for two-- and three--mesons with associated $E^{(2)}_{\text{n.i.}}$, $E^{(2+1)}_{\text{n.i.}}$ and $E^{(3)}_{\text{n.i.}}$ lying below the $\pi\pi\pi\pi$--threshold are possible. This can make identifying an `additional' level more challenging in these irreps. However, in the $[111]\,A_2$ irrep we can clearly see an additional energy level on each volume relative to the number expected from counting the non-interacting two-meson energies. We also observe an `avoided level crossing' where the $\pi_{000} \omega_{111}$ non-interacting energy crosses $a_t E_\mathsf{cm} \sim 0.25$, another hint that we may have a narrow resonance in this energy region.

\begin{figure*}[tb]
	\centering
    \includegraphics[trim={0cm 0 0cm 0},clip,width=0.5\textwidth]{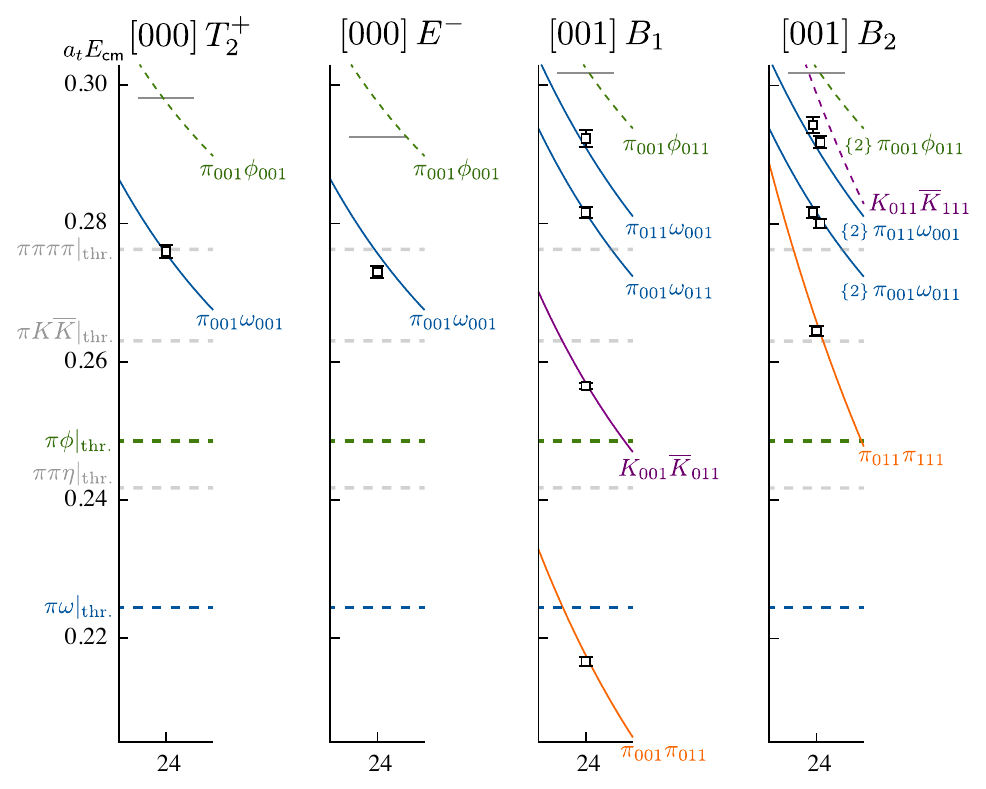}
	\caption{As Figure~\ref{Fig:spec-A2} but for irreps~$[000]\,T_2^+$,~$[000]\,E^-$,~$[001]\,B_1$ and~$[001]\,B_2$ on the largest lattice volume. Dashed curves show non-interacting two-meson energies where the corresponding operator was not included in the basis.}
	\label{Fig:spec-Bs}
\end{figure*}

In Figure~\ref{Fig:spec-Bs}, we present finite-volume spectra on the largest lattice volume for irreps $[000]\,T_2^+$,~$[000]\,E^-$,~$[001]\,B_1$ and~$[001]\,B_2$ where the lowest contributing spin is $J=2$. We observe very little deviation of the extracted energy levels from non-interacting $\pi\omega$ energies, suggesting that the $\pi\omega$ scattering amplitudes in $J\geq 2$ partial-waves are very small in this energy region. We also find levels in $[001]B_1$ and $[001]B_2$ consistent with non-interacting $\pi\pi$ energies and with dominant overlaps onto $\pi\pi$ operators. This is in line with the results of Ref.~\cite{Dudek:2012xn} where the $\pi\pi \{  \prescript{1\!}{}{F}_3  \}$ amplitude ($J^P=3^-$) was found to be consistent with zero in this energy region. We also find a level in $[001]B_1$ consistent with the non-interacting $K\overline{K}$ energy and with dominant overlap onto $K\overline{K}$ operators, suggesting that the opening of the $K\overline{K}$ threshold does not enhance the scattering in $J^P=3^-$.


\section{Scattering Analysis \label{Sec:Scattering_Analysis}}


Finite-volume energy levels and infinite-volume scattering amplitudes are related through a quantisation condition derived by L\"{u}scher~\cite{Luscher:1985dn,Luscher:1986pf,Luscher:1990ux} and extended by many others~\cite{Rummukainen:1995vs,He:2005ey,Christ:2005gi,Kim:2005gf,Guo:2012hv,Hansen:2012tf,Briceno:2012yi,Briceno:2014oea,Gockeler:2012yj} to accommodate the most general case of two particle scattering. The quantisation condition, subduced into lattice irrep $\vec{P}\Lambda$, can be expressed as the determinant of a matrix in the space of intrinsic spin $S = S_1 \oplus S_2$, orbital angular momenta $\ell$, total angular momenta $J$, the embedding number $n$ of a particular partial-wave in lattice irrep $\vec{P}\Lambda$ and hadron-hadron channel $a$. Written compactly, following the notation of Ref.~\cite{Woss:2018irj},
\begin{equation}\label{Eq:luescher}
\text{det}_{\{\ell J n a\}}\big[\bm{1}+ i \bm{\rho} \cdot \bm{t}\cdot \big(\bm{1}+i\bm{\overline{\mathcal{M}}}\big)\big]=0,
\end{equation}
where the determinant over intrinsic spin is trivial in the current case as $S$ takes only the value $1$ for vector-pseudoscalar scattering. Here $\bm{t}(E_{\mathsf{cm}})$ is the scattering t-matrix,\footnote{related to the unitary S-matrix via $\bm{S}=\bm{1}+2i\sqrt{\bm{\rho}}\cdot \bm{t} \cdot \sqrt{\bm{\rho}}$} diagonal in $J$ with components $t_{\ell J a,\ell' J b}$. The diagonal matrix of phase-space factors $\bm{\rho}(E_{\mathsf{cm}})$ has components
\begin{equation*}
\rho_{\ell J a,\ell' J' b} = \delta_{\ell\ell'}  \, \delta_{JJ'}  \, \delta_{ab}\, \frac{2\, k^{(a)} }{E_{\mathsf{cm}}}
\end{equation*}
where $k^{(a)}$ is the {\cm}-frame momentum for hadron-hadron channel $a$,
\[
k^{\!(a)} \!=\! 
\frac{1}{2E_{\mathsf{cm}}}\!
\left[E_{\mathsf{cm}}^2 \!-\!  \left(m^{\!(a)}_{1} \!\!+\! m^{\!(a)}_{2}\right)^2\right]^{\frac{1}{2}} \!
\left[E_{\mathsf{cm}}^2 \!-\!  \left(m^{\!(a)}_{1} \!\!-\! m^{\!(a)}_{2}\right)^2\right]^{\frac{1}{2}}.
\]
Both $\bm{t}$ and $\bm{\rho}$, being infinite-volume quantities, are diagonal in embedding number $n$ and we have dropped this index for brevity. Lastly, $\bm{\overline{\mathcal{M}}}(E_{\mathsf{cm}},L)$ is a matrix of known functions, diagonal in hadron-hadron channel, describing the kinematics of the system in a finite cubic volume, with components $\overline{\mathcal{M}}_{n \ell J a,n' \ell' J' b}$.

The subduced quantisation condition in Equation~\ref{Eq:luescher} reflects the little-group symmetry. The finite-volume spectrum calculated in irrep $\vec{P}\Lambda$ depends upon the various partial-wave amplitudes present in that irrep (see, for example, Tables \ref{Tab:PW} and \ref{Tab:high_pw}). In this way computing spectra in multiple irreps offers additional constraints on scattering. Further details can be found in Appendix~C of Ref.~\cite{Woss:2018irj}.

Equation~\ref{Eq:luescher} is limited to describing two-body scattering -- developments in the pursuit of a corresponding three-body formalism~\cite{Briceno:2012rv,Hansen:2014eka,Hansen:2015zga,Polejaeva:2012ut,Briceno:2017tce,Hammer:2017kms,Mai:2017bge,Doring:2018xxx,Briceno:2018aml,Briceno:2018mlh,Mai:2018djl,Hansen:2019nir,Blanton:2019igq} have seen significant recent progress towards a general quantisation condition, but they are not yet mature at the level where we could apply them in the case considered in this paper. We therefore mainly restrict our attention to the two-body channels, $\pi\omega$ and $\pi\phi$, and in Section~\ref{Sec:Systematic_Analysis} we estimate the systematic effects of neglecting the three-body channels $\pi\pi\eta$ and $\pi K \overline{K}$ in the energy region considered, finding them to be small.


In the case of elastic scattering, where a single meson-meson channel appears in a single partial-wave, the \mbox{$t$-matrix} can be expressed in terms of a single real energy-dependent phase-shift $\delta(E_{\mathsf{cm}})$, where
 \begin{equation}\label{Eq:elastic_t_matrix}
 t(E_{\mathsf{cm}}) = \frac{1}{\rho(E_{\mathsf{cm}})}\, e^{i\delta(E_{\mathsf{cm}})}\sin{\delta(E_{\mathsf{cm}})}.
 \end{equation}
In this case we can invert Eq.~\ref{Eq:luescher} to obtain a one-to-one relation between $E_{\mathsf{cm}}$ and $\delta(E_{\mathsf{cm}})$. Given a set of discrete energy levels below the inelastic threshold, we can hence obtain a set of phase-shift points. In the case considered in this paper, there is never rigorously elastic scattering -- as soon as the $\pi \omega$ threshold opens, in $J^P=1^+$ there are always two coupled partial-waves, $\threeSone$ and $\threeDone$.  However, at low energies the angular momentum suppression of the $D$-wave may make the system effectively elastic in $S$-wave.

For scattering with more than one partial-wave or hadron-hadron channel, there is no longer a one-to-one relation between $E_{\mathsf{cm}}$ and elements of the $t$-matrix and we choose to make progress by parameterizing the energy dependence of $\bm{t}(E_{\mathsf{cm}})$. In order to calculate the scattering amplitudes using Eq.~\ref{Eq:luescher} we follow the successful approach detailed in~\cite{Dudek:2012gj}. In brief, taking an appropriate parameterization of the scattering $t$-matrix, we calculate the energy spectrum in each irrep using Eq.~\ref{Eq:luescher}. By varying the free parameters in the parameterization, we find the best description of the finite volume spectra by minimizing a $\chi^2$, described in Eq.~9 of Ref.~\cite{Dudek:2012xn}, measuring the agreement between finite-volume spectra obtained in the lattice calculations and those found by solving Eq.~\ref{Eq:luescher} with the parameterized $t$-matrix. To ensure that we have not introduced bias by any particular choice of $t$-matrix parameterization, we repeat the analysis for a range of parameterization forms, establishing which features of the resulting amplitudes are robust. 

A very convenient approach to building parameterizations of the $t$-matrix is to work in terms of a real symmetric $K$-matrix, $\bm{K}(s)$, where $s = E_\mathsf{cm}^2$,
\begin{align}\label{Eq:Kmat}
\big[t^{-1}(s)\big]_{\ell J a, \ell' J b} &= 
\frac{1}{\left(2k^{\!(a)}\right)^\ell}
\big[K^{-1}(s)\big]_{\ell J a, \ell' J b}
\frac{1}{\left(   2k^{\!(b)} \right)^{\ell'}} \nonumber \\
&\quad\quad+\delta_{\ell \ell'}  \, I_{ab}(s),
\end{align}
and $I_{ab}(s)=I_a(s)\, \delta_{ab}$ is a matrix diagonal in hadron-hadron channel. Unitarity of the $S$-matrix is guaranteed if $\text{Im}\,I_a(s)=-\rho_a(s)$ above threshold in channel $a$ and zero below. A simple choice is $I_a(s)=-i\rho_a(s)$. Alternatively, the Chew-Mandelstam prescription~\cite{Chew:1960iv} defines $\text{Re}\,I_{a}(s)$ through a dispersive integral featuring $\rho_a(s)$ -- this has improved analytic structure as we transition across thresholds and move away from the real energy axis. A detailed discussion of our implementation can be found in Ref.~\cite{Wilson:2014cna}. 

One parameterization we utilize expresses the components of $\bm{K}^{-1}(s)$ as polynomials in $s$,
\begin{equation}\label{Eq:kinv}
\big[K^{-1}(s)\big]_{\ell J a, \ell' J b} = \sum_{n=0}^N c^{(n)}_{\ell J a, \ell' J b}\,\cdot \, s^n\, ,
\end{equation}
where $\bm{c}^{(n)}$ is a real symmetric matrix. Flexibility in this form comes from varying $N$ and allowing parameter freedom in different combinations of $c^{(n)}_{\ell J a, \ell' J b}$ coefficients.

An alternative approach is to parameterize the components of $\bm{K}(s)$ directly, using a parameterization of the form
\begin{equation}\label{Eq:general_k_matrix}
K_{\ell J a, \ell' J b}(s) = \frac{g_{\ell J a}(s) \, g_{\ell' J b}(s)}{m^2 - s} + \sum_{n=0}^N \gamma^{(n)}_{\ell J a, \ell' J b} \,\cdot \, s^n \, ,
\end{equation}
where $m$ is a real parameter, $g_{\ell J a}(s)$ is some real polynomial in $s$, and $\bm{\gamma}^{(n)}$ is a symmetric matrix of real parameters. These forms assume nothing about a nearby resonance or bound state but the pole can efficiently describe such behavior where it is present. These and similar $K$-matrix parameterizations have been successfully used in previous lattice QCD calculations of resonant and non-resonant scattering~\cite{Woss:2018irj,Moir:2016srx,Briceno:2017qmb,Dudek:2014qha,Wilson:2014cna,Dudek:2016cru,Wilson:2015dqa}.

As an explicit example, one that we will make use of later, consider a $K$-matrix parameterization suitable for describing the dynamically-coupled $J^P=1^+$ channels $\piomegaS$, $\piomegaD$ and $\piphiS$.\footnote{In principle, we should also consider $\pi\phi$ in the $\threeDone$ partial-wave; however, suppression due to the centrifugal barrier factor, compounded with strong OZI suppression of $\pi\phi$, suggests it will be negligibly small and we find later in Section~\ref{Sec:Systematic_Analysis} that the amplitude is consistent with zero in the energy region we consider.} One possible choice, with 7 free parameters, is
\vspace{-0.5cm}
\begin{widetext}
\begin{align}\label{Eq:explicit_kmatrix}
\bm{K}(s) = \nonumber \\
\frac{1}{m^2 - s} 
& \left( \begin{matrix} 
g^2_{\piomegaSsub}  					& g_{\piomegaSsub}\, g_{\piomegaDsub} 	& g_{\piomegaSsub} \, g_{\piphisub} \\  
g_{\piomegaSsub} \, g_{\piomegaDsub} 	& g^2_{\piomegaDsub} 					& g_{\piomegaDsub} \, g_{\piphisub}  \\ 
g_{\piomegaSsub}\, g_{\piphisub}  		& g_{\piomegaDsub} \, g_{\piphisub}  	& g^2_{\piphisub}  
\end{matrix}   \right) 
+ \left( \begin{matrix} \gamma^{(0)}_{\piomegaSsub,\piomegaSsub} & \gamma^{(0)}_{\piomegaSsub,\piomegaDsub}  & 0 \\  \gamma^{(0)}_{\piomegaSsub,\piomegaDsub} & 0 & 0 \\  0 & 0 & \gamma^{(0)}_{\piphisub,\piphisub} \end{matrix}   \right),
\end{align}
\end{widetext}
where this form allows mixing between $\pi\omega$ and $\pi\phi$ channels only through $g_{\piphisub}$.

To include additional partial-waves that contribute as a consequence of the finite-volume but which do not mix in an infinite-volume,~i.e.~those with distinct $J^P$ as seen in Table~\ref{Tab:PW} for irreps $[000]\,T_1^+$ and $\vec{P}\,A_2$, we write the $t$-matrix in block-diagonal form with each block corresponding to a $J^P$. We refer the reader to Ref.~\cite{Woss:2018irj} for more details.

Statistical uncertainties on the scattering parameters and parameter correlations are determined by calculating the second derivatives of the correlated $\chi^2$ at its minimum. We make a conservative estimate of systematic uncertainties on each scattering parameter due to the uncertainties on stable hadron masses and the anisotropy by repeating the $\chi^2$ minimization fitting procedure at all the various combinations of $\xi \pm \delta \xi$ and $m_i \pm \delta m_i$.\footnote{Values of the anisotropy, masses and uncertainties are given in Section~\ref{Sec:lattice_setup}.} For each of these minimizations, we keep the finite-volume energies, $E_{\mathsf{cm}}$, their corresponding uncertainties, $\delta E_{\mathsf{cm}}$, and correlations between energy levels fixed, where
\begin{align}\label{Eq:errs}
a_tE_{\mathsf{cm}}&=f\big(a_tE_{\mathsf{lat}},\xi\big)=\sqrt{ \big(a_tE_{\mathsf{lat}}\big)^2 - \frac{1}{\xi^2}\bigg(\frac{2\pi}{L/a_s}\bigg)^2  \big|\vec{n}\big|^2} \nonumber \\ 
a_t\,\delta\! E_{\mathsf{cm}}&= \sqrt{\bigg(\frac{\partial f}{\partial \big(a_tE_{\mathsf{lat}}\big)    }\bigg)^2   \big(a_t\, \delta\! E_{\mathsf{lat}} \big)^2 +  \bigg(\frac{\partial f}{\partial \xi}\bigg)^2 \delta \xi^2} \, ,
\end{align}
and $E_{\mathsf{lat}}$ is the energy in the lattice frame. For each scattering parameter, the largest change in the central value is quoted as its systematic uncertainty.

Utilizing the approach outlined above, we now determine scattering amplitudes starting with a single partial wave using energy levels below $\pi\phi$ threshold, and progressing to a larger set of partial waves using the full set of energy levels.

\subsection{``Elastic'' $\piomegaS$ scattering \label{Sec:one_channel}}


Below $\pi\phi$ threshold, the kinematically-open hadron channels are the two-body $\pi\omega$ and three-body $\pi\pi\eta$. We expect $\pi\pi\eta$ to become an important channel near the lowest $E^{(2+1)}_{\text{n.i.}}$ where the $\rho$ resonance enhances $\pi\pi$ as discussed in Section~\ref{Sec:three-meson_ops}. Below this energy, we expect the need to have a $P$-wave to get overall $J^P=1^+$ will strongly suppress the amplitude. The lowest {$E^{(2+1)}_{\text{n.i.}}$} in each of the irreps we consider is typically much higher in energy than $\pi\phi$ threshold, and so we will initially propose that we can ignore $\pi\pi \eta$.

In this energy region only slightly above $\pi\omega$ threshold, the centrifugal barrier suppresses contributions of higher-partial waves, $t_{\ell J,\ell' J} \sim k_{\mathsf{cm}}^{\ell+\ell'}$, such that we expect the $\threeDone$ contributions to the coupled $\threeSone$, $\threeDone$ partial-waves to be rather small. Similarly, $\pi\omega$ scattering amplitudes in other partial-waves that appear in these irreps due to the finite-volume, as shown in Table~\ref{Tab:PW}, are expected to be suppressed relative to the $\threeSone$ amplitude and to have no significant resonant enhancement below $\pi \phi$ threshold. It follows that we can attempt an ``elastic'' analysis in terms of pure $\piomegaS \to \piomegaS$ scattering at low energy. 

\begin{figure}[tb]
	\centering
	\includegraphics[trim={0cm 0cm 0 0},clip,width=0.45\textwidth]{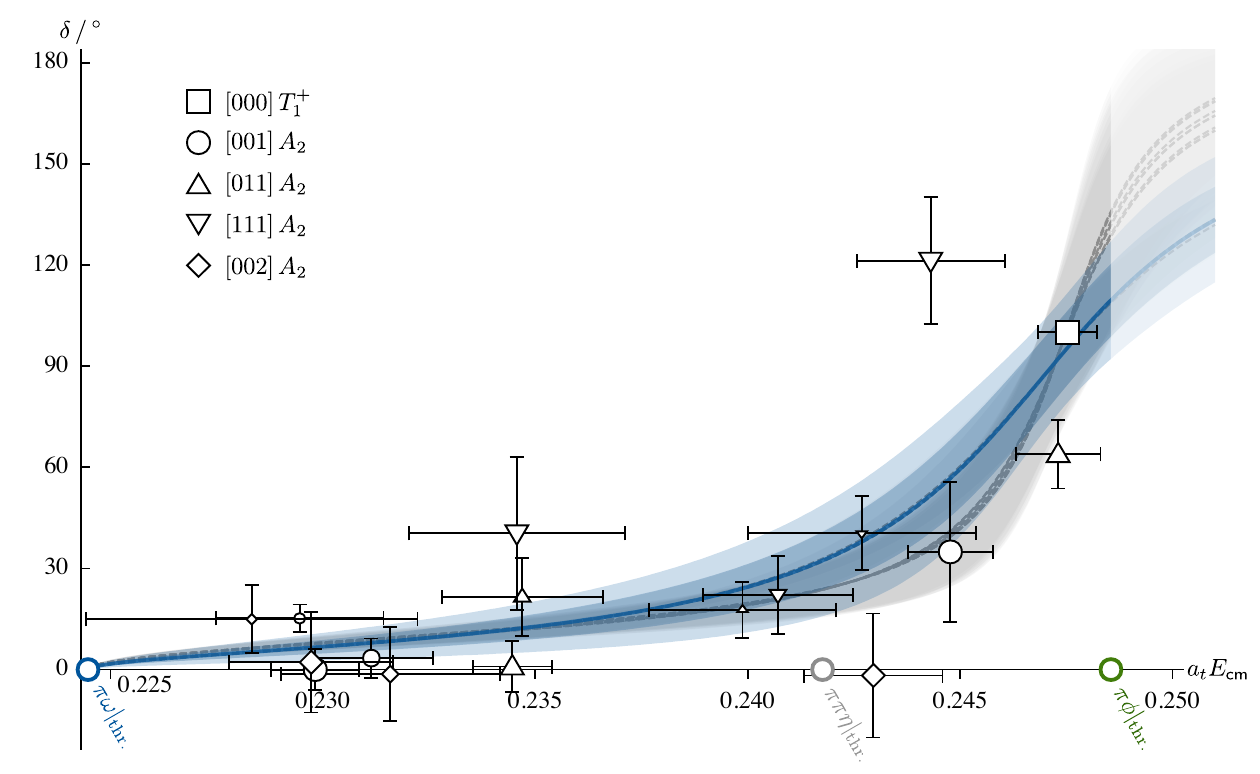}
	\caption{$\piomegaS$ elastic phase-shift assuming no $\threeDone$ amplitude. The blue line shows the reference amplitude given in Eq.~\ref{Fit:A} with the blue bands reflecting the statistical (inner) plus systematic (outer) uncertainty. Gray lines and bands correspond to a range parameterizations presented in Table~\ref{Tab:below_piphi_Swave} of Appendix~\ref{App:Scattering} with only the statistical uncertainties shown. The point size (small to large) of the discrete phase-shift point encodes the lattice volume (small to large).}
	\label{Fig:A_phase_shift}
\end{figure}

We use 20 levels, all at least $1\sigma$ below the $\pi\phi$ threshold. Specifically, for each irrep these correspond to the lowest level on each of the $(L/a_s)=16$ and $20$ volumes and the lowest two levels on the $(L/a_s)=24$ volume\footnote{On the $(L/a_s)=24$ volume, of the two levels close to $\pi\phi$ threshold, the slightly lower level is included but the slightly higher level, essentially a decoupled $\pi\phi$ energy level as indicated by the histograms in Figure~\ref{Fig:spec-000-T1p}, is excluded.}, shown as the black points below $\pi\phi$ threshold in Figure~\ref{Fig:spec-A2}. The resulting discrete phase-shift points are plotted in Figure~\ref{Fig:A_phase_shift}, where we see that the trend is for them to increase toward a value close to $90^\circ$ as they approach the energy cut-off at $\pi \phi$ threshold. This is certainly consistent with a resonance located somewhere near to that energy.

Instead of extracting discrete phase-shift points, we can also fit the spectrum using energy-dependent parameterizations of elastic scattering; a selection of choices which describe the finite-volume spectra well are included as gray curves in Figure~\ref{Fig:A_phase_shift} with the details of the parameterizations presented in Appendix~\ref{App:Scattering}. One description, chosen as a reference amplitude and plotted as the blue curve in Figure~\ref{Fig:A_phase_shift}, is given by,
\begin{equation}\label{Eq:elas_param}
K(s) = \frac{g^2_{\piomegaSsub}}{m^2 - s} \, ,
\end{equation}
using the Chew-Mandelstam prescription for $I(s)$ with $\text{Re}\,I(s=m^2) = 0$ -- see Appendix B of Ref.~\cite{Wilson:2014cna}. The best fit description of the finite-volume spectrum is 
\begin{center}
	\begin{tabular}{rll}
		$m =$ & $(0.2472 \pm 0.0007 \pm 0.0003 ) \cdot a_t^{-1}$   &
		\multirow{2}{*}{ $\begin{bmatrix} 1     &  -0.04  \\[1.3ex]
	    & 1   \end{bmatrix}$ } \\[1.3ex]
		$g_{\piomegaSsub} =$  & $(0.068 \pm 0.009 \pm 0.010) \cdot a_t^{-1}$ &  \\[1.3ex]
		\multicolumn{3}{c}{$\qquad\chi^2/N_{\text{dof}}=\frac{15.1}{20-2}=0.84$,}
	\end{tabular}
\end{center}
\vspace{-0.5cm}
\begin{equation}\label{Fit:A}\end{equation}
where the first uncertainty is statistical and the second is systematic as discussed above, and where the matrix shows the correlations between the parameters.

\subsection{Dynamically-coupled $\piomegaS$, $\piomegaD$ scattering \label{Sec:two_channel}}


Now we relax the assumption of negligible $\piomegaD$ contributions and perform a coupled-channel analysis on the dynamically-coupled $\piomegaS$ and $\piomegaD$ system, restricted to the same low energy region below $\pi\phi$ threshold as in Section~\ref{Sec:one_channel}. Motivated by the suggestion of resonant behavior in the $\piomegaS$ phase-shift in the previous section, we should allow for a resonance to have a $\piomegaD$ coupling as this could significantly enhance the $\piomegaD$ contribution above what might be expected on the basis of angular momentum suppression at threshold.

\begin{figure}[tb]
	\centering
    \includegraphics[trim={0 0 0 0},clip,width=0.5\textwidth]{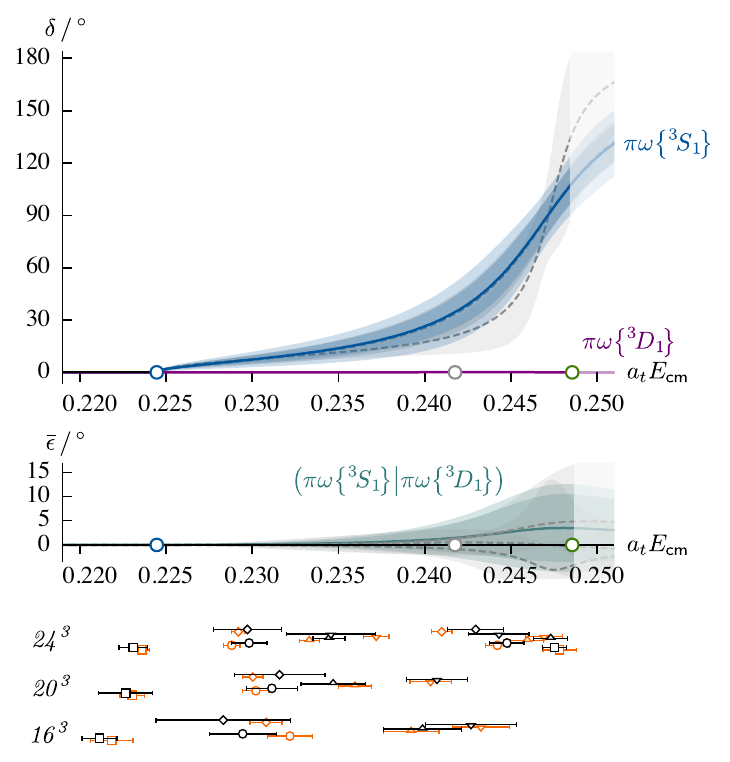}
	\caption{\textbf{Upper}: $\piomegaS$ (blue) and $\piomegaD$ (purple) phase-shifts for the reference amplitude in Eq.~\ref{Fit:B} with the bands reflecting the statistical (inner) plus systematic (outer) uncertainties. In gray are parameterizations given in Table~\ref{Tab:below_piphi_S+Dwave} of Appendix~\ref{App:Scattering} with only statistical uncertainties shown. \textbf{Middle}: As upper but for the mixing-angle, $\bar{\epsilon}(\piomegaS|\piomegaD)$. \textbf{Lower}: Black points are the finite-volume energy levels used to constrain the fit and orange points are the energy levels calculated using Eq.~\ref{Eq:luescher} for the reference amplitude in Eq.~\ref{Fit:B}.}
	\label{Fig:B_phase_shift}
\end{figure}

An example of a two-channel parameterization capable of describing the finite-volume spectra is
\begin{align*}
\bm{K}(s) = 
\frac{1}{m^2 - s} 
& \left( \begin{matrix} 
g^2_{\piomegaSsub}  					& g_{\piomegaSsub}\, g_{\piomegaDsub} \\  
g_{\piomegaSsub} \, g_{\piomegaDsub} 	& g^2_{\piomegaDsub} 				 
\end{matrix}   \right), 
\end{align*}
using the Chew-Mandelstam prescription for $I(s)$ with $\text{Re}\,I(s=m^2) = 0$. The best-fit parameters are found to be
\begin{widetext}
	\begin{center}
		\begin{tabular}{rll}
			$m =$                         & $(0.2471 \pm 0.0007 \pm 0.0004 ) \cdot a_t^{-1}$   &
			\multirow{3}{*}{ $\begin{bmatrix} 1 & -0.04  & 0.00  \\[1.3ex]
				& 1     &  0.49  \\[1.3ex]
				&       & 1   \end{bmatrix}$ } \\[1.3ex]
			$g_{\piomegaSsub} = $                  & $(0.071 \pm 0.011 \pm 0.010) \cdot a_t^{-1}$   & \\[1.3ex]
			$g_{\piomegaDsub} = $ & $(0.45 \pm 0.91 \pm 0.28)\cdot a_t$   & \\[1.3ex]
			\multicolumn{2}{r}{$\chi^2/N_{\text{dof}}=\frac{14.9}{20-3}=0.87$.}
		\end{tabular}
	\end{center}
	\vspace{-1.2cm}
	\begin{equation}\label{Fit:B}\end{equation}
\end{widetext}
The parameters $m$ and $g_{\piomegaSsub}$ are compatible with those of the reference amplitude in Eq.~\ref{Fit:A} and we find $g_{\piomegaDsub}$ to be consistent with zero within uncertainties. In Figure~\ref{Fig:B_phase_shift} we present the $\piomegaS$ and $\piomegaD$ phase-shifts and the $\bar{\epsilon}(\piomegaS|\piomegaD)$ mixing-angle as defined in the Stapp-parameterization~\cite{Stapp:1956mz} and given in Eq.~\ref{Eq:Stapp} of Appendix~\ref{App:Generalised-Stapp}. A number of different $K$-matrix parameterizations were explored and are plotted as the gray curves in Figure~\ref{Fig:B_phase_shift} and listed in Table~\ref{Tab:below_piphi_S+Dwave} of Appendix~\ref{App:Scattering}. We observe that all descriptions exhibit a $\piomegaS$ phase-shift compatible with the behavior seen in Section~\ref{Sec:one_channel}, a $\piomegaD$ phase-shift that is very small, and a mixing-angle that is consistent with zero within a modest uncertainty over this energy range.

\subsection{Coupled $\piomegaS$, $\piomegaD$, $\piphiS$ scattering \label{Sec:three_channel}}


We now consider scattering amplitudes in an energy region up to the $\pi\pi\pi\pi$--threshold. In this region, $\pi\omega$, $\pi\pi\eta$, $\pi\phi$, $\pi K \overline{K}$ and $\pi\pi\sigma$ are all kinematically open, but we expect the three-body channels to have only a small effect. By using only energy levels below the lowest $E^{(2+1)}_{\text{n.i.}}$ or $E^{(3)}_{\text{n.i.}}$ in each irrep, and excluding any energy levels which show significant sensitivity to the presence of $\rho\eta$,~$K^*\overline{K}$ and~$a_0\pi$ operators, we propose that we can effectively neglect the effect of three-body channels. In Section~\ref{Sec:Systematic_Analysis} we will explore possible effects of relaxing this assumption. We proceed with a total of $36$ energy levels -- all the black points shown in Figure~\ref{Fig:spec-A2}.

Both $\pi\omega$ and $\pi\phi$ are vector-pseudoscalar channels dynamically-coupled in $\threeSone$ and $\threeDone$ partial-waves. However, considering the centrifugal barrier for the heavier threshold and the lack of mixing observed in the histograms presented in Figure~\ref{Fig:spec-000-T1p}, we assume that $\pi\phi\big\{\!\threeDone\big\}$ will have negligible impact at low energies. Subsequently, we are left with a system of three coupled channels:~$\piomegaS$, $\piomegaD$ and $\piphiS$. Many other partial-waves can contribute to the finite-volume spectra as can be seen from Tables~\ref{Tab:PW} and~\ref{Tab:high_pw}, but, as discussed in Section~\ref{Sec:operator_bases}, we expect these to be negligibly small and we will explicitly show this in Section~\ref{Sec:Systematic_Analysis}.


To parameterize the energy dependence of the three-channel $t$-matrix, we use $K$-matrices of the form in Eq.~\ref{Eq:general_k_matrix} restricted to linear expansions in $g_{\ell J a}(s)$ and $\bm{\gamma}$ -- the parameterizations used are presented in full in Table~\ref{Tab:below_isobars_params} of Appendix~\ref{App:Scattering}. It should be noted that, while use of the $K$-matrix guarantees unitarity, it does not guarantee good analytic properties. Indeed, we found that some parameterizations, which successfully describe the finite-volume spectra, have $t$-matrix pole singularities at complex energies on the physical sheet. Such poles are forbidden by causality, and these parameterizations must be rejected as giving rise to unphysical solutions. A list of such parameterizations is provided in Table~0.2 in the Supplemental Material and the resulting amplitudes are omitted from the figures in what follows.

A somewhat minimal parameterization,
\begin{widetext}
\begin{align}\label{Eq:C}
\bm{K}(s) = 
\frac{1}{m^2 - s} 
 \left( \begin{matrix} 
g^2_{\piomegaSsub}  					& g_{\piomegaSsub}\, g_{\piomegaDsub} 	& 0 \\  
g_{\piomegaSsub} \, g_{\piomegaDsub} 	& g^2_{\piomegaDsub} 					& 0  \\ 
0 		& 0  	&0 
\end{matrix}   \right) 
+ \left( \begin{matrix} \gamma^{(0)}_{\piomegaSsub,\piomegaSsub} & 0 & 0 \\  0 & 0 & 0 \\  0 & 0 & \gamma^{(0)}_{\piphisub,\piphisub} \end{matrix}   \right),
\end{align}
used with the Chew-Mandelstam prescription with $\text{Re}\,I_{a}(s=m^2) = 0$, proves to be capable of the describing the finite-volume spectra. The best-fit parameters are
\vspace{0.5cm}
	\begin{center}
		\begin{tabular}{rll}
			$m =$                         & $(0.2465 \pm 0.0007 \pm 0.0001 ) \cdot a_t^{-1}$   &
			\multirow{6}{*}{ $\begin{bmatrix} 1 &  -0.05 & 0.05 & -0.01 & -0.23 \\[1.3ex]
				& 1     &  0.70  & -0.54 & -0.06 \\[1.3ex]
				&       & 1   & -0.39 & -0.06 \\[1.3ex]
				&       &   & 1 & 0.22 \\[1.3ex]
				&       &    & & 1 \end{bmatrix}$ } \\[1.1ex]
			$g_{\piomegaSsub} =$                  & $(0.106 \pm 0.007 \pm 0.007) \cdot a_t^{-1}$   & \\[1.1ex]
			$g_{\piomegaDsub} =$                  & $(1.08 \pm 0.47 \pm 0.28) \cdot a_t$   & \\[1.1ex]
			$\gamma^{(0)}_{\piomegaSsub,\piomegaSsub} = $ & $-0.35 \pm 0.19 \pm 0.18 $   & \\[1.1ex]
			$\gamma^{(0)}_{\piphiSsub,\piphiSsub} = $ & $0.90 \pm 0.24 \pm 0.27 $   & \\[1.1ex]
			\multicolumn{3}{c}{$\chi^2/N_{\text{dof}}=\frac{36.8}{36-5}=1.19$.}
		\end{tabular}
	\end{center}
	\vspace{-0.5cm}
	\begin{equation} \label{Fit:C}\end{equation}
\end{widetext}
We found no improvement in the description of the finite-volume spectra by including freedom in $g_{\piphisub}$ and subsequently fixed this parameter to be zero in the reference amplitude.

%

There is no established method to minimally display the $S$-matrix in three-channel scattering. Plotting the real and imaginary parts of the elements of the $S$-matrix contains redundancy as it does not account for the constraints provided by unitarity. Plotting the magnitudes via $\rho_a \rho_b \, \big| t_{ab} \big|^2$ has the advantage of being closely related to a differential cross-section, but discards important phase information. In the two channel case, the Stapp parameterization is minimal with regard to unitarity and reduces to single-channel phase-shifts when the channels decouple, but to our knowledge there is not a generalization to more channels that reduces to the two-channel Stapp parameterization. In Appendix~\ref{App:Generalised-Stapp} we provide such a generalization to $n$-channels where, if $k$ are decoupled, the scattering $S$-matrix naturally block diagonalises into an $(n-k)$ coupled-channel block and a diagonal block containing $k$ decoupled phase-shifts.

\begin{figure}[tb]
	\centering
	\includegraphics[trim={0 0 0 0},clip,width=0.5\textwidth]{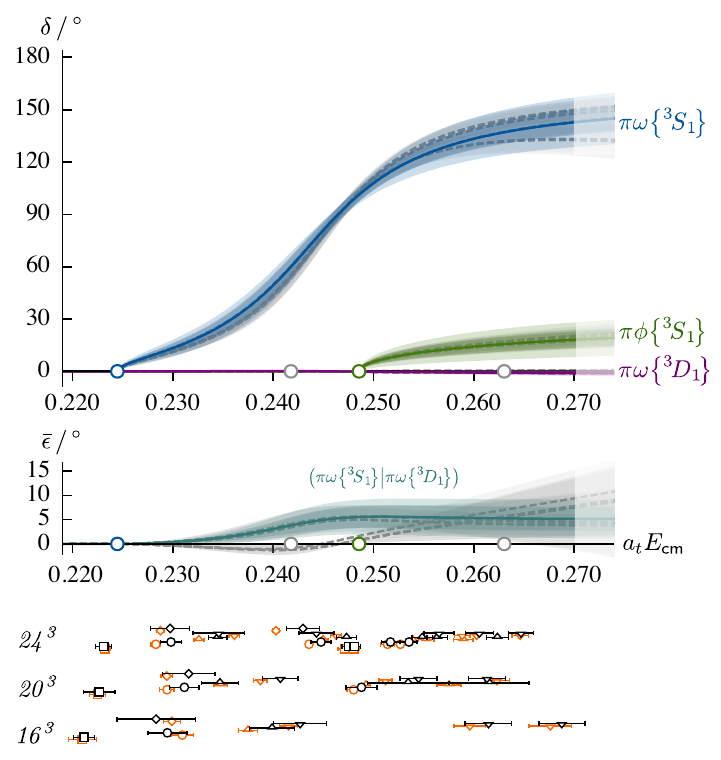}
	\caption{
	\textbf{Upper}: As in Figure~\ref{Fig:B_phase_shift} but for the $\piomegaS$ (blue), $\piomegaD$ (purple) and $\piphiS$ (green) phase-shifts for the reference amplitude in Eqs.~\ref{Eq:C} and \ref{Fit:C}, and for other parameterizations presented in Table~\ref{Tab:below_isobars_params} of Appendix~\ref{App:Scattering} (gray). 
	\textbf{Middle}: As upper but for the mixing-angle $\bar{\epsilon}(\piomegaS|\piomegaD)$. The other mixing-angles, $\bar{\epsilon}(\piomegaS|\piphiS)$ and $\bar{\epsilon}(\piomegaD|\piphiS)$, are extremely small and consistent with zero for all parameterizations and are not plotted. 
	\textbf{Lower}: The energy levels used to constrain the scattering amplitude (black) and their corresponding description by the amplitude in Eqs.~\ref{Eq:C} and \ref{Fit:C} (orange).
	}
	\label{Fig:C_phase_shift_angles}
\end{figure}

The phase-shifts and mixing-angles are plotted in Figure~\ref{Fig:C_phase_shift_angles} for the amplitude in Eqs.~\ref{Eq:C} and \ref{Fit:C} (colored curves) and the many other parameterizations listed in Table~\ref{Tab:below_isobars_params} of Appendix~\ref{App:Scattering} (gray curves). We observe that the behavior of the $\piomegaS$ phase-shift is in close agreement with the results of Section~\ref{Sec:two_channel}, and the $\piomegaD$ phase-shift is once again very small. The $\piphiS$ phase-shift shows a small positive tendency indicative of a weak attraction. The mixing-angle $\bar{\epsilon}(\piomegaS|\piomegaD)$ is small but likely non-zero, while the mixing angles $\bar{\epsilon}(\piomegaS|\piphiS)$ and $\bar{\epsilon}(\piomegaD|\piphiS)$ are around two orders of magnitude smaller and statistically consistent with zero everywhere.

\begin{figure}[tb]
	\centering
    \includegraphics[trim={0 0 0 0},clip,width=0.5\textwidth]{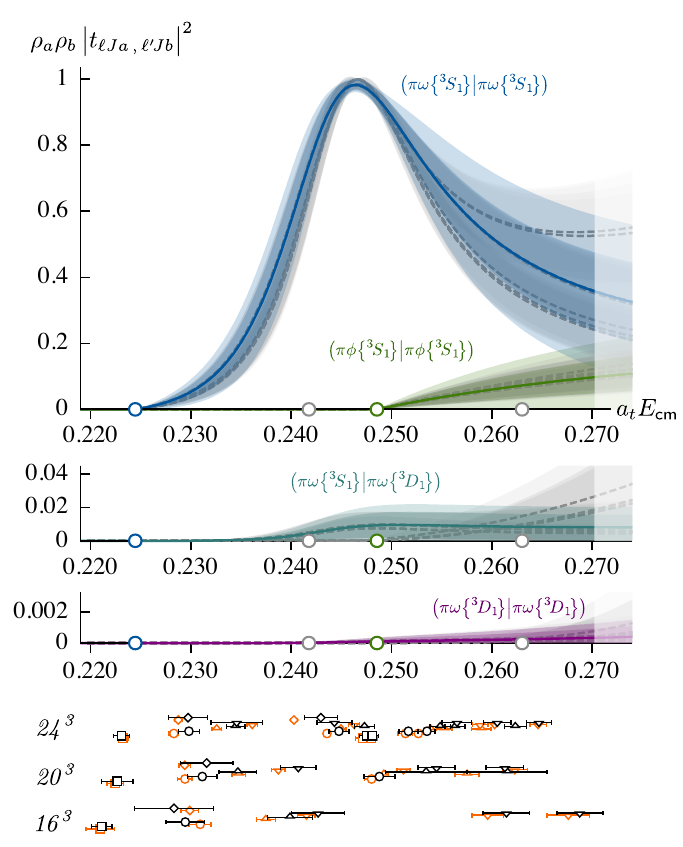}
	\caption{As Figure~\ref{Fig:C_phase_shift_angles} but for $\rho_a \rho_b |t_{\ell J a,\ell' J b}|^2$. Colored curves illustrate the reference amplitude in Eqs.~\ref{Eq:C} and \ref{Fit:C} with bands reflecting the statistical (inner) plus systematic (outer) uncertainty. Other parameterizations presented in Table~\ref{Tab:below_isobars_params} of Appendix~\ref{App:Scattering} are in gray with bands reflecting only the statistical uncertainties. $\rho_a \rho_b |t_{\ell J a,\ell' J b}|^2$ not plotted are significantly smaller than those shown and consistent with zero.}
	\label{Fig:rho_t_sq}
\end{figure}

The same amplitudes are plotted as $\rho_a \rho_b |t_{\ell J a,\ell' J b}|^2$ in Figure~\ref{Fig:rho_t_sq}. We observe a significant bump-like enhancement in the $\piomegaS \!\to\! \piomegaS$ process which would be a canonical indication for a resonance in a scattering cross-section measurement.

\begin{figure*}[tb]
	\centering
\includegraphics[trim={0cm 0 0cm 0},clip,width=1\textwidth]{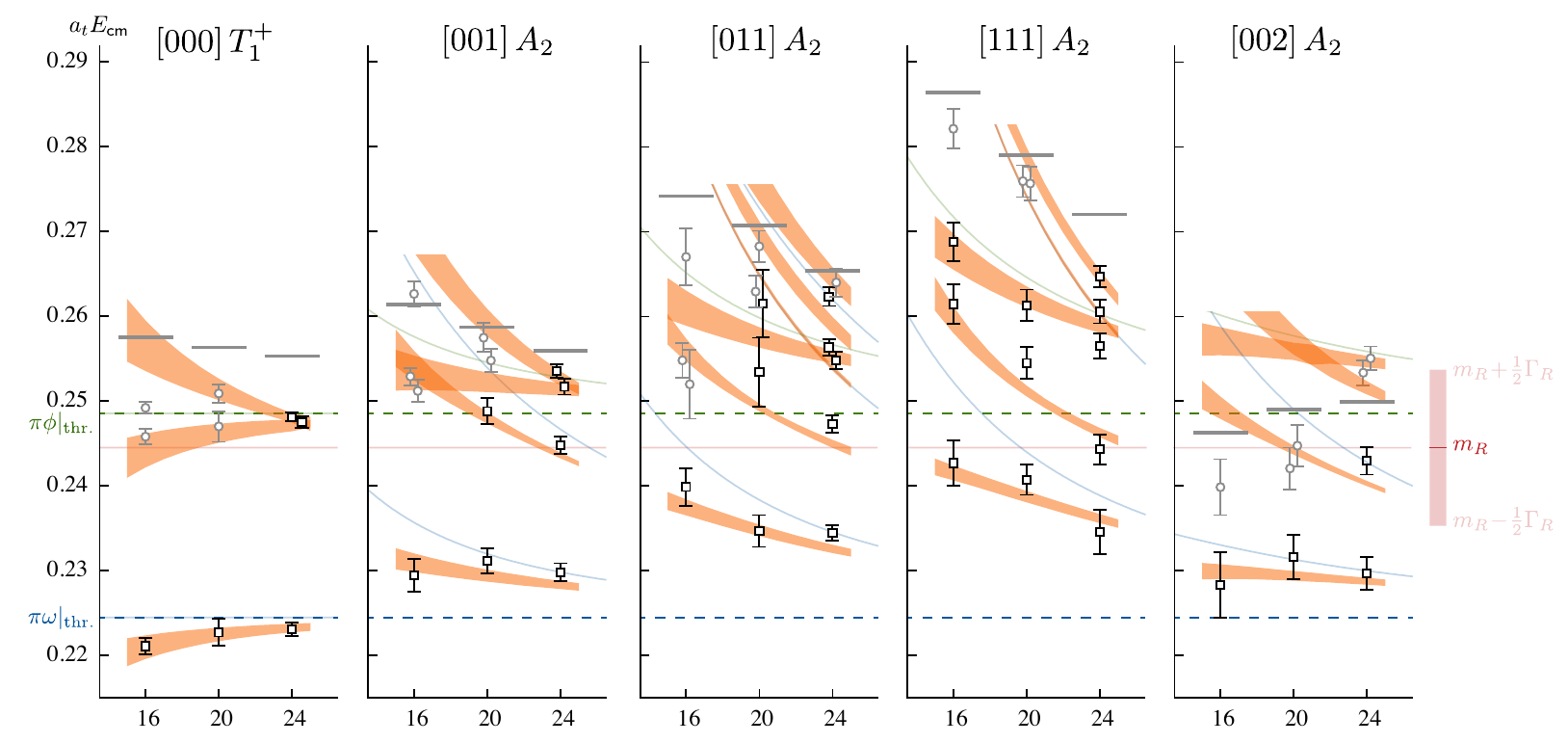}
	\caption{As Figure~\ref{Fig:spec-A2} but including, as orange bands, the energy levels calculated from the reference amplitude in Eqs.~\ref{Eq:C} and \ref{Fit:C} using Eq.~\ref{Eq:luescher} as a function $L/a_s$. The thickness of the bands reflect the combined statistical and systematic uncertainties. The vertical red band on the right of the figure indicates the position of the resonant pole of $m_R$ and width $\Gamma_R$ as determined in Section~\ref{Sec:Pole_Analysis}. The red horizontal line at the resonant mass is shown in each irrep to guide the eye.
	}
	\label{Fig:fv-spec}
\end{figure*}

In Figure~\ref{Fig:fv-spec} we present the energies calculated using Eq.~\ref{Eq:luescher} with the reference amplitude of Eqs.~\ref{Eq:C} and \ref{Fit:C} which, as suggested by the small $\chi^2$, are seen to be in good agreement with the lattice finite-volume energy levels. Notably, for levels \emph{not included} in the fits, shown in gray, the predicted spectra on the $(L/a_s)=20, 24$ volumes appear to be mainly in reasonable agreement, while on the $(L/a_s)=16$ volume there is a larger discrepancy. This may be attributed to more significant contributions from three-meson amplitudes on smaller volumes, further supported by the observation that there is a much larger variation in the spectrum in the $[000]\,T_1^+$ irrep on the smaller volume when three-meson like operators are removed -- see Figure~\ref{Fig:spec-000-T1p}.

A final comment concerns the effect on the scattering results of the uncertainty placed on the anisotropy due to the observed dependence on vector-meson helicity in Section~\ref{Sec:lattice_setup}. Unlike in the $\rho\pi$ isospin-2 case presented in Ref.~\cite{Woss:2018irj}, where the weak nature of the scattering led to the anisotropy uncertainty being the largest systematic effect, here the interactions are strong and the anisotropy uncertainty contributes relatively little as can be seen from the relative sizes of the inner and outer bands in Figures~\ref{Fig:C_phase_shift_angles} and~\ref{Fig:rho_t_sq}.

To summarise, the characteristic ``bump'' we found in the scattering magnitudes in Figure~\ref{Fig:rho_t_sq} and the clearly observed avoided level crossing in the $[111]A_2$ spectrum seen in Figure~\ref{Fig:spec-A2} strongly suggests a resonance. To demonstrate this rigorously, we proceed to determine the pole singularities of our scattering amplitudes.

\section{Pole analysis for coupled-channel amplitudes\label{Sec:Pole_Analysis}}


At each threshold, unitarity necessitates a branch point singularity and the corresponding branch cut divides the complex $s$-plane into two Riemann sheets, so for $n$ open thresholds there are $2^n$ sheets. Riemann sheets can be labelled by the sign of the imaginary component of the \textsf{cm}-frame momentum $k^{(a)}$ in each hadron channel $a$. We identify the \textit{physical} sheet, where physical scattering occurs just above the real energy axis, as having $\text{Im}(k^{(a)}) > 0$ for all $a$. Sheets with other sign combinations are referred to as \textit{unphysical}, and it is on these sheets that \emph{pole singularities} corresponding to resonances lie, in complex-conjugate pairs, off the real energy axis. Poles off the real axis on the physical sheet indicate causality violating amplitudes and signal an unacceptable description of the scattering process.

For poles off the real axis, we define the real and imaginary parts of the pole singularity at $s=s_0$ in terms of the mass $m_R$ and the width $\Gamma_R$ of a resonance respectively, by $\sqrt{s_0}=m_R \pm \frac{i}{2} \Gamma_R$. For narrow resonances, with a single dominant decay mode, these definitions of the resonance mass and width agree well with the location and full-width at half-maximum of the ``bump'' seen in scattering cross sections. The advantage of associating the pole singularity with the resonance is that this definition is still useful in complicated coupled-channel cases, such as those seen in the lattice calculations of the $a_0$~\cite{Dudek:2016cru} and $f_0$~\cite{Briceno:2017qmb}, where the resonance does not appear as a clear isolated bump for real energies. 

In the current case, the hadron-hadron channels $\pi\omega$ and $\pi\phi$ lead to four sheets, $( \text{sign}(\text{Im}\,k_{\pi\omega}),\, \text{sign}(\text{Im}\,k_{\pi\phi})) = \big\{ \mathsf{I}(+,+),\, \mathsf{II}(-, +),\, \mathsf{III}(-,-),\, \mathsf{IV}(+,-) \big\}$.  Close to the $\pi\phi$ threshold, all of sheets $\mathsf{II}$ (lower half-plane), $\mathsf{III}$ (lower half-plane) and $\mathsf{IV}$ (upper half-plane) are close to physical scattering. A single resonance can appear as a pole in slightly different positions on multiple sheets -- some discussion of this in the context of a simple coupled-channel amplitude model can be found in Ref.~\cite{Dudek:2016cru}.

For complex energies close to a pole singularity at $s_0$, the scattering $t$-matrix can be written in the factorised form
\begin{equation}\label{Eq:t-mat-pole}
t_{\ell J a, \ell' J b}(s \sim s_0) \sim \frac{c_{\ell J a}\,c_{\ell' J b}}{s_0-s},
\end{equation}
where the complex valued couplings $c_{\ell J a}$ reflect the strength of the resonance coupling to channel $a\{^3\ell_J\}$. For each coupled hadron-hadron channel, the coupling is determined only up to a sign which gives no change to the physics. In the current case this leads to a sign ambiguity between the $\pi\omega$ and $\pi\phi$ couplings, but conversely the relative sign between the $\threeSone$ and $\threeDone$ partial-waves in $\pi\omega$ \emph{can} be unambiguously determined and physically would lead to different angular decay shapes depending on its value. In Ref.~\cite{Woss:2018irj}, it was shown that in a finite volume, moving-frame spectra are required to constrain this sign.



For each amplitude parameterization we considered, using the best-fit values of parameters, we perform a search across all Riemann sheets over a large range of complex $s$, finding any pole singularities present and determining the couplings by factorizing the residue of the pole. Uncertainties on the pole positions and couplings are estimated by appropriately propagating through the uncertainties and correlations on the fit parameters. For the case of the reference amplitude presented in Eqs.~\ref{Eq:C} and \ref{Fit:C}, poles were found in complex conjugate pairs on sheet $\mathsf{II}$ at
\begin{align}\label{Eq:poles}
a_t\sqrt{s_0}_\mathsf{II} &= 0.2435(13)(10) \pm \tfrac{i}{2} \, 0.0175(20)(19), 
\end{align}
where the first uncertainty is statistical and the second is systematic. A complex conjugate pair of poles was also found on sheet $\mathsf{III}$ in agreement with Eq.~\ref{Eq:poles} up to the precision shown. The couplings for the pole in the lower half-plane are  
\begin{align}\label{Eq:Couplings_II_III}
a_t c(\piomegaS)_\mathsf{II} &= 0.106(6)(6)\exp[-i\,\pi\, 0.078(28)(26) ] \nonumber \\
a_t c(\piomegaD)_\mathsf{II} &= 0.010(4)(3)\exp[-i\,\pi\, 0.181(26)(24) ], 
\end{align}
and {$c(\piphiS)_\mathsf{II}$} is exactly zero, a result of the choice of reference amplitude. Considered as a ratio we have
\begin{align}
\Big|c(\piomegaD)_\mathsf{II}/c(\piomegaS)_\mathsf{II} \Big| &= 0.091(37)(20) \nonumber \\[0.4ex]
\arg \big[{c(\piomegaD)_\mathsf{II}/c(\piomegaS)_\mathsf{II}} \big] &= {-\pi\, 0.103(26)(24)}. \nonumber
\end{align}
That the poles on sheets $\mathsf{II}$ and $\mathsf{III}$ are in essentially the same position is a consequence of the $\pi\phi$ channel being almost completely decoupled from the $\pi\omega$ channel as discussed in Section~\ref{Sec:finite_volume_spectra}. 

\begin{figure}[htb]
	\centering
	\includegraphics[trim={0cm 0cm 0cm 0cm},clip,width=0.5\textwidth]{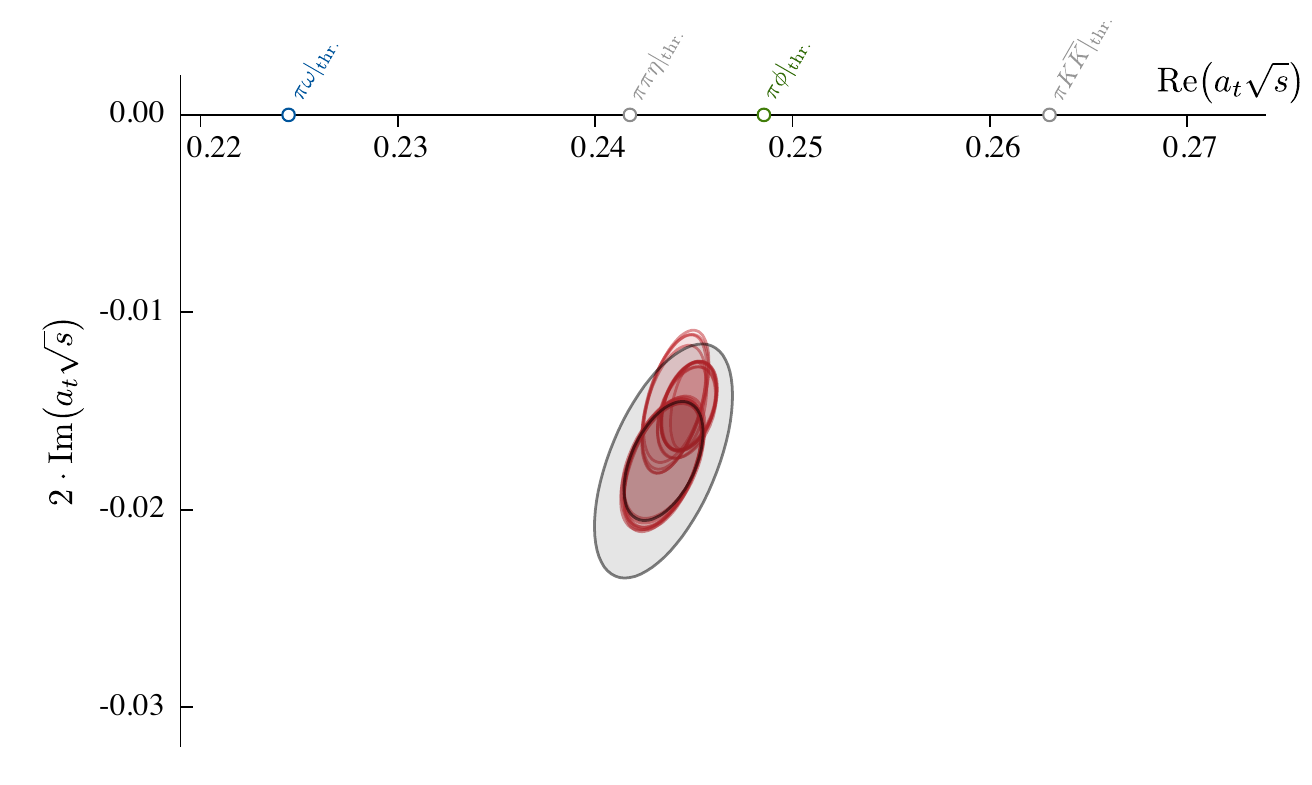}
	\includegraphics[trim={0cm 0cm 0cm 0cm},clip,width=0.5\textwidth]{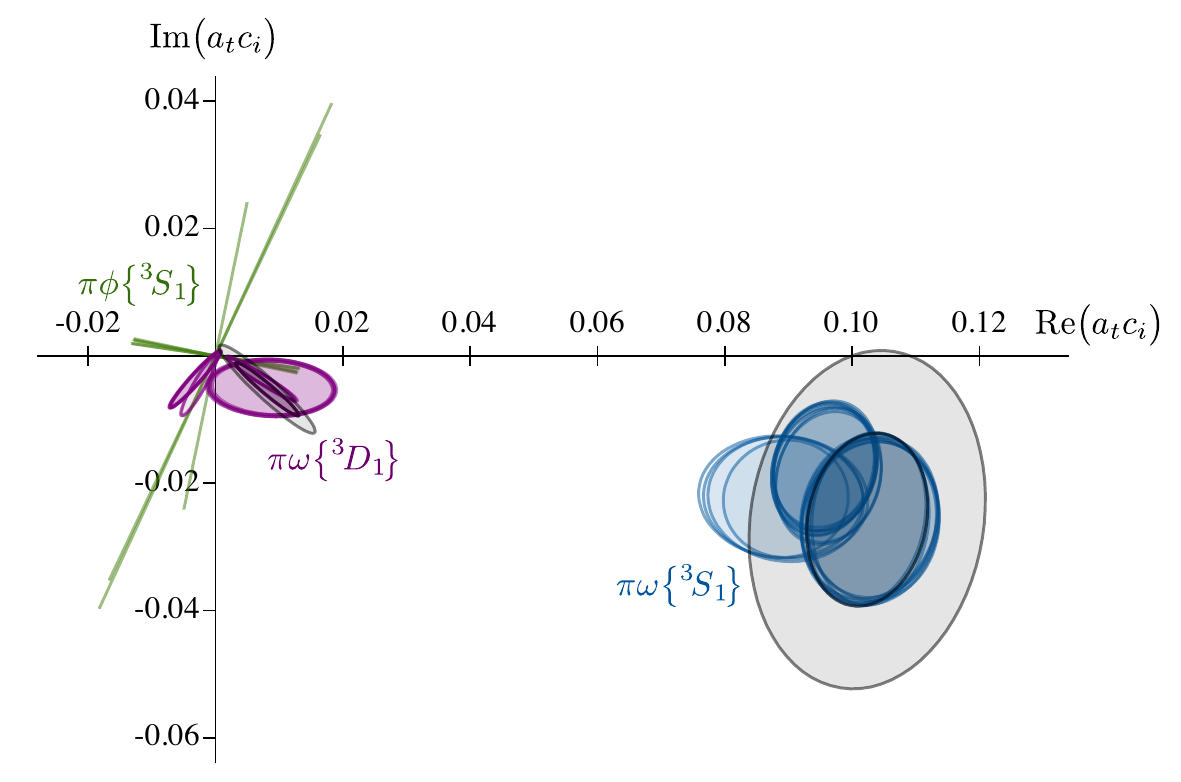}
	\caption{\textbf{Top}: Lower half-plane sheet $\mathsf{II}$ poles. Red ellipses reflect the statistical uncertainties, oriented to account for correlations between the real and imaginary parts, for poles from all the parameterizations shown in Table~\ref{Tab:below_isobars_params} of Appendix~\ref{App:Scattering}. Black ellipses correspond to the reference amplitude in Eq.~\ref{Fit:C} reflecting the statistical (inner) plus systematic (outer) uncertainties. \textbf{Bottom}: As top but for the corresponding couplings, $c(\piomegaS)_\mathsf{II}$ (blue), $c(\piomegaD)_\mathsf{II}$ (purple) and $c(\piphiS)_\mathsf{II}$ (green). Black ellipses again correspond to the couplings of the reference amplitude in Eq.~\ref{Fit:C} where $c(\piphiS)_\mathsf{II}=0$.}
	\label{Fig:poles}
\end{figure}

For each three-channel parameterization presented in Table~\ref{Tab:below_isobars_params} of Appendix~\ref{App:Scattering}, we found poles and couplings broadly consistent with those given above. We show these in Figure~\ref{Fig:poles}, observing that the scatter over different parameterizations is in this case not significantly larger than the uncertainty on the reference amplitude.

\section{Systematic Tests \label{Sec:Systematic_Analysis}}


To test the robustness of the extracted scattering amplitudes and the determination of the resonant pole and couplings, we consider two sources of potential systematic uncertainties due to possibilities we have so far neglected. First, we examine the partial-waves that mix as a consequence of the finite-volume, which we neglected based on observations discussed in Section~\ref{Sec:finite_volume_spectra}, and the $\pi\phi\{\threeDone\!\}$ amplitude which we asserted was negligible. Second, we examine the dependence of the energy levels on the $\pi\omega\{\threeDone\}$, $\pi\omega\{\threePzero\}$ and $\pi\omega\{\threePtwo\}$ parameters to demonstrate that we are able to constrain these amplitudes. Lastly, we make a crude estimate of the possible size of effects due to the neglected three-body channels.

\subsection{Additional Partial-Waves}\label{Subsec:additional_pw}
We first consider the $\pi\omega\{\threePzero\}$ and $\pi\omega\{\threePtwo\}$ amplitudes that enter in the $\vec{P}\,A_2$ irreps as shown in Table~\ref{Tab:PW}. Since a $P$-wave has less threshold suppression than a $D$-wave, we might expect these waves to be at least as important as $\piomegaD$, though they are not expected to be resonant at such low energies.
Augmenting the reference amplitude as defined in Eq.~\ref{Eq:C}, we allow a non-zero amplitude in the $\pi\omega\{\threePzero\}$ and $\pi\omega\{\threePtwo\}$ channels by including a constant $\gamma$-term for each in the $K$-matrix and for these additional channels we set $\text{Re}\,I_{a}(s=(m_\pi \!+\! m_\omega)^2) = 0$ in the Chew-Mandelstam phase-space. The resulting $t$-matrix is block diagonal in $J^P$ reflecting the fact that this mixing is a result of the reduced symmetry on the lattice. We fit to the same 36 energy levels as in Section~\ref{Sec:three_channel} and, allowing all parameters to vary, find
\begin{center}
	\begin{tabular}{rll}
		$m =$                         & $(0.2466 \pm 0.0007) \cdot a_t^{-1}$   & \\[1.1ex]
		$g_{\piomegaSsub} =$                  & $(0.105 \pm 0.007) \cdot a_t^{-1}$   & \\[1.1ex]
		$g_{\piomegaDsub} =$                  & $(1.12 \pm 0.46) \cdot a_t$   & \\[1.1ex]
		$\gamma^{(0)}_{\piomegaSsub,\piomegaSsub} = $ & $-0.34 \pm 0.19 $   & \\[1.1ex]
		$\gamma^{(0)}_{\piomegaSsub,\piphiSsub} = $ & $0.79 \pm 0.25 $   & \\[1.1ex]
		$\gamma^{(0)}_{\pi\omega\{\!\threePzero\!\},\pi\omega\{\!\threePzero\!\}} = $ & $(-8 \pm 21 ) \cdot a_t^2$   & \\[1.1ex]
		$\gamma^{(0)}_{\pi\omega\{\!\threePtwo\!\},\pi\omega\{\!\threePtwo\!\}} = $ & $(-10 \pm 12) \cdot a_t^2$   & \\[1.1ex]
		\multicolumn{3}{c}{$\chi^2/N_{\text{dof}}=\frac{34.4}{36-7}=1.19$  \,,}
	\end{tabular}
	\begin{equation} \label{Fit:D}\end{equation}
\end{center}
where correlations between the $\piomegaS$, $\piomegaD$ and $\piphiS$ parameters are compatible with those shown in Eq.~\ref{Fit:C}, and correlations between these and $\pi\omega\{\threePzero\}$ and $\pi\omega\{\threePtwo\}$ parameters are small. We observe that the amplitudes in both $\pi\omega\{\threePzero\}$ and $\pi\omega\{\threePtwo\}$ are consistent with zero. A similar approach allowing for $\pi\omega\{\threeDtwo\}$ and $\pi\omega\{\threeDthree\}$ parameter freedom finds no evidence for large amplitudes as one would expect given the larger angular momentum suppression and lack of low-energy resonances with $J^{PC}=2^{+-}$ and $3^{+-}$. 

In order to investigate the possible effect of the previously excluded $\pi\phi\{\threeDone\}$, we take the reference amplitude in Eq.~\ref{Eq:C} and extend it to include a constant diagonal $\gamma$-term in $\pi\phi\{\threeDone\}$ in the $K$-matrix.  Once again fitting to the 36 energy levels and allowing all parameters to vary, we find the $\pi\phi\{\threeDone\}$ parameter to be consistent with zero, as expected, with all other parameters compatible with those presented in Eq.~\ref{Fit:C}.

\subsection{Spectrum dependence on $\pi\omega\{^3P_0\}$, $\pi\omega\{^3P_2\}$ and $\piomegaD$}
\begin{figure*}[tb]
		\centering
	    \includegraphics[trim={0 0 0 0},clip,width=1\textwidth]{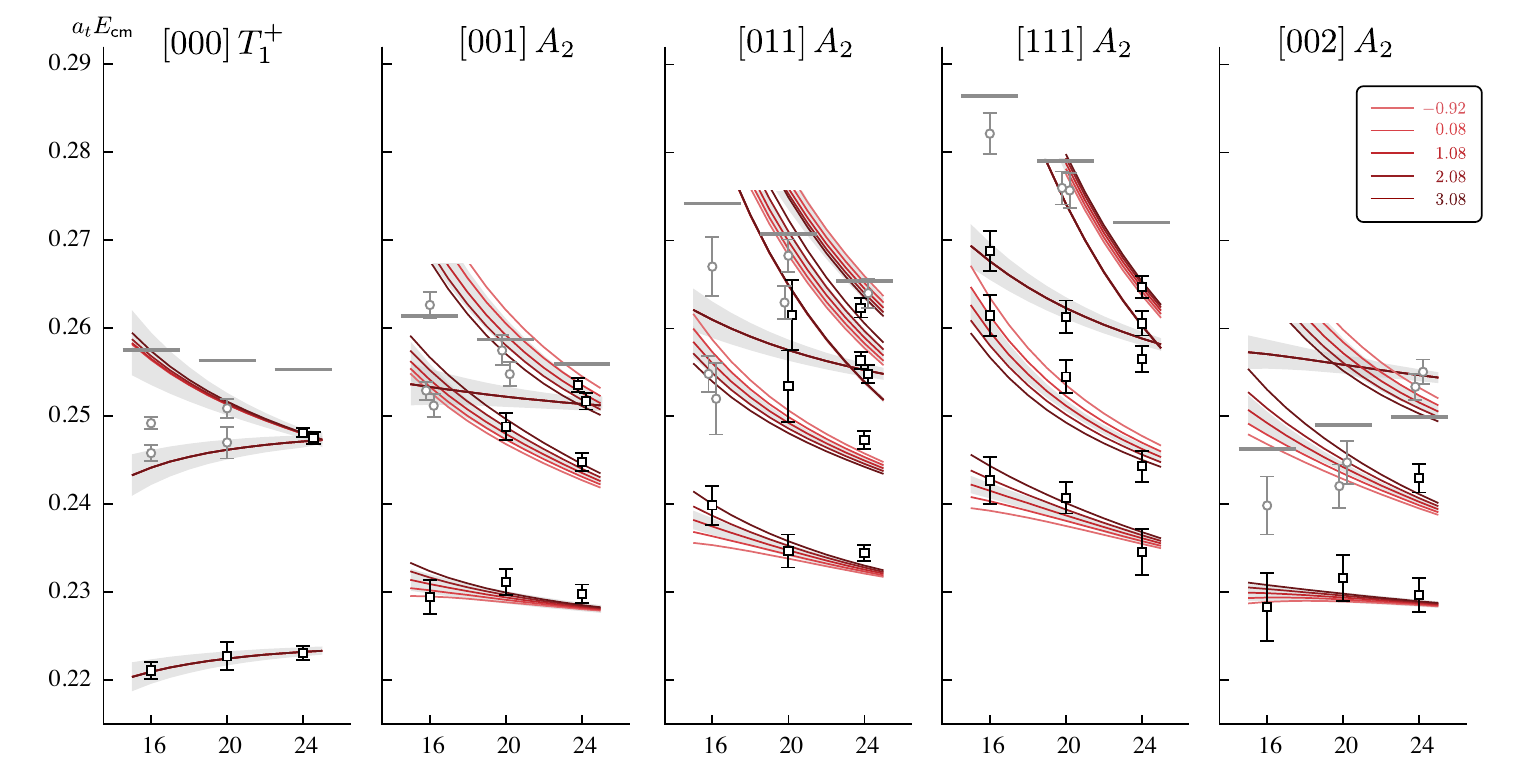}
	\caption{Sensitivity of the finite-volume spectra to $g_{\pi\omega\{^3D_1\}}$. Lighter to darker red curves reflect smaller to larger values of $g_{\pi\omega\{^3D_1\}}$ as shown in the key. The central curves corresponds to $g_{\pi\omega\{^3D_1\}}=1.08$, i.e.~the mean value in the reference amplitude in Eq.~\ref{Fit:C}. The gray bands reflect the combined statistical and systematic uncertainties of Eq.~\ref{Fit:C}. The horizontal axes are in units of $L/a_s$.}
	\label{Fig:g3d1_var}
\end{figure*}

It is worth illustrating at this stage how particular energy levels in the finite-volume spectra depend upon the strength in $\piomegaD$ and the $\pi\omega$ $P$-waves. For $\piomegaD$ this is shown in Figure~\ref{Fig:g3d1_var}, where the curves present the finite-volume energy spectrum for the reference amplitude in Eqs.~\ref{Eq:C} and \ref{Fit:C}, varying the value of $g_{\piomegaDsub}$ while keeping all other parameters fixed. In each irrep, we see a level near the lowest $\pi\phi$ non-interacting energy which appears to be independent of the value of $g_{\piomegaDsub}$, as expected given the near complete decoupling of $\pi\phi$. Most other levels show significant dependence on $g_{\piomegaDsub}$, indicating that the lattice computed levels are providing constraint on the $D$-wave strength, but there are some notable exceptions.  In irreps $[011]\,A_2$ and $[111]\,A_2$, there are levels observed to be consistent with the two-fold degenerate non-interacting $\pi\omega$ energies, which show no visible dependence on $g_{\piomegaDsub}$. 

\begin{figure*}[tb]
	\centering
	\includegraphics[trim={0 0cm 0 0cm},clip,width=1\textwidth]{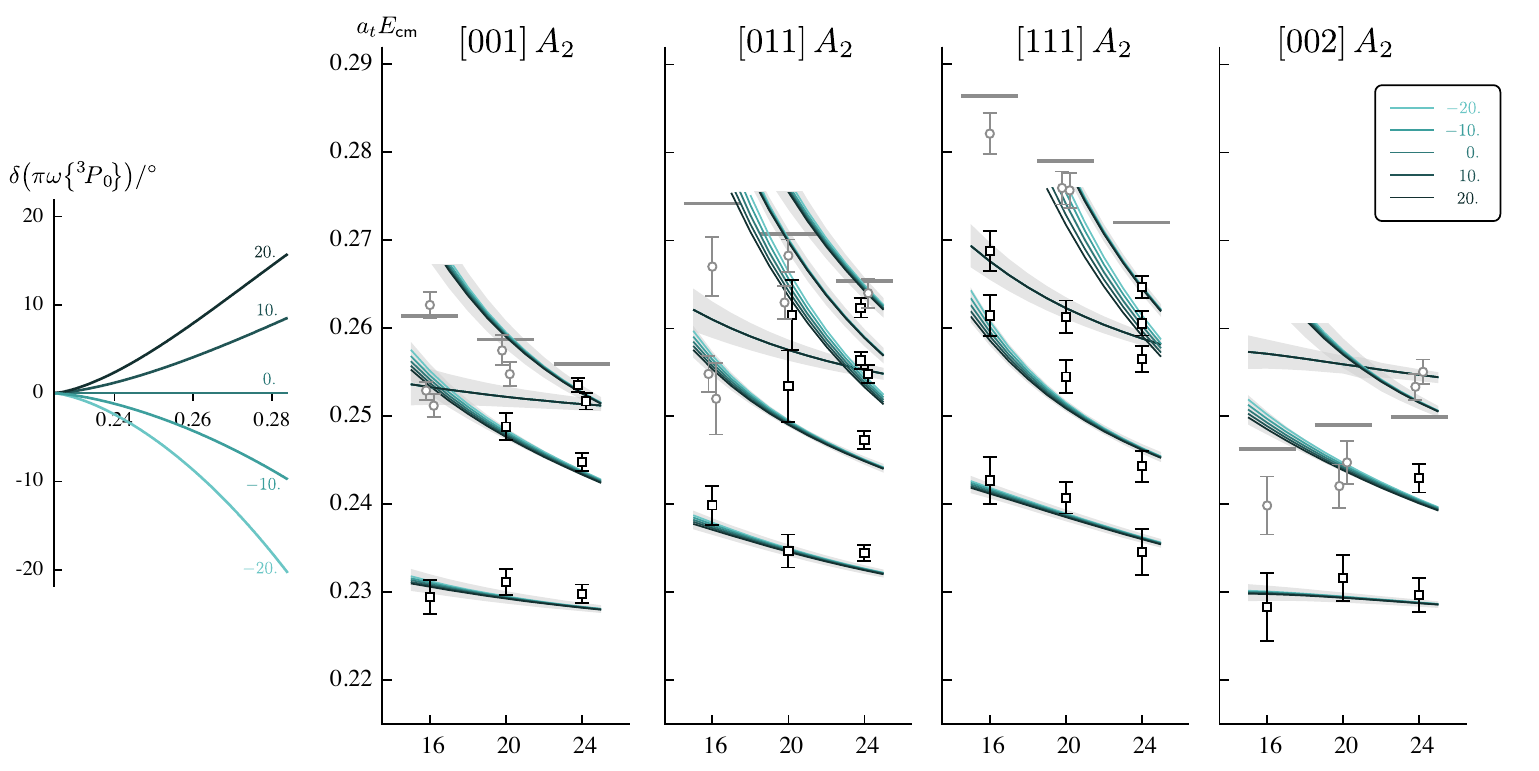}
	\caption{As Figure~\ref{Fig:g3d1_var} but for varying $\gamma^{(0)}_{\pi\omega\{\!\threePzero\!\},\pi\omega\{\!\threePzero\!\}}$. The central curves corresponds to $\gamma^{(0)}_{\pi\omega\{\!\threePzero\!\},\pi\omega\{\!\threePzero\!\}}=0$. The phase-shifts on the left reflect the strengths of the $\pi\omega\{\threePzero\}$ amplitudes.}
	\label{Fig:3p0_var}
\end{figure*}

\begin{figure*}[tb]
	\centering
	\includegraphics[trim={0 0cm 0 0cm},clip,width=1\textwidth]{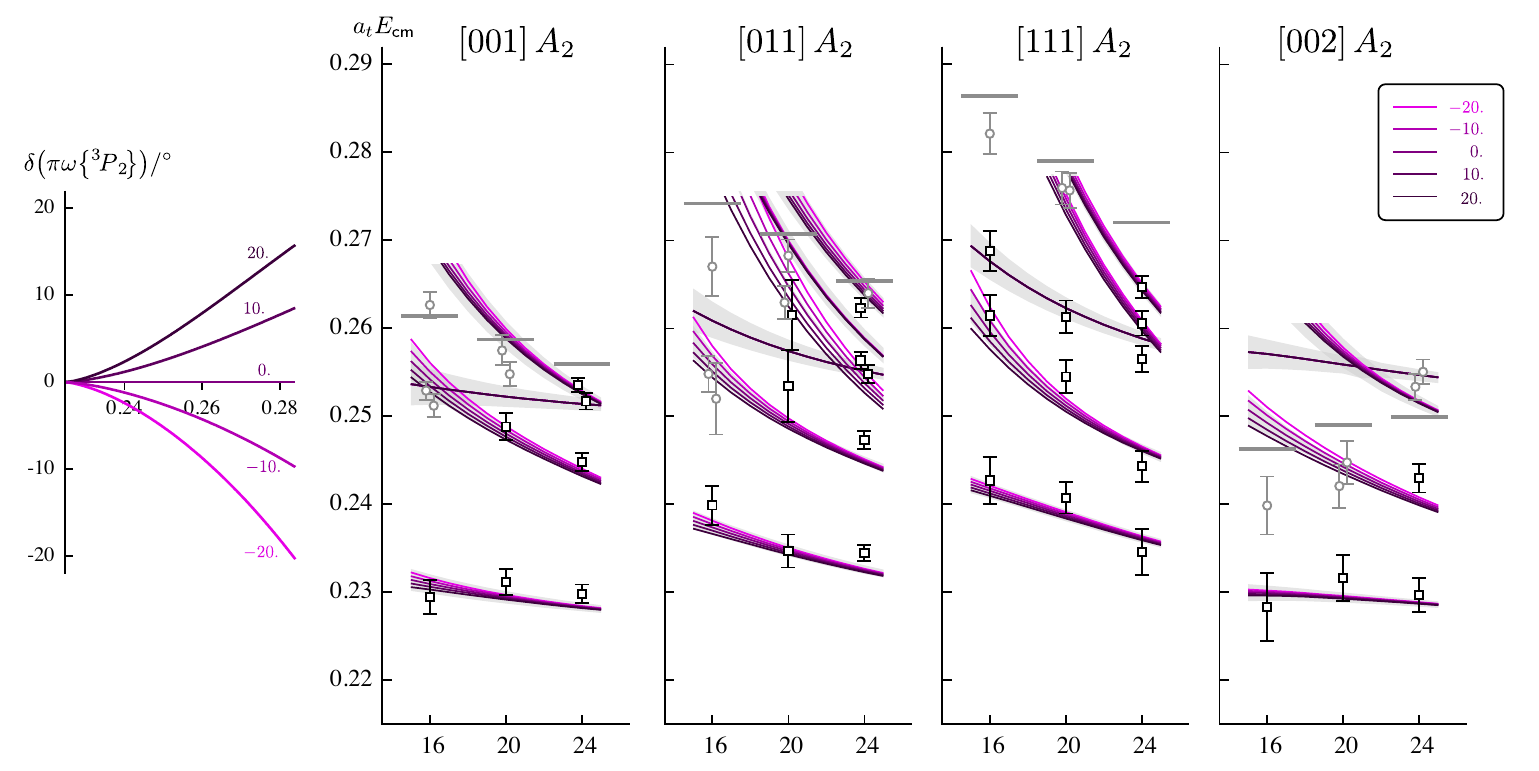}
	\caption{As Figure~\ref{Fig:3p0_var} but for varying $\gamma^{(0)}_{\pi\omega\{\!\threePtwo\!\},\pi\omega\{\!\threePtwo\!\}}$.}
	\label{Fig:3p2_var}
\end{figure*}

Interestingly, the position of these same levels proves to be strongly dependent on the amplitude strength in the $\pi\omega\{\threePzero\}$ and $\pi\omega\{\threePtwo\}$ partial-waves, so the lattice computed energies allow us to confidently limit the amplitude of these $P$-waves to be very small in this energy region. Figures~\ref{Fig:3p0_var} and~\ref{Fig:3p2_var} show the analogue of Figure~\ref{Fig:g3d1_var} but for varying $\pi\omega\{\threePzero\}$ and $\pi\omega\{\threePtwo\}$ channel parameters respectively. In these two cases, the reference amplitude in Eq.~\ref{Eq:C} is augmented, as described in Section~\ref{Subsec:additional_pw}, to include a constant $\gamma$-term in the $K$-matrix for channels $\pi\omega\{\threePzero\}$ and $\pi\omega\{\threePtwo\}$. 

%

\subsection{Three-body channels}\label{Subsec:5chan}
For the light-quark masses used in this calculation, the resonant behavior is found to occur between the relatively low-lying $\pi\pi\eta$ threshold and the somewhat higher-lying $\pi K\overline{K}$ threshold. As such, we might worry that these channels could have a significant impact on the physics in this region. We previously presented some arguments for why we do not expect this to be the case, but noted that in Figure~\ref{Fig:spec-000-T1p} there appeared to be deviations in the finite-volume spectra depending on whether or not three-meson operators were included in the bases, most notably in the smallest, $(L/a_s)=16$, volume. As a precaution, we ensured that we only made use of those energy levels which lie below the lowest {$E^{(2+1)}_{\text{n.i.}}$} value and which show no significant dependence on the presence/absence of $\rho\eta$,~$K^*\overline{K}$ or~$a_0\pi$ operators. 

In this section, we attempt to quantify the size of possible contributions from the three-body sector on our scattering amplitudes and resonance pole by treating the scattering system as though $\pi\pi$ in $\pi\pi\eta$ can be completely replaced by a \emph{stable} $\rho$ (with a fixed mass $a_t m_\rho = 0.1509$) and $\pi K$ in $\pi K \overline{K}$ can be completely replaced by a stable $K^*$ ($a_t m_{K^*} = 0.1648$). In this way we augment our scattering matrix with two extra channels $\rho \eta \big\{\! \threeSone\! \big\}$ and $K^* \overline{K} \big\{ \!\threeSone \!\big\}$.

This approach cannot be expected to completely describe the finite-volume spectra because, for example, whenever the $\rho$ has non-zero momentum, we expect there to be \emph{more than one} corresponding energy level, as indicated by Figure 1 in Ref.~\cite{Dudek:2012xn} -- a stable $\rho$ model cannot capture this and will not even give the right number of energy levels in the `three-body' spectrum.  However, for the $\rho$ at rest the nearest non-interacting $\pi\pi$ energy is much higher, and there is effectively only one finite-volume level which lies very close to the $\rho$ resonance mass. In this case, the stable $\rho$ may be a reasonable first approximation to the true three-body physics. 

\begin{figure*}[tb]
		\centering
	    \includegraphics[trim={0 0 0 0},clip,width=1.05\textwidth]{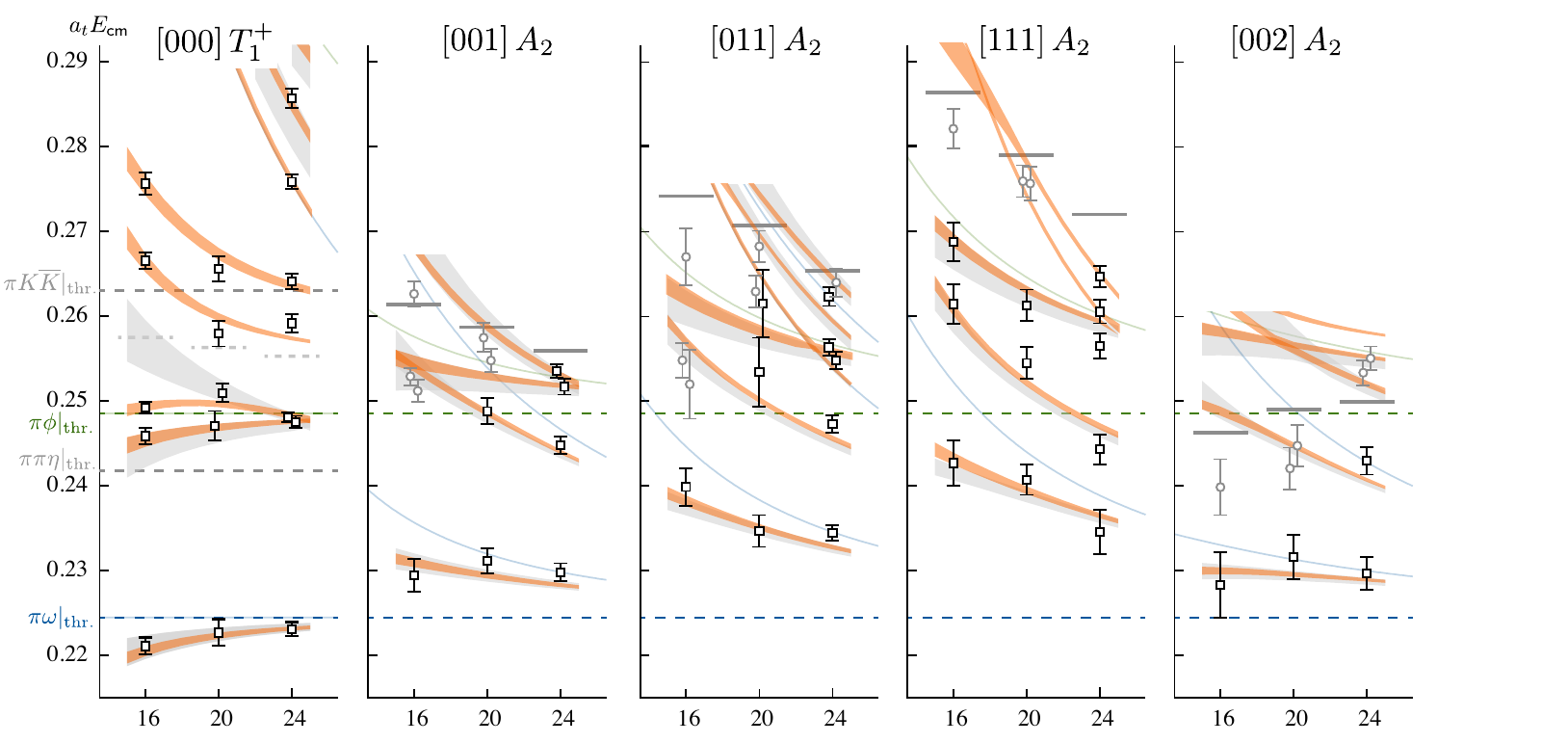}
	\caption{As Figure~\ref{Fig:fv-spec} but for the amplitude in Eq.~\ref{Fit:E}. Orange bands reflect only the statistical uncertainty on the scattering parameters. The gray bands are transcribed from Figure~\ref{Fig:fv-spec}.}
	\label{Fig:fv-spec-5chan}
\end{figure*}

For $[000]\, T_1^+$, the relevant low-lying three-meson like operators are of the form $\rho_{000}\eta_{000}$ and $K^*_{000}\overline{K}_{000}$ as shown in Table~\ref{Tab:ops_b1_rest}. We will therefore restrict our analysis to the three volumes of this irrep and include, in addition to the 36 energy levels with which we have constrained the amplitude in Section~\ref{Sec:three_channel}, the remaining energy levels shown in Figure~\ref{Fig:spec-000-T1p}, giving a total of 48 levels to constrain five coupled channels. Taking the reference amplitude in Eq.~\ref{Eq:C}, augmented to include a `pole plus constant' term in $\rho\eta \big\{ \! \threeSone \! \big\}$ and $K^*\overline{K} \big\{ \!  \threeSone  \! \big\}$, we find best-fit parameters
\begin{center}
	\begin{tabular}{rll}
		$m =$                         & $(0.2485 \pm 0.0008) \cdot a_t^{-1}$   & \\[1.1ex]
		$g_{\piomegaSsub} =$                  & $(0.14 \pm 0.01) \cdot a_t^{-1}$   & \\[1.1ex]
		$g_{\piomegaDsub} =$                  & $(1.8 \pm 0.5) \cdot a_t$   & \\[1.1ex]
		$g_{\rho\eta\{ \! \threeSone \!\}} =$                  & $(0.0 \pm 0.1) \cdot a_t^{-1}$   & \\[1.1ex]
		$g_{K^*\! \overline{K}\{\!\threeSone\!\}} =$                  & $(0.20 \pm 0.01) \cdot a_t^{-1}$   & \\[2.4ex]
		$\gamma^{(0)}_{\piomegaSsub,\piomegaSsub} = $ & $-0.52 \pm 0.16 $   & \\[1.1ex]
		$\gamma^{(0)}_{\piphiSsub, \piphiSsub} = $ & $0.64 \pm 0.17 $   & \\[1.1ex]
		$\gamma^{(0)}_{\rho\eta\{ \! \threeSone \!\},\rho\eta\{ \! \threeSone \!\}} = $ & $-1.82 \pm 0.13 $   & \\[1.1ex]
		$\gamma^{(0)}_{K^*\! \overline{K}\{\!\threeSone\!\},K^*\! \overline{K}\{\!\threeSone\!\}} = $ & $1.27 \pm 0.52 $   & \\[3.4ex]
		\multicolumn{3}{c}{$\chi^2/N_{\text{dof}}=\frac{46.6}{48-9}=1.19$ \, .}
	\end{tabular}
\end{center}
\vspace{-1cm}
\begin{equation} \label{Fit:E}\end{equation}
The $\piomegaS$, $\piomegaD$ and $\piphiS$ parameters are in reasonable agreement with those found for the reference amplitude in Eq.~\ref{Fit:C}. We show in Figure~\ref{Fig:fv-spec-5chan} the finite-volume spectra calculated through Eq.~\ref{Eq:luescher}, analogous to Figure~\ref{Fig:fv-spec}. We observe that the dependence of the finite-volume energy levels in moving-frame irreps, lying below the lowest $E_\text{n.i.}^{(2+1)}$, on the new `three-body' part of the amplitude is very slight. However, there is improved agreement in $[000]T_1^+$ where the previously excluded levels, in particular on the $(L/a_s)=16$ volume, are now described quite well. We argue that this shows our original selection criteria, giving the 36 energy levels across all irreps, is sound and leads to a robust determination of the scattering $t$-matrix. Utilizing the generalized $n$-channel Stapp-parameterization, we give the $\piomegaS$, $\piomegaD$, $\piphiS$, $\rho\eta \big\{ \!\threeSone \big\}$ and $K^* \overline{K}\big\{ \!\threeSone \big\}$ coupled-channel phase-shifts and mixing-angles in Appendix~\ref{App:Generalised-Stapp}.

As a final test of the effects of the $\rho\eta\{\threeSone\}$ and $K^*\overline{K}\{\threeSone\}$ channels, we find the resonance pole and corresponding couplings. There are 16 Riemann sheets and several `mirror poles', but the closest pole is located at 
\begin{align}\label{Eq:poles2}
a_t\sqrt{s_0}_\mathsf{II} &= 0.2448(12) - \tfrac{i}{2} \, 0.0215(21),
\end{align}
which agrees within uncertainties with the pole position found in Section~\ref{Sec:Pole_Analysis}. The corresponding couplings are,
\begin{align}\label{Eq:Couplings2}
a_t c(\pi\omega  \{\threeSone\})_\mathsf{II} &= 0.117(7) \exp[-i\,\pi \, 0.084(20)] \nonumber \\
a_t c(\pi\omega  \{\threeDone\})_\mathsf{II} &= 0.016(4) \exp[-i\,\pi \, 0.182(22)] \nonumber \\
a_t c(\rho\eta   \{\threeSone\})_\mathsf{II} &= 0.003(52) \nonumber \\
a_t c(K^*\overline{K} \{\threeSone\})_\mathsf{II} &= 0.166(8) \exp[-i\,\pi \, 0.043(12)], 
\end{align}
where we exclude the meaningless phase on $a_t c(\rho\eta   \{\threeSone\})_\mathsf{II}$ as the magnitude is consistent with zero and where $c(\pi\phi\{^3S_1\})_\mathsf{II} = 0$ by choice of amplitude. The coupling to $\rho\eta\{\threeSone\}$ is small but has a large uncertainty, while the coupling to $K^*\overline{K}\{\threeSone\}$ is larger.\footnote{We might expect the $K^* \overline{K}$ coupling to be comparable to the $\pi \omega$ coupling because in an `OZI rule' obeying framework they differ only in the flavor of $q\bar{q}$ pair creation needed to allow the resonance to decay.}

We conclude that although we cannot currently rigorously handle three-body contributions due to $\pi\pi \eta$ and $\pi K \overline{K}$, we do not see any evidence to suggest that they significantly affect the results reported in this paper.

\section{Interpretation \label{Sec:Interpretation}}


All $J^P=1^+$ amplitude parameterizations used, that prove to be capable of describing the finite-volume spectra in the energy region we are considering, had the same characteristic resonant bump in the $\piomegaS$ to $\piomegaS$ amplitude squared, with little strength in the diagonal $\piomegaD$ and $\piphiS$ elements. The off-diagonal amplitudes were all found to be relatively small. In every case we found that the bump is associated with a complex conjugate pair of poles on sheets $\mathsf{II}$ and $\mathsf{III}$, which we interpret as the effect of a single resonance.

As in previous calculations, to quote results in physical units, we choose to set the scale using the $\Omega$-baryon mass measured on these lattices, $a_t m_\Omega = 0.2951$~\cite{Edwards:2011jj}, and the physical $\Omega$-baryon mass, $m_\Omega^{\text{phys}}=1672\text{ MeV}$~\cite{PhysRevD.98.030001}. This gives ${a_t^{-1}={m_\Omega^{\text{phys}}}/({a_tm_\Omega}) = 5666 \text{ MeV}  }$ and stable hadron masses $m_\pi \approx 391$ MeV, $m_K \approx 549$ MeV, $m_\eta \approx 587$ MeV, $m_\omega \approx 881$ MeV and $m_\phi \approx 1017$ MeV.

\begin{figure}[tb]
	\centering
	\includegraphics[trim={0cm 0 0cm 0},clip,width=0.5\textwidth]{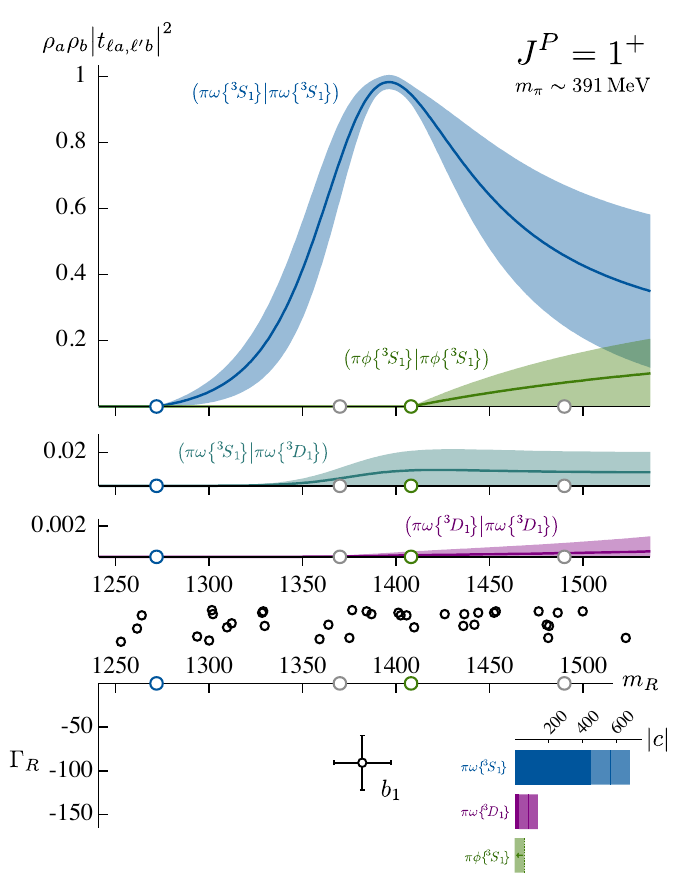}
	\caption{\textbf{Top}: The scattering amplitudes-squared, $\rho_a \rho_b |t_{\ell J a,\ell' J b}|^2$, transcribed from Figure~\ref{Fig:rho_t_sq} with the energy axis converted to physical units. Below the amplitudes are the energy levels used to constrain the amplitudes (black points). \textbf{Bottom:} The best estimate of the resonant pole position, where uncertainties combine statistical and systematic uncertainties with variations across parameterizations. The histograms show the best estimate of the magnitude of each coupling with the lightly-shaded region reflecting the combined uncertainties. The $\pi\phi\{^3S_1\}$ coupling is an estimate of the upper bound.}
	\label{Fig:summary_plot}
\end{figure}

Using this scale setting, we summarise the scattering amplitudes resulting from this work in Figure~\ref{Fig:summary_plot}, expressing all quantities in physical units.  We find a $b_1$ resonant pole of mass ${m_R = 1382(15)\text{ MeV}}$ and width ${\Gamma_R = 91(31)\text{ MeV}}$, where the uncertainties are a conservative estimate from a combination of statistical and systematic uncertainties and encompass variation over different parameterizations. Similarly, we find for the couplings, 
\begin{align}
\big|c_{\piomegaSsub} \big|&=564(114)\text{ MeV} \nonumber \\
\big|c_{\piomegaDsub} \big|&=81(56)\text{ MeV}, \nonumber \\
\big|c_{\piphiSsub} \big|&=59(41)\text{ MeV}.  \nonumber
\end{align}

\begin{figure}[tb]
	\centering
	\includegraphics[trim={0cm 0 0cm 0},clip,width=0.5\textwidth]{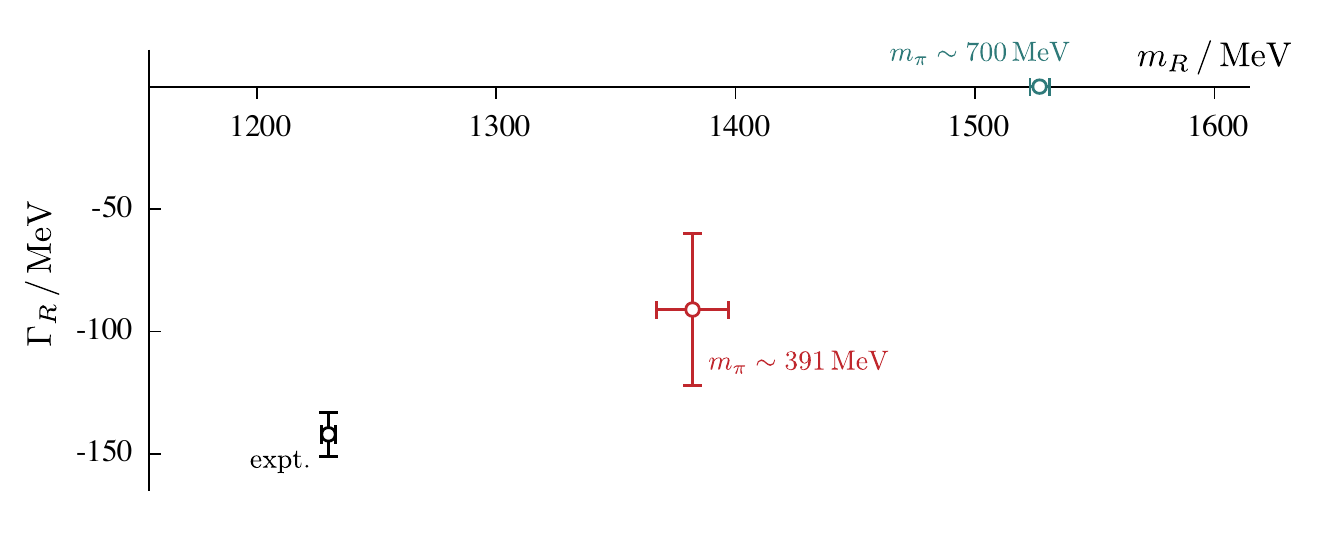}
	\caption{The $b_1$ pole position for various pion masses. Blue shows the ground-state mass of the axial-vector octet from a lattice calculation with $m_\pi \approx 700 \text{ MeV}$~\cite{Wilson:2015dqa}, red shows the estimate from this work with $m_\pi \approx 391 \text{ MeV}$ and black is the experimentally determined mass and width of the $b_1$ resonance~\cite{PhysRevD.98.030001}.}
	\label{Fig:expt}
\end{figure}

In Figure~\ref{Fig:expt} we plot the position of the pole found in this calculation compared to the experimental $b_1$ resonance, with mass $m_{b_1} = 1230(3)\text{ MeV}$ and width $\Gamma_{b_1} = 142(9)\text{ MeV}$~\cite{PhysRevD.98.030001}, and a lattice calculation at the $\text{SU}(3)_F$ point with $m_\pi\approx 700$ MeV~\cite{Dudek:2010wm}.  In the latter calculation, the $b_1$ forms part of an axial-vector octet with mass around $1525$ MeV; the pseudoscalar-vector threshold corresponding to $\pi\omega$ is at roughly $1695$ MeV, and thus the $b_1$ is stable at this pion mass. We observe that the trajectory of the pole with varying pion mass appears to be similar to that of the $\rho$ meson shown in Ref.~\cite{Wilson:2015dqa}, as may be expected for a reasonably narrow resonance.

Since we find the $b_1$ to be a narrow resonance a moderate distance above $\pi\omega$ threshold, it is reasonable to compute theoretical `branching fractions' for its decay to $\pi\omega$. For channels $a\{^3\ell_J\}$ these are given by~\cite{PhysRevD.98.030001},
\begin{equation}\label{Eq:Br_fr}
\text{Br}\Big( R \rightarrow a\{^3\ell_J\}  \Big) \equiv \frac{1}{\Gamma_R}\cdot \frac{|c_{\ell J a}|^2}{m_R} \rho_a(m_R).
\end{equation}
As mentioned in Ref.~\cite{Briceno:2017qmb}, the sum of these partial branching fractions does not necessarily give unity. We obtain
\begin{align}\label{Eq:Br_fr_result}
\text{Br} \big(b_1 \rightarrow \piomegaS \big) &\sim 93\% \nonumber \\
\text{Br} \big(b_1 \rightarrow \piomegaD \big) &\sim 2\% \nonumber 
\end{align}
where using the definition in Eq.~\ref{Eq:Br_fr} the $\pi\phi\{^3S_1\}$ branching fraction is zero as the channel is kinematically closed (${m_R < m_\pi + m_\phi}$).

A crude extrapolation of the couplings to the physical value of the light quark masses comes if we assume them to be independent of the light quark masses once the threshold behavior is removed. Such a behavior is not guaranteed, but has been observed in lattice calculations of the $\rho$~\cite{Wilson:2015dqa,Andersen:2018mau,Leskovec:2018tju,Alexandrou:2017mpi} and $K^*$~\cite{Prelovsek:2013ela,Bali:2015gji,Brett:2018jqw,Rendon:2018fem} couplings at various values of $m_\pi$. Considering
\begin{equation}\label{Eq:scaling}
\left| \frac{c^{\text{phys.}}_{\pi\omega\{^3\ell_J\}}} { \big(k_{\pi\omega}^{\text{phys.}} \big)^\ell } \right| = 
\left|\frac{c_{\pi\omega\{^3\ell_J\} } } {   \big(k_{\pi\omega} \big)^\ell } \right|,
\end{equation}
where the {\cm}-frame momentum is evaluated at the resonance pole position, and where we use the values presented above on the right-hand side, and the experimental $b_1$ mass to compute $k_{\pi\omega}^{\text{phys.}}$, gives a prediction of {$\big|c^{\text{phys.}}_{\pi\omega\{^3D_1\}} \big|= 146(101)\text{ MeV}$}. Subsequently, we obtain an estimate for the ratio of couplings at the physical pion mass of,
\begin{equation}\label{Eq:coupl}
\left|\frac{c^{\text{phys.}}_{\piomegaDsub} }{c^{\text{phys.}}_{\piomegaSsub} }\right| = 0.27(20).
\end{equation}
The PDG~\cite{PhysRevD.98.030001} reports a ratio of $D$-wave to $S$-wave amplitudes for the $b_1$ resonance of magnitude $0.277(27)$, which is not computed at the complex pole position and therefore not precisely the same quantity as we quote.

\section{Summary \label{Sec:Summary}}


This paper has reported on the first lattice QCD calculation of coupled $\pi\omega$, $\pi\phi$ scattering, the first time coupled pseudoscalar-vector scattering amplitudes have been computed. This large-scale calculation made use of a significant number of operators resembling single, two and three-meson constructions to extract finite-volume spectra which were used to constrain the coupled-channel scattering amplitudes.

Analysis of the obtained finite-volume spectra required consideration of coupled $\threeSone - \threeDone$ partial-waves in $\pi \omega$ scattering. A clear $b_1$ resonance was observed, visible as a rapid increase in the $\piomegaS$ phase-shift through $90^\circ$ or correspondingly as a bump in the magnitude of the ${\piomegaS \to \piomegaS}$ $t$-matrix element. More rigorously, we found pole singularities on unphysical Riemann sheets relatively close to the real energy axis with couplings that are large for the $\piomegaS$ final state, significantly smaller for $\piomegaD$ and compatible with zero for $\pi\phi$. The mass and width of the $b_1$ resonance found in this calculation, with light-quark masses such that $m_\pi \approx 391$ MeV, appear to be compatible with a smooth interpolation between a stable state for much larger quark mass, and the experimental resonance at lower quark mass.

We explored the role of three-body channels by including operators in our bases whose construction resembles a meson coupled to a two-body resonance, utilizing earlier calculations of meson-meson scattering channels~\cite{Dudek:2014qha,Dudek:2012xn,Dudek:2016cru}. There is no sufficiently-mature finite-volume formalism capable of rigorously incorporating three-body scattering channels here. However, as a systematic test, the finite-volume formalism, which in principle can handle any number of coupled meson-meson channels, was applied in a limited study of five coupled-channels -- $\big( \piomegaS, \piomegaD, \piphiS, \rho\eta \big\{ \!\threeSone \big\}, K^* \overline{K}\big\{ \!\threeSone \big\} \big)$. Our investigations suggested that the three-body channels have a negligible effect in this particular case of a low-lying $b_1$ resonance. Furthermore, observations were made of how particular finite-volume energy levels depend upon the various partial-waves which `mix' due to the cubic nature of the lattice boundary. 

In order to provide a way to minimally present $n$-channel scattering on the real energy axis, a generalization of the two-channel Stapp parameterization was presented in which a unitary $S$-matrix is expressed in terms of $n$ phase-shifts and $n(n-1)/2$ mixing-angles. This parameterization was used to present the three-channel $\piomegaS, \piomegaD, \piphiS$ $J^P=1^+$ scattering matrix in which the $b_1$ resonance appears. The construction provided conveniently reduces to the Stapp form in the case that one channel decouples from the others (as approximately found here).

As expected, no $I^G=1^+$ resonances are observed with a mass comparable to the $b_1$ in $J^P = 0^-, 2^-$. Notably, no resonating behavior is observed in a largely decoupled $\pi \phi$ channel, suggesting the absence of a $Z_s$ state which might be proposed as an analogue of the $Z_c$ state claimed in $\pi J/\psi$.

This work has advanced lattice techniques for studying coupled-channel scattering involving hadrons with non-zero spin and operators which effectively interpolate three hadrons. Looking forward, once a three-hadron scattering formalism is practical to use, a future calculation would enable the rigorous determination of the $\pi\pi\eta$ and $\pi K\overline{K}$ scattering amplitudes. Furthermore, utilizing such a formalism would allow the calculation of the $G$-parity-negative axial-vector, the $a_1$, which has a dominant decay to the pseudoscalar-vector meson pair $\pi \rho$, for which the $\rho$ is unstable at this pion mass, and would make for an interesting comparison. Moving on from the simplest low-lying resonances, and as the light-quark mass approaches its physical value, it becomes more important to reliably determine such three-hadron scattering processes.

\begin{acknowledgments}
We thank our colleagues within the Hadron Spectrum Collaboration, with particular thanks to R.~Brice\~no and M.T.~Hansen for useful discussions.
AJW is supported by the U.K. Science and Technology Facilities Council (STFC).  AJW and CET acknowledge support from STFC [grant number ST/P000681/1]. 
JJD acknowledges support from the U.S. Department of Energy contract DE-SC0018416.
JJD and RGE acknowledge support from the U.S. Department of Energy contract DE-AC05-06OR23177, under which Jefferson Science Associates, LLC, manages and operates Jefferson Lab. 
DJW acknowledges support from a Royal Society--Science Foundation Ireland University Research Fellowship Award UF160419.

The software codes
{\tt Chroma}~\cite{Edwards:2004sx} and {\tt QUDA}~\cite{Clark:2009wm,Babich:2010mu} were used. 
The authors acknowledge support from the U.S. Department of Energy, Office of Science, Office of Advanced Scientific Computing Research and Office of Nuclear Physics, Scientific Discovery through Advanced Computing (SciDAC) program.
Also acknowledged is support from the U.S. Department of Energy Exascale Computing Project.
 
This work was performed using the Cambridge Service for Data Driven Discovery (CSD3), part of which is operated by the University of Cambridge Research Computing on behalf of the STFC DiRAC HPC Facility (www.dirac.ac.uk). The DiRAC component of CSD3 was funded by BEIS capital funding via STFC capital grants ST/P002307/1 and ST/R002452/1 and STFC operations grant ST/R00689X/1. DiRAC is part of the National e-Infrastructure.
This work was performed using the Darwin Supercomputer of the University of Cambridge High Performance Computing Service (www.hpc.cam.ac.uk), provided by Dell Inc. using Strategic Research Infrastructure Funding from the Higher Education Funding Council for England and funding from the Science and Technology Facilities Council.
This work was also performed on clusters at Jefferson Lab under the USQCD Collaboration and the LQCD ARRA Project.
This research was supported in part under an ALCC award, and used resources of the Oak Ridge Leadership Computing Facility at the Oak Ridge National Laboratory, which is supported by the Office of Science of the U.S. Department of Energy under Contract No. DE-AC05-00OR22725.
This research used resources of the National Energy Research Scientific Computing Center (NERSC), a DOE Office of Science User Facility supported by the Office of Science of the U.S. Department of Energy under Contract No. DE-AC02-05CH11231.
The authors acknowledge the Texas Advanced Computing Center (TACC) at The University of Texas at Austin for providing HPC resources.
Gauge configurations were generated using resources awarded from the U.S. Department of Energy INCITE program at the Oak Ridge Leadership Computing Facility, the NERSC, the NSF Teragrid at the TACC and the Pittsburgh Supercomputer Center, as well as at Jefferson Lab.
\end{acknowledgments}

\bibliographystyle{apsrev4-1}
\bibliography{paper}

\begin{thebibliography}{69}%
\makeatletter
\providecommand \@ifxundefined [1]{%
 \@ifx{#1\undefined}
}%
\providecommand \@ifnum [1]{%
 \ifnum #1\expandafter \@firstoftwo
 \else \expandafter \@secondoftwo
 \fi
}%
\providecommand \@ifx [1]{%
 \ifx #1\expandafter \@firstoftwo
 \else \expandafter \@secondoftwo
 \fi
}%
\providecommand \natexlab [1]{#1}%
\providecommand \enquote  [1]{``#1''}%
\providecommand \bibnamefont  [1]{#1}%
\providecommand \bibfnamefont [1]{#1}%
\providecommand \citenamefont [1]{#1}%
\providecommand \href@noop [0]{\@secondoftwo}%
\providecommand \href [0]{\begingroup \@sanitize@url \@href}%
\providecommand \@href[1]{\@@startlink{#1}\@@href}%
\providecommand \@@href[1]{\endgroup#1\@@endlink}%
\providecommand \@sanitize@url [0]{\catcode `\\12\catcode `\$12\catcode
  `\&12\catcode `\#12\catcode `\^12\catcode `\_12\catcode `\%12\relax}%
\providecommand \@@startlink[1]{}%
\providecommand \@@endlink[0]{}%
\providecommand \url  [0]{\begingroup\@sanitize@url \@url }%
\providecommand \@url [1]{\endgroup\@href {#1}{\urlprefix }}%
\providecommand \urlprefix  [0]{URL }%
\providecommand \Eprint [0]{\href }%
\providecommand \doibase [0]{http://dx.doi.org/}%
\providecommand \selectlanguage [0]{\@gobble}%
\providecommand \bibinfo  [0]{\@secondoftwo}%
\providecommand \bibfield  [0]{\@secondoftwo}%
\providecommand \translation [1]{[#1]}%
\providecommand \BibitemOpen [0]{}%
\providecommand \bibitemStop [0]{}%
\providecommand \bibitemNoStop [0]{.\EOS\space}%
\providecommand \EOS [0]{\spacefactor3000\relax}%
\providecommand \BibitemShut  [1]{\csname bibitem#1\endcsname}%
\let\auto@bib@innerbib\@empty
\bibitem [{\citenamefont {Luscher}(1986{\natexlab{a}})}]{Luscher:1985dn}%
  \BibitemOpen
  \bibfield  {author} {\bibinfo {author} {\bibfnamefont {M.}~\bibnamefont
  {Luscher}},\ }\href {\doibase 10.1007/BF01211589} {\bibfield  {journal}
  {\bibinfo  {journal} {Commun. Math. Phys.}\ }\textbf {\bibinfo {volume}
  {104}},\ \bibinfo {pages} {177} (\bibinfo {year}
  {1986}{\natexlab{a}})}\BibitemShut {NoStop}%
\bibitem [{\citenamefont {Luscher}(1986{\natexlab{b}})}]{Luscher:1986pf}%
  \BibitemOpen
  \bibfield  {author} {\bibinfo {author} {\bibfnamefont {M.}~\bibnamefont
  {Luscher}},\ }\href {\doibase 10.1007/BF01211097} {\bibfield  {journal}
  {\bibinfo  {journal} {Commun. Math. Phys.}\ }\textbf {\bibinfo {volume}
  {105}},\ \bibinfo {pages} {153} (\bibinfo {year}
  {1986}{\natexlab{b}})}\BibitemShut {NoStop}%
\bibitem [{\citenamefont {Luscher}\ and\ \citenamefont
  {Wolff}(1990)}]{Luscher:1990ck}%
  \BibitemOpen
  \bibfield  {author} {\bibinfo {author} {\bibfnamefont {M.}~\bibnamefont
  {Luscher}}\ and\ \bibinfo {author} {\bibfnamefont {U.}~\bibnamefont
  {Wolff}},\ }\href {\doibase 10.1016/0550-3213(90)90540-T} {\bibfield
  {journal} {\bibinfo  {journal} {Nucl. Phys.}\ }\textbf {\bibinfo {volume}
  {B339}},\ \bibinfo {pages} {222} (\bibinfo {year} {1990})}\BibitemShut
  {NoStop}%
\bibitem [{\citenamefont {Luscher}(1991)}]{Luscher:1990ux}%
  \BibitemOpen
  \bibfield  {author} {\bibinfo {author} {\bibfnamefont {M.}~\bibnamefont
  {Luscher}},\ }\href {\doibase 10.1016/0550-3213(91)90366-6} {\bibfield
  {journal} {\bibinfo  {journal} {Nucl. Phys.}\ }\textbf {\bibinfo {volume}
  {B354}},\ \bibinfo {pages} {531} (\bibinfo {year} {1991})}\BibitemShut
  {NoStop}%
\bibitem [{\citenamefont {Brice\~no}(2014)}]{Briceno:2014oea}%
  \BibitemOpen
  \bibfield  {author} {\bibinfo {author} {\bibfnamefont {R.~A.}\ \bibnamefont
  {Brice\~no}},\ }\href {\doibase 10.1103/PhysRevD.89.074507} {\bibfield
  {journal} {\bibinfo  {journal} {Phys. Rev.}\ }\textbf {\bibinfo {volume}
  {D89}},\ \bibinfo {pages} {074507} (\bibinfo {year} {2014})},\ \Eprint
  {http://arxiv.org/abs/1401.3312} {arXiv:1401.3312 [hep-lat]} \BibitemShut
  {NoStop}%
\bibitem [{\citenamefont {Briceño}\ and\ \citenamefont
  {Davoudi}(2013{\natexlab{a}})}]{Briceno:2012yi}%
  \BibitemOpen
  \bibfield  {author} {\bibinfo {author} {\bibfnamefont {R.~A.}\ \bibnamefont
  {Briceño}}\ and\ \bibinfo {author} {\bibfnamefont {Z.}~\bibnamefont
  {Davoudi}},\ }\href {\doibase 10.1103/PhysRevD.88.094507} {\bibfield
  {journal} {\bibinfo  {journal} {Phys. Rev.}\ }\textbf {\bibinfo {volume}
  {D88}},\ \bibinfo {pages} {094507} (\bibinfo {year} {2013}{\natexlab{a}})},\
  \Eprint {http://arxiv.org/abs/1204.1110} {arXiv:1204.1110 [hep-lat]}
  \BibitemShut {NoStop}%
\bibitem [{\citenamefont {Christ}\ \emph {et~al.}(2005)\citenamefont {Christ},
  \citenamefont {Kim},\ and\ \citenamefont {Yamazaki}}]{Christ:2005gi}%
  \BibitemOpen
  \bibfield  {author} {\bibinfo {author} {\bibfnamefont {N.~H.}\ \bibnamefont
  {Christ}}, \bibinfo {author} {\bibfnamefont {C.}~\bibnamefont {Kim}}, \ and\
  \bibinfo {author} {\bibfnamefont {T.}~\bibnamefont {Yamazaki}},\ }\href
  {\doibase 10.1103/PhysRevD.72.114506} {\bibfield  {journal} {\bibinfo
  {journal} {Phys. Rev.}\ }\textbf {\bibinfo {volume} {D72}},\ \bibinfo {pages}
  {114506} (\bibinfo {year} {2005})},\ \Eprint
  {http://arxiv.org/abs/hep-lat/0507009} {arXiv:hep-lat/0507009 [hep-lat]}
  \BibitemShut {NoStop}%
\bibitem [{\citenamefont {Guo}\ \emph {et~al.}(2013)\citenamefont {Guo},
  \citenamefont {Dudek}, \citenamefont {Edwards},\ and\ \citenamefont
  {Szczepaniak}}]{Guo:2012hv}%
  \BibitemOpen
  \bibfield  {author} {\bibinfo {author} {\bibfnamefont {P.}~\bibnamefont
  {Guo}}, \bibinfo {author} {\bibfnamefont {J.}~\bibnamefont {Dudek}}, \bibinfo
  {author} {\bibfnamefont {R.}~\bibnamefont {Edwards}}, \ and\ \bibinfo
  {author} {\bibfnamefont {A.~P.}\ \bibnamefont {Szczepaniak}},\ }\href
  {\doibase 10.1103/PhysRevD.88.014501} {\bibfield  {journal} {\bibinfo
  {journal} {Phys. Rev.}\ }\textbf {\bibinfo {volume} {D88}},\ \bibinfo {pages}
  {014501} (\bibinfo {year} {2013})},\ \Eprint {http://arxiv.org/abs/1211.0929}
  {arXiv:1211.0929 [hep-lat]} \BibitemShut {NoStop}%
\bibitem [{\citenamefont {Kim}\ \emph {et~al.}(2005)\citenamefont {Kim},
  \citenamefont {Sachrajda},\ and\ \citenamefont {Sharpe}}]{Kim:2005gf}%
  \BibitemOpen
  \bibfield  {author} {\bibinfo {author} {\bibfnamefont {C.~H.}\ \bibnamefont
  {Kim}}, \bibinfo {author} {\bibfnamefont {C.~T.}\ \bibnamefont {Sachrajda}},
  \ and\ \bibinfo {author} {\bibfnamefont {S.~R.}\ \bibnamefont {Sharpe}},\
  }\href {\doibase 10.1016/j.nuclphysb.2005.08.029} {\bibfield  {journal}
  {\bibinfo  {journal} {Nucl. Phys.}\ }\textbf {\bibinfo {volume} {B727}},\
  \bibinfo {pages} {218} (\bibinfo {year} {2005})},\ \Eprint
  {http://arxiv.org/abs/hep-lat/0507006} {arXiv:hep-lat/0507006 [hep-lat]}
  \BibitemShut {NoStop}%
\bibitem [{\citenamefont {He}\ \emph {et~al.}(2005)\citenamefont {He},
  \citenamefont {Feng},\ and\ \citenamefont {Liu}}]{He:2005ey}%
  \BibitemOpen
  \bibfield  {author} {\bibinfo {author} {\bibfnamefont {S.}~\bibnamefont
  {He}}, \bibinfo {author} {\bibfnamefont {X.}~\bibnamefont {Feng}}, \ and\
  \bibinfo {author} {\bibfnamefont {C.}~\bibnamefont {Liu}},\ }\href {\doibase
  10.1088/1126-6708/2005/07/011} {\bibfield  {journal} {\bibinfo  {journal}
  {JHEP}\ }\textbf {\bibinfo {volume} {07}},\ \bibinfo {pages} {011} (\bibinfo
  {year} {2005})},\ \Eprint {http://arxiv.org/abs/hep-lat/0504019}
  {arXiv:hep-lat/0504019 [hep-lat]} \BibitemShut {NoStop}%
\bibitem [{\citenamefont {Rummukainen}\ and\ \citenamefont
  {Gottlieb}(1995)}]{Rummukainen:1995vs}%
  \BibitemOpen
  \bibfield  {author} {\bibinfo {author} {\bibfnamefont {K.}~\bibnamefont
  {Rummukainen}}\ and\ \bibinfo {author} {\bibfnamefont {S.~A.}\ \bibnamefont
  {Gottlieb}},\ }\href {\doibase 10.1016/0550-3213(95)00313-H} {\bibfield
  {journal} {\bibinfo  {journal} {Nucl. Phys.}\ }\textbf {\bibinfo {volume}
  {B450}},\ \bibinfo {pages} {397} (\bibinfo {year} {1995})},\ \Eprint
  {http://arxiv.org/abs/hep-lat/9503028} {arXiv:hep-lat/9503028 [hep-lat]}
  \BibitemShut {NoStop}%
\bibitem [{\citenamefont {Gockeler}\ \emph {et~al.}(2012)\citenamefont
  {Gockeler}, \citenamefont {Horsley}, \citenamefont {Lage}, \citenamefont
  {Meissner}, \citenamefont {Rakow}, \citenamefont {Rusetsky}, \citenamefont
  {Schierholz},\ and\ \citenamefont {Zanotti}}]{Gockeler:2012yj}%
  \BibitemOpen
  \bibfield  {author} {\bibinfo {author} {\bibfnamefont {M.}~\bibnamefont
  {Gockeler}}, \bibinfo {author} {\bibfnamefont {R.}~\bibnamefont {Horsley}},
  \bibinfo {author} {\bibfnamefont {M.}~\bibnamefont {Lage}}, \bibinfo {author}
  {\bibfnamefont {U.~G.}\ \bibnamefont {Meissner}}, \bibinfo {author}
  {\bibfnamefont {P.~E.~L.}\ \bibnamefont {Rakow}}, \bibinfo {author}
  {\bibfnamefont {A.}~\bibnamefont {Rusetsky}}, \bibinfo {author}
  {\bibfnamefont {G.}~\bibnamefont {Schierholz}}, \ and\ \bibinfo {author}
  {\bibfnamefont {J.~M.}\ \bibnamefont {Zanotti}},\ }\href {\doibase
  10.1103/PhysRevD.86.094513} {\bibfield  {journal} {\bibinfo  {journal} {Phys.
  Rev.}\ }\textbf {\bibinfo {volume} {D86}},\ \bibinfo {pages} {094513}
  (\bibinfo {year} {2012})},\ \Eprint {http://arxiv.org/abs/1206.4141}
  {arXiv:1206.4141 [hep-lat]} \BibitemShut {NoStop}%
\bibitem [{\citenamefont {Dudek}\ \emph {et~al.}(2016)\citenamefont {Dudek},
  \citenamefont {Edwards},\ and\ \citenamefont {Wilson}}]{Dudek:2016cru}%
  \BibitemOpen
  \bibfield  {author} {\bibinfo {author} {\bibfnamefont {J.~J.}\ \bibnamefont
  {Dudek}}, \bibinfo {author} {\bibfnamefont {R.~G.}\ \bibnamefont {Edwards}},
  \ and\ \bibinfo {author} {\bibfnamefont {D.~J.}\ \bibnamefont {Wilson}}
  (\bibinfo {collaboration} {Hadron Spectrum}),\ }\href {\doibase
  10.1103/PhysRevD.93.094506} {\bibfield  {journal} {\bibinfo  {journal} {Phys.
  Rev.}\ }\textbf {\bibinfo {volume} {D93}},\ \bibinfo {pages} {094506}
  (\bibinfo {year} {2016})},\ \Eprint {http://arxiv.org/abs/1602.05122}
  {arXiv:1602.05122 [hep-ph]} \BibitemShut {NoStop}%
\bibitem [{\citenamefont {Briceño}\ \emph
  {et~al.}(2017{\natexlab{a}})\citenamefont {Briceño}, \citenamefont {Dudek},
  \citenamefont {Edwards},\ and\ \citenamefont {Wilson}}]{Briceno:2017qmb}%
  \BibitemOpen
  \bibfield  {author} {\bibinfo {author} {\bibfnamefont {R.~A.}\ \bibnamefont
  {Briceño}}, \bibinfo {author} {\bibfnamefont {J.~J.}\ \bibnamefont {Dudek}},
  \bibinfo {author} {\bibfnamefont {R.~G.}\ \bibnamefont {Edwards}}, \ and\
  \bibinfo {author} {\bibfnamefont {D.~J.}\ \bibnamefont {Wilson}},\
  }\href@noop {} {\  (\bibinfo {year} {2017}{\natexlab{a}})},\ \Eprint
  {http://arxiv.org/abs/1708.06667} {arXiv:1708.06667 [hep-lat]} \BibitemShut
  {NoStop}%
\bibitem [{\citenamefont {Tanabashi}\ \emph {et~al.}(2018)\citenamefont
  {Tanabashi} \emph {et~al.}}]{PhysRevD.98.030001}%
  \BibitemOpen
  \bibfield  {author} {\bibinfo {author} {\bibfnamefont {M.}~\bibnamefont
  {Tanabashi}} \emph {et~al.} (\bibinfo {collaboration} {Particle Data
  Group}),\ }\href {\doibase 10.1103/PhysRevD.98.030001} {\bibfield  {journal}
  {\bibinfo  {journal} {Phys. Rev. D}\ }\textbf {\bibinfo {volume} {98}},\
  \bibinfo {pages} {030001} (\bibinfo {year} {2018})}\BibitemShut {NoStop}%
\bibitem [{\citenamefont {Nozar}\ \emph {et~al.}(2002)\citenamefont {Nozar}
  \emph {et~al.}}]{Nozar:2002br}%
  \BibitemOpen
  \bibfield  {author} {\bibinfo {author} {\bibfnamefont {M.}~\bibnamefont
  {Nozar}} \emph {et~al.} (\bibinfo {collaboration} {E852}),\ }\href {\doibase
  10.1016/S0370-2693(02)02194-9} {\bibfield  {journal} {\bibinfo  {journal}
  {Phys. Lett.}\ }\textbf {\bibinfo {volume} {B541}},\ \bibinfo {pages} {35}
  (\bibinfo {year} {2002})},\ \Eprint {http://arxiv.org/abs/hep-ex/0206026}
  {arXiv:hep-ex/0206026 [hep-ex]} \BibitemShut {NoStop}%
\bibitem [{\citenamefont {Woss}\ \emph {et~al.}(2018)\citenamefont {Woss},
  \citenamefont {Thomas}, \citenamefont {Dudek}, \citenamefont {Edwards},\ and\
  \citenamefont {Wilson}}]{Woss:2018irj}%
  \BibitemOpen
  \bibfield  {author} {\bibinfo {author} {\bibfnamefont {A.}~\bibnamefont
  {Woss}}, \bibinfo {author} {\bibfnamefont {C.~E.}\ \bibnamefont {Thomas}},
  \bibinfo {author} {\bibfnamefont {J.~J.}\ \bibnamefont {Dudek}}, \bibinfo
  {author} {\bibfnamefont {R.~G.}\ \bibnamefont {Edwards}}, \ and\ \bibinfo
  {author} {\bibfnamefont {D.~J.}\ \bibnamefont {Wilson}},\ }\href@noop {} {\
  (\bibinfo {year} {2018})},\ \Eprint {http://arxiv.org/abs/1802.05580}
  {arXiv:1802.05580 [hep-lat]} \BibitemShut {NoStop}%
\bibitem [{\citenamefont {Dudek}\ \emph {et~al.}(2011)\citenamefont {Dudek},
  \citenamefont {Edwards}, \citenamefont {Joo}, \citenamefont {Peardon},
  \citenamefont {Richards},\ and\ \citenamefont {Thomas}}]{Dudek:2011tt}%
  \BibitemOpen
  \bibfield  {author} {\bibinfo {author} {\bibfnamefont {J.~J.}\ \bibnamefont
  {Dudek}}, \bibinfo {author} {\bibfnamefont {R.~G.}\ \bibnamefont {Edwards}},
  \bibinfo {author} {\bibfnamefont {B.}~\bibnamefont {Joo}}, \bibinfo {author}
  {\bibfnamefont {M.~J.}\ \bibnamefont {Peardon}}, \bibinfo {author}
  {\bibfnamefont {D.~G.}\ \bibnamefont {Richards}}, \ and\ \bibinfo {author}
  {\bibfnamefont {C.~E.}\ \bibnamefont {Thomas}},\ }\href {\doibase
  10.1103/PhysRevD.83.111502} {\bibfield  {journal} {\bibinfo  {journal} {Phys.
  Rev.}\ }\textbf {\bibinfo {volume} {D83}},\ \bibinfo {pages} {111502}
  (\bibinfo {year} {2011})},\ \Eprint {http://arxiv.org/abs/1102.4299}
  {arXiv:1102.4299 [hep-lat]} \BibitemShut {NoStop}%
\bibitem [{\citenamefont {Dudek}\ \emph
  {et~al.}(2013{\natexlab{a}})\citenamefont {Dudek}, \citenamefont {Edwards},
  \citenamefont {Guo},\ and\ \citenamefont {Thomas}}]{Dudek:2013yja}%
  \BibitemOpen
  \bibfield  {author} {\bibinfo {author} {\bibfnamefont {J.~J.}\ \bibnamefont
  {Dudek}}, \bibinfo {author} {\bibfnamefont {R.~G.}\ \bibnamefont {Edwards}},
  \bibinfo {author} {\bibfnamefont {P.}~\bibnamefont {Guo}}, \ and\ \bibinfo
  {author} {\bibfnamefont {C.~E.}\ \bibnamefont {Thomas}} (\bibinfo
  {collaboration} {Hadron Spectrum}),\ }\href {\doibase
  10.1103/PhysRevD.88.094505} {\bibfield  {journal} {\bibinfo  {journal} {Phys.
  Rev.}\ }\textbf {\bibinfo {volume} {D88}},\ \bibinfo {pages} {094505}
  (\bibinfo {year} {2013}{\natexlab{a}})},\ \Eprint
  {http://arxiv.org/abs/1309.2608} {arXiv:1309.2608 [hep-lat]} \BibitemShut
  {NoStop}%
\bibitem [{\citenamefont {Lang}\ \emph {et~al.}(2014)\citenamefont {Lang},
  \citenamefont {Leskovec}, \citenamefont {Mohler},\ and\ \citenamefont
  {Prelovsek}}]{Lang:2014tia}%
  \BibitemOpen
  \bibfield  {author} {\bibinfo {author} {\bibfnamefont {C.~B.}\ \bibnamefont
  {Lang}}, \bibinfo {author} {\bibfnamefont {L.}~\bibnamefont {Leskovec}},
  \bibinfo {author} {\bibfnamefont {D.}~\bibnamefont {Mohler}}, \ and\ \bibinfo
  {author} {\bibfnamefont {S.}~\bibnamefont {Prelovsek}},\ }\href {\doibase
  10.1007/JHEP04(2014)162} {\bibfield  {journal} {\bibinfo  {journal} {JHEP}\
  }\textbf {\bibinfo {volume} {04}},\ \bibinfo {pages} {162} (\bibinfo {year}
  {2014})},\ \Eprint {http://arxiv.org/abs/1401.2088} {arXiv:1401.2088
  [hep-lat]} \BibitemShut {NoStop}%
\bibitem [{\citenamefont {McNeile}\ and\ \citenamefont
  {Michael}(2006)}]{McNeile:2006bz}%
  \BibitemOpen
  \bibfield  {author} {\bibinfo {author} {\bibfnamefont {C.}~\bibnamefont
  {McNeile}}\ and\ \bibinfo {author} {\bibfnamefont {C.}~\bibnamefont
  {Michael}} (\bibinfo {collaboration} {UKQCD}),\ }\href {\doibase
  10.1103/PhysRevD.73.074506} {\bibfield  {journal} {\bibinfo  {journal} {Phys.
  Rev.}\ }\textbf {\bibinfo {volume} {D73}},\ \bibinfo {pages} {074506}
  (\bibinfo {year} {2006})},\ \Eprint {http://arxiv.org/abs/hep-lat/0603007}
  {arXiv:hep-lat/0603007 [hep-lat]} \BibitemShut {NoStop}%
\bibitem [{\citenamefont {Briceño}\ \emph
  {et~al.}(2017{\natexlab{b}})\citenamefont {Briceño}, \citenamefont
  {Hansen},\ and\ \citenamefont {Sharpe}}]{Briceno:2017tce}%
  \BibitemOpen
  \bibfield  {author} {\bibinfo {author} {\bibfnamefont {R.~A.}\ \bibnamefont
  {Briceño}}, \bibinfo {author} {\bibfnamefont {M.~T.}\ \bibnamefont
  {Hansen}}, \ and\ \bibinfo {author} {\bibfnamefont {S.~R.}\ \bibnamefont
  {Sharpe}},\ }\href {\doibase 10.1103/PhysRevD.95.074510} {\bibfield
  {journal} {\bibinfo  {journal} {Phys. Rev.}\ }\textbf {\bibinfo {volume}
  {D95}},\ \bibinfo {pages} {074510} (\bibinfo {year} {2017}{\natexlab{b}})},\
  \Eprint {http://arxiv.org/abs/1701.07465} {arXiv:1701.07465 [hep-lat]}
  \BibitemShut {NoStop}%
\bibitem [{\citenamefont {Briceño}\ \emph {et~al.}(2019)\citenamefont
  {Briceño}, \citenamefont {Hansen},\ and\ \citenamefont
  {Sharpe}}]{Briceno:2018aml}%
  \BibitemOpen
  \bibfield  {author} {\bibinfo {author} {\bibfnamefont {R.~A.}\ \bibnamefont
  {Briceño}}, \bibinfo {author} {\bibfnamefont {M.~T.}\ \bibnamefont
  {Hansen}}, \ and\ \bibinfo {author} {\bibfnamefont {S.~R.}\ \bibnamefont
  {Sharpe}},\ }\href {\doibase 10.1103/PhysRevD.99.014516} {\bibfield
  {journal} {\bibinfo  {journal} {Phys. Rev.}\ }\textbf {\bibinfo {volume}
  {D99}},\ \bibinfo {pages} {014516} (\bibinfo {year} {2019})},\ \Eprint
  {http://arxiv.org/abs/1810.01429} {arXiv:1810.01429 [hep-lat]} \BibitemShut
  {NoStop}%
\bibitem [{\citenamefont {Briceño}\ \emph {et~al.}(2018)\citenamefont
  {Briceño}, \citenamefont {Hansen},\ and\ \citenamefont
  {Sharpe}}]{Briceno:2018mlh}%
  \BibitemOpen
  \bibfield  {author} {\bibinfo {author} {\bibfnamefont {R.~A.}\ \bibnamefont
  {Briceño}}, \bibinfo {author} {\bibfnamefont {M.~T.}\ \bibnamefont
  {Hansen}}, \ and\ \bibinfo {author} {\bibfnamefont {S.~R.}\ \bibnamefont
  {Sharpe}},\ }\href {\doibase 10.1103/PhysRevD.98.014506} {\bibfield
  {journal} {\bibinfo  {journal} {Phys. Rev.}\ }\textbf {\bibinfo {volume}
  {D98}},\ \bibinfo {pages} {014506} (\bibinfo {year} {2018})},\ \Eprint
  {http://arxiv.org/abs/1803.04169} {arXiv:1803.04169 [hep-lat]} \BibitemShut
  {NoStop}%
\bibitem [{\citenamefont {Hammer}\ \emph {et~al.}(2017)\citenamefont {Hammer},
  \citenamefont {Pang},\ and\ \citenamefont {Rusetsky}}]{Hammer:2017kms}%
  \BibitemOpen
  \bibfield  {author} {\bibinfo {author} {\bibfnamefont {H.~W.}\ \bibnamefont
  {Hammer}}, \bibinfo {author} {\bibfnamefont {J.~Y.}\ \bibnamefont {Pang}}, \
  and\ \bibinfo {author} {\bibfnamefont {A.}~\bibnamefont {Rusetsky}},\ }\href
  {\doibase 10.1007/JHEP10(2017)115} {\bibfield  {journal} {\bibinfo  {journal}
  {JHEP}\ }\textbf {\bibinfo {volume} {10}},\ \bibinfo {pages} {115} (\bibinfo
  {year} {2017})},\ \Eprint {http://arxiv.org/abs/1707.02176} {arXiv:1707.02176
  [hep-lat]} \BibitemShut {NoStop}%
\bibitem [{\citenamefont {Mai}\ and\ \citenamefont
  {Döring}(2017)}]{Mai:2017bge}%
  \BibitemOpen
  \bibfield  {author} {\bibinfo {author} {\bibfnamefont {M.}~\bibnamefont
  {Mai}}\ and\ \bibinfo {author} {\bibfnamefont {M.}~\bibnamefont {Döring}},\
  }\href {\doibase 10.1140/epja/i2017-12440-1} {\bibfield  {journal} {\bibinfo
  {journal} {Eur. Phys. J.}\ }\textbf {\bibinfo {volume} {A53}},\ \bibinfo
  {pages} {240} (\bibinfo {year} {2017})},\ \Eprint
  {http://arxiv.org/abs/1709.08222} {arXiv:1709.08222 [hep-lat]} \BibitemShut
  {NoStop}%
\bibitem [{\citenamefont {Mai}\ and\ \citenamefont
  {Doring}(2019)}]{Mai:2018djl}%
  \BibitemOpen
  \bibfield  {author} {\bibinfo {author} {\bibfnamefont {M.}~\bibnamefont
  {Mai}}\ and\ \bibinfo {author} {\bibfnamefont {M.}~\bibnamefont {Doring}},\
  }\href {\doibase 10.1103/PhysRevLett.122.062503} {\bibfield  {journal}
  {\bibinfo  {journal} {Phys. Rev. Lett.}\ }\textbf {\bibinfo {volume} {122}},\
  \bibinfo {pages} {062503} (\bibinfo {year} {2019})},\ \Eprint
  {http://arxiv.org/abs/1807.04746} {arXiv:1807.04746 [hep-lat]} \BibitemShut
  {NoStop}%
\bibitem [{\citenamefont {Hansen}\ and\ \citenamefont
  {Sharpe}(2019)}]{Hansen:2019nir}%
  \BibitemOpen
  \bibfield  {author} {\bibinfo {author} {\bibfnamefont {M.~T.}\ \bibnamefont
  {Hansen}}\ and\ \bibinfo {author} {\bibfnamefont {S.~R.}\ \bibnamefont
  {Sharpe}},\ }\href@noop {} {\  (\bibinfo {year} {2019})},\ \Eprint
  {http://arxiv.org/abs/1901.00483} {arXiv:1901.00483 [hep-lat]} \BibitemShut
  {NoStop}%
\bibitem [{\citenamefont {Blanton}\ \emph {et~al.}(2019)\citenamefont
  {Blanton}, \citenamefont {Romero-López},\ and\ \citenamefont
  {Sharpe}}]{Blanton:2019igq}%
  \BibitemOpen
  \bibfield  {author} {\bibinfo {author} {\bibfnamefont {T.~D.}\ \bibnamefont
  {Blanton}}, \bibinfo {author} {\bibfnamefont {F.}~\bibnamefont
  {Romero-López}}, \ and\ \bibinfo {author} {\bibfnamefont {S.~R.}\
  \bibnamefont {Sharpe}},\ }\href {\doibase 10.1007/JHEP03(2019)106} {\bibfield
   {journal} {\bibinfo  {journal} {JHEP}\ }\textbf {\bibinfo {volume} {03}},\
  \bibinfo {pages} {106} (\bibinfo {year} {2019})},\ \Eprint
  {http://arxiv.org/abs/1901.07095} {arXiv:1901.07095 [hep-lat]} \BibitemShut
  {NoStop}%
\bibitem [{\citenamefont {Dudek}\ \emph {et~al.}(2010)\citenamefont {Dudek},
  \citenamefont {Edwards}, \citenamefont {Peardon}, \citenamefont {Richards},\
  and\ \citenamefont {Thomas}}]{Dudek:2010wm}%
  \BibitemOpen
  \bibfield  {author} {\bibinfo {author} {\bibfnamefont {J.~J.}\ \bibnamefont
  {Dudek}}, \bibinfo {author} {\bibfnamefont {R.~G.}\ \bibnamefont {Edwards}},
  \bibinfo {author} {\bibfnamefont {M.~J.}\ \bibnamefont {Peardon}}, \bibinfo
  {author} {\bibfnamefont {D.~G.}\ \bibnamefont {Richards}}, \ and\ \bibinfo
  {author} {\bibfnamefont {C.~E.}\ \bibnamefont {Thomas}},\ }\href {\doibase
  10.1103/PhysRevD.82.034508} {\bibfield  {journal} {\bibinfo  {journal} {Phys.
  Rev.}\ }\textbf {\bibinfo {volume} {D82}},\ \bibinfo {pages} {034508}
  (\bibinfo {year} {2010})},\ \Eprint {http://arxiv.org/abs/1004.4930}
  {arXiv:1004.4930 [hep-ph]} \BibitemShut {NoStop}%
\bibitem [{\citenamefont {Ablikim}\ \emph {et~al.}(2019)\citenamefont {Ablikim}
  \emph {et~al.}}]{Ablikim:2018ofc}%
  \BibitemOpen
  \bibfield  {author} {\bibinfo {author} {\bibfnamefont {M.}~\bibnamefont
  {Ablikim}} \emph {et~al.} (\bibinfo {collaboration} {BESIII}),\ }\href
  {\doibase 10.1103/PhysRevD.99.011101} {\bibfield  {journal} {\bibinfo
  {journal} {Phys. Rev.}\ }\textbf {\bibinfo {volume} {D99}},\ \bibinfo {pages}
  {011101} (\bibinfo {year} {2019})},\ \Eprint
  {http://arxiv.org/abs/1801.10384} {arXiv:1801.10384 [hep-ex]} \BibitemShut
  {NoStop}%
\bibitem [{\citenamefont {Liu}\ \emph {et~al.}(2013)\citenamefont {Liu} \emph
  {et~al.}}]{Liu:2013dau}%
  \BibitemOpen
  \bibfield  {author} {\bibinfo {author} {\bibfnamefont {Z.~Q.}\ \bibnamefont
  {Liu}} \emph {et~al.} (\bibinfo {collaboration} {Belle}),\ }\href {\doibase
  10.1103/PhysRevLett.110.252002} {\bibfield  {journal} {\bibinfo  {journal}
  {Phys. Rev. Lett.}\ }\textbf {\bibinfo {volume} {110}},\ \bibinfo {pages}
  {252002} (\bibinfo {year} {2013})},\ \Eprint {http://arxiv.org/abs/1304.0121}
  {arXiv:1304.0121 [hep-ex]} \BibitemShut {NoStop}%
\bibitem [{\citenamefont {Ablikim}\ \emph {et~al.}(2013)\citenamefont {Ablikim}
  \emph {et~al.}}]{Ablikim:2013mio}%
  \BibitemOpen
  \bibfield  {author} {\bibinfo {author} {\bibfnamefont {M.}~\bibnamefont
  {Ablikim}} \emph {et~al.} (\bibinfo {collaboration} {BESIII}),\ }\href
  {\doibase 10.1103/PhysRevLett.110.252001} {\bibfield  {journal} {\bibinfo
  {journal} {Phys. Rev. Lett.}\ }\textbf {\bibinfo {volume} {110}},\ \bibinfo
  {pages} {252001} (\bibinfo {year} {2013})},\ \Eprint
  {http://arxiv.org/abs/1303.5949} {arXiv:1303.5949 [hep-ex]} \BibitemShut
  {NoStop}%
\bibitem [{\citenamefont {Moore}\ and\ \citenamefont
  {Fleming}(2006)}]{Moore:2005dw}%
  \BibitemOpen
  \bibfield  {author} {\bibinfo {author} {\bibfnamefont {D.~C.}\ \bibnamefont
  {Moore}}\ and\ \bibinfo {author} {\bibfnamefont {G.~T.}\ \bibnamefont
  {Fleming}},\ }\href {\doibase 10.1103/PhysRevD.73.014504,
  10.1103/PhysRevD.74.079905} {\bibfield  {journal} {\bibinfo  {journal} {Phys.
  Rev.}\ }\textbf {\bibinfo {volume} {D73}},\ \bibinfo {pages} {014504}
  (\bibinfo {year} {2006})},\ \bibinfo {note} {[Erratum: Phys.
  Rev.D74,079905(2006)]},\ \Eprint {http://arxiv.org/abs/hep-lat/0507018}
  {arXiv:hep-lat/0507018 [hep-lat]} \BibitemShut {NoStop}%
\bibitem [{\citenamefont {Thomas}\ \emph {et~al.}(2012)\citenamefont {Thomas},
  \citenamefont {Edwards},\ and\ \citenamefont {Dudek}}]{Thomas:2011rh}%
  \BibitemOpen
  \bibfield  {author} {\bibinfo {author} {\bibfnamefont {C.~E.}\ \bibnamefont
  {Thomas}}, \bibinfo {author} {\bibfnamefont {R.~G.}\ \bibnamefont {Edwards}},
  \ and\ \bibinfo {author} {\bibfnamefont {J.~J.}\ \bibnamefont {Dudek}},\
  }\href {\doibase 10.1103/PhysRevD.85.014507, 10.1103/PhysRevD.85.039901}
  {\bibfield  {journal} {\bibinfo  {journal} {Phys. Rev.}\ }\textbf {\bibinfo
  {volume} {D85}},\ \bibinfo {pages} {014507} (\bibinfo {year} {2012})},\
  \Eprint {http://arxiv.org/abs/1107.1930} {arXiv:1107.1930 [hep-lat]}
  \BibitemShut {NoStop}%
\bibitem [{\citenamefont {Michael}(1985)}]{Michael:1985ne}%
  \BibitemOpen
  \bibfield  {author} {\bibinfo {author} {\bibfnamefont {C.}~\bibnamefont
  {Michael}},\ }\href {\doibase 10.1016/0550-3213(85)90297-4} {\bibfield
  {journal} {\bibinfo  {journal} {Nucl. Phys.}\ }\textbf {\bibinfo {volume}
  {B259}},\ \bibinfo {pages} {58} (\bibinfo {year} {1985})}\BibitemShut
  {NoStop}%
\bibitem [{\citenamefont {Dudek}\ \emph {et~al.}(2008)\citenamefont {Dudek},
  \citenamefont {Edwards}, \citenamefont {Mathur},\ and\ \citenamefont
  {Richards}}]{Dudek:2007wv}%
  \BibitemOpen
  \bibfield  {author} {\bibinfo {author} {\bibfnamefont {J.~J.}\ \bibnamefont
  {Dudek}}, \bibinfo {author} {\bibfnamefont {R.~G.}\ \bibnamefont {Edwards}},
  \bibinfo {author} {\bibfnamefont {N.}~\bibnamefont {Mathur}}, \ and\ \bibinfo
  {author} {\bibfnamefont {D.~G.}\ \bibnamefont {Richards}},\ }\href {\doibase
  10.1103/PhysRevD.77.034501} {\bibfield  {journal} {\bibinfo  {journal} {Phys.
  Rev.}\ }\textbf {\bibinfo {volume} {D77}},\ \bibinfo {pages} {034501}
  (\bibinfo {year} {2008})},\ \Eprint {http://arxiv.org/abs/0707.4162}
  {arXiv:0707.4162 [hep-lat]} \BibitemShut {NoStop}%
\bibitem [{\citenamefont {Wilson}\ \emph
  {et~al.}(2015{\natexlab{a}})\citenamefont {Wilson}, \citenamefont {Briceño},
  \citenamefont {Dudek}, \citenamefont {Edwards},\ and\ \citenamefont
  {Thomas}}]{Wilson:2015dqa}%
  \BibitemOpen
  \bibfield  {author} {\bibinfo {author} {\bibfnamefont {D.~J.}\ \bibnamefont
  {Wilson}}, \bibinfo {author} {\bibfnamefont {R.~A.}\ \bibnamefont
  {Briceño}}, \bibinfo {author} {\bibfnamefont {J.~J.}\ \bibnamefont {Dudek}},
  \bibinfo {author} {\bibfnamefont {R.~G.}\ \bibnamefont {Edwards}}, \ and\
  \bibinfo {author} {\bibfnamefont {C.~E.}\ \bibnamefont {Thomas}},\ }\href
  {\doibase 10.1103/PhysRevD.92.094502} {\bibfield  {journal} {\bibinfo
  {journal} {Phys. Rev.}\ }\textbf {\bibinfo {volume} {D92}},\ \bibinfo {pages}
  {094502} (\bibinfo {year} {2015}{\natexlab{a}})},\ \Eprint
  {http://arxiv.org/abs/1507.02599} {arXiv:1507.02599 [hep-ph]} \BibitemShut
  {NoStop}%
\bibitem [{\citenamefont {Cheung}\ \emph {et~al.}(2017)\citenamefont {Cheung},
  \citenamefont {Thomas}, \citenamefont {Dudek},\ and\ \citenamefont
  {Edwards}}]{Cheung:2017tnt}%
  \BibitemOpen
  \bibfield  {author} {\bibinfo {author} {\bibfnamefont {G.~K.~C.}\
  \bibnamefont {Cheung}}, \bibinfo {author} {\bibfnamefont {C.~E.}\
  \bibnamefont {Thomas}}, \bibinfo {author} {\bibfnamefont {J.~J.}\
  \bibnamefont {Dudek}}, \ and\ \bibinfo {author} {\bibfnamefont {R.~G.}\
  \bibnamefont {Edwards}} (\bibinfo {collaboration} {Hadron Spectrum}),\ }\href
  {\doibase 10.1007/JHEP11(2017)033} {\bibfield  {journal} {\bibinfo  {journal}
  {JHEP}\ }\textbf {\bibinfo {volume} {11}},\ \bibinfo {pages} {033} (\bibinfo
  {year} {2017})},\ \Eprint {http://arxiv.org/abs/1709.01417} {arXiv:1709.01417
  [hep-lat]} \BibitemShut {NoStop}%
\bibitem [{\citenamefont {Padmanath}\ \emph {et~al.}(2015)\citenamefont
  {Padmanath}, \citenamefont {Lang},\ and\ \citenamefont
  {Prelovsek}}]{Padmanath:2015era}%
  \BibitemOpen
  \bibfield  {author} {\bibinfo {author} {\bibfnamefont {M.}~\bibnamefont
  {Padmanath}}, \bibinfo {author} {\bibfnamefont {C.~B.}\ \bibnamefont {Lang}},
  \ and\ \bibinfo {author} {\bibfnamefont {S.}~\bibnamefont {Prelovsek}},\
  }\href {\doibase 10.1103/PhysRevD.92.034501} {\bibfield  {journal} {\bibinfo
  {journal} {Phys. Rev.}\ }\textbf {\bibinfo {volume} {D92}},\ \bibinfo {pages}
  {034501} (\bibinfo {year} {2015})},\ \Eprint
  {http://arxiv.org/abs/1503.03257} {arXiv:1503.03257 [hep-lat]} \BibitemShut
  {NoStop}%
\bibitem [{\citenamefont {Dudek}\ \emph {et~al.}(2012)\citenamefont {Dudek},
  \citenamefont {Edwards},\ and\ \citenamefont {Thomas}}]{Dudek:2012gj}%
  \BibitemOpen
  \bibfield  {author} {\bibinfo {author} {\bibfnamefont {J.~J.}\ \bibnamefont
  {Dudek}}, \bibinfo {author} {\bibfnamefont {R.~G.}\ \bibnamefont {Edwards}},
  \ and\ \bibinfo {author} {\bibfnamefont {C.~E.}\ \bibnamefont {Thomas}},\
  }\href {\doibase 10.1103/PhysRevD.86.034031} {\bibfield  {journal} {\bibinfo
  {journal} {Phys. Rev.}\ }\textbf {\bibinfo {volume} {D86}},\ \bibinfo {pages}
  {034031} (\bibinfo {year} {2012})},\ \Eprint {http://arxiv.org/abs/1203.6041}
  {arXiv:1203.6041 [hep-ph]} \BibitemShut {NoStop}%
\bibitem [{\citenamefont {Dudek}\ \emph
  {et~al.}(2013{\natexlab{b}})\citenamefont {Dudek}, \citenamefont {Edwards},\
  and\ \citenamefont {Thomas}}]{Dudek:2012xn}%
  \BibitemOpen
  \bibfield  {author} {\bibinfo {author} {\bibfnamefont {J.~J.}\ \bibnamefont
  {Dudek}}, \bibinfo {author} {\bibfnamefont {R.~G.}\ \bibnamefont {Edwards}},
  \ and\ \bibinfo {author} {\bibfnamefont {C.~E.}\ \bibnamefont {Thomas}}
  (\bibinfo {collaboration} {Hadron Spectrum}),\ }\href {\doibase
  10.1103/PhysRevD.87.034505, 10.1103/PhysRevD.90.099902} {\bibfield  {journal}
  {\bibinfo  {journal} {Phys. Rev.}\ }\textbf {\bibinfo {volume} {D87}},\
  \bibinfo {pages} {034505} (\bibinfo {year} {2013}{\natexlab{b}})},\ \bibinfo
  {note} {[Erratum: Phys. Rev.D90,no.9,099902(2014)]},\ \Eprint
  {http://arxiv.org/abs/1212.0830} {arXiv:1212.0830 [hep-ph]} \BibitemShut
  {NoStop}%
\bibitem [{\citenamefont {Dudek}\ \emph {et~al.}(2014)\citenamefont {Dudek},
  \citenamefont {Edwards}, \citenamefont {Thomas},\ and\ \citenamefont
  {Wilson}}]{Dudek:2014qha}%
  \BibitemOpen
  \bibfield  {author} {\bibinfo {author} {\bibfnamefont {J.~J.}\ \bibnamefont
  {Dudek}}, \bibinfo {author} {\bibfnamefont {R.~G.}\ \bibnamefont {Edwards}},
  \bibinfo {author} {\bibfnamefont {C.~E.}\ \bibnamefont {Thomas}}, \ and\
  \bibinfo {author} {\bibfnamefont {D.~J.}\ \bibnamefont {Wilson}} (\bibinfo
  {collaboration} {Hadron Spectrum}),\ }\href {\doibase
  10.1103/PhysRevLett.113.182001} {\bibfield  {journal} {\bibinfo  {journal}
  {Phys. Rev. Lett.}\ }\textbf {\bibinfo {volume} {113}},\ \bibinfo {pages}
  {182001} (\bibinfo {year} {2014})},\ \Eprint {http://arxiv.org/abs/1406.4158}
  {arXiv:1406.4158 [hep-ph]} \BibitemShut {NoStop}%
\bibitem [{\citenamefont {Briceño}\ \emph
  {et~al.}(2017{\natexlab{c}})\citenamefont {Briceño}, \citenamefont {Dudek},
  \citenamefont {Edwards},\ and\ \citenamefont {Wilson}}]{Briceno:2016mjc}%
  \BibitemOpen
  \bibfield  {author} {\bibinfo {author} {\bibfnamefont {R.~A.}\ \bibnamefont
  {Briceño}}, \bibinfo {author} {\bibfnamefont {J.~J.}\ \bibnamefont {Dudek}},
  \bibinfo {author} {\bibfnamefont {R.~G.}\ \bibnamefont {Edwards}}, \ and\
  \bibinfo {author} {\bibfnamefont {D.~J.}\ \bibnamefont {Wilson}},\ }\href
  {\doibase 10.1103/PhysRevLett.118.022002} {\bibfield  {journal} {\bibinfo
  {journal} {Phys. Rev. Lett.}\ }\textbf {\bibinfo {volume} {118}},\ \bibinfo
  {pages} {022002} (\bibinfo {year} {2017}{\natexlab{c}})},\ \Eprint
  {http://arxiv.org/abs/1607.05900} {arXiv:1607.05900 [hep-ph]} \BibitemShut
  {NoStop}%
\bibitem [{\citenamefont {Moir}\ \emph {et~al.}(2016)\citenamefont {Moir},
  \citenamefont {Peardon}, \citenamefont {Ryan}, \citenamefont {Thomas},\ and\
  \citenamefont {Wilson}}]{Moir:2016srx}%
  \BibitemOpen
  \bibfield  {author} {\bibinfo {author} {\bibfnamefont {G.}~\bibnamefont
  {Moir}}, \bibinfo {author} {\bibfnamefont {M.}~\bibnamefont {Peardon}},
  \bibinfo {author} {\bibfnamefont {S.~M.}\ \bibnamefont {Ryan}}, \bibinfo
  {author} {\bibfnamefont {C.~E.}\ \bibnamefont {Thomas}}, \ and\ \bibinfo
  {author} {\bibfnamefont {D.~J.}\ \bibnamefont {Wilson}},\ }\href {\doibase
  10.1007/JHEP10(2016)011} {\bibfield  {journal} {\bibinfo  {journal} {JHEP}\
  }\textbf {\bibinfo {volume} {10}},\ \bibinfo {pages} {011} (\bibinfo {year}
  {2016})},\ \Eprint {http://arxiv.org/abs/1607.07093} {arXiv:1607.07093
  [hep-lat]} \BibitemShut {NoStop}%
\bibitem [{\citenamefont {Wilson}\ \emph
  {et~al.}(2015{\natexlab{b}})\citenamefont {Wilson}, \citenamefont {Dudek},
  \citenamefont {Edwards},\ and\ \citenamefont {Thomas}}]{Wilson:2014cna}%
  \BibitemOpen
  \bibfield  {author} {\bibinfo {author} {\bibfnamefont {D.~J.}\ \bibnamefont
  {Wilson}}, \bibinfo {author} {\bibfnamefont {J.~J.}\ \bibnamefont {Dudek}},
  \bibinfo {author} {\bibfnamefont {R.~G.}\ \bibnamefont {Edwards}}, \ and\
  \bibinfo {author} {\bibfnamefont {C.~E.}\ \bibnamefont {Thomas}},\ }\href
  {\doibase 10.1103/PhysRevD.91.054008} {\bibfield  {journal} {\bibinfo
  {journal} {Phys. Rev.}\ }\textbf {\bibinfo {volume} {D91}},\ \bibinfo {pages}
  {054008} (\bibinfo {year} {2015}{\natexlab{b}})},\ \Eprint
  {http://arxiv.org/abs/1411.2004} {arXiv:1411.2004 [hep-ph]} \BibitemShut
  {NoStop}%
\bibitem [{\citenamefont {Edwards}\ \emph {et~al.}(2008)\citenamefont
  {Edwards}, \citenamefont {Joo},\ and\ \citenamefont {Lin}}]{Edwards:2008ja}%
  \BibitemOpen
  \bibfield  {author} {\bibinfo {author} {\bibfnamefont {R.~G.}\ \bibnamefont
  {Edwards}}, \bibinfo {author} {\bibfnamefont {B.}~\bibnamefont {Joo}}, \ and\
  \bibinfo {author} {\bibfnamefont {H.-W.}\ \bibnamefont {Lin}},\ }\href
  {\doibase 10.1103/PhysRevD.78.054501} {\bibfield  {journal} {\bibinfo
  {journal} {Phys. Rev.}\ }\textbf {\bibinfo {volume} {D78}},\ \bibinfo {pages}
  {054501} (\bibinfo {year} {2008})},\ \Eprint {http://arxiv.org/abs/0803.3960}
  {arXiv:0803.3960 [hep-lat]} \BibitemShut {NoStop}%
\bibitem [{\citenamefont {Lin}\ \emph {et~al.}(2009)\citenamefont {Lin} \emph
  {et~al.}}]{Lin:2008pr}%
  \BibitemOpen
  \bibfield  {author} {\bibinfo {author} {\bibfnamefont {H.-W.}\ \bibnamefont
  {Lin}} \emph {et~al.} (\bibinfo {collaboration} {Hadron Spectrum}),\ }\href
  {\doibase 10.1103/PhysRevD.79.034502} {\bibfield  {journal} {\bibinfo
  {journal} {Phys. Rev.}\ }\textbf {\bibinfo {volume} {D79}},\ \bibinfo {pages}
  {034502} (\bibinfo {year} {2009})},\ \Eprint {http://arxiv.org/abs/0810.3588}
  {arXiv:0810.3588 [hep-lat]} \BibitemShut {NoStop}%
\bibitem [{\citenamefont {Peardon}\ \emph {et~al.}(2009)\citenamefont
  {Peardon}, \citenamefont {Bulava}, \citenamefont {Foley}, \citenamefont
  {Morningstar}, \citenamefont {Dudek}, \citenamefont {Edwards}, \citenamefont
  {Joo}, \citenamefont {Lin}, \citenamefont {Richards},\ and\ \citenamefont
  {Juge}}]{Peardon:2009gh}%
  \BibitemOpen
  \bibfield  {author} {\bibinfo {author} {\bibfnamefont {M.}~\bibnamefont
  {Peardon}}, \bibinfo {author} {\bibfnamefont {J.}~\bibnamefont {Bulava}},
  \bibinfo {author} {\bibfnamefont {J.}~\bibnamefont {Foley}}, \bibinfo
  {author} {\bibfnamefont {C.}~\bibnamefont {Morningstar}}, \bibinfo {author}
  {\bibfnamefont {J.}~\bibnamefont {Dudek}}, \bibinfo {author} {\bibfnamefont
  {R.~G.}\ \bibnamefont {Edwards}}, \bibinfo {author} {\bibfnamefont
  {B.}~\bibnamefont {Joo}}, \bibinfo {author} {\bibfnamefont {H.-W.}\
  \bibnamefont {Lin}}, \bibinfo {author} {\bibfnamefont {D.~G.}\ \bibnamefont
  {Richards}}, \ and\ \bibinfo {author} {\bibfnamefont {K.~J.}\ \bibnamefont
  {Juge}} (\bibinfo {collaboration} {Hadron Spectrum}),\ }\href {\doibase
  10.1103/PhysRevD.80.054506} {\bibfield  {journal} {\bibinfo  {journal} {Phys.
  Rev.}\ }\textbf {\bibinfo {volume} {D80}},\ \bibinfo {pages} {054506}
  (\bibinfo {year} {2009})},\ \Eprint {http://arxiv.org/abs/0905.2160}
  {arXiv:0905.2160 [hep-lat]} \BibitemShut {NoStop}%
\bibitem [{\citenamefont {Hansen}\ and\ \citenamefont
  {Sharpe}(2012)}]{Hansen:2012tf}%
  \BibitemOpen
  \bibfield  {author} {\bibinfo {author} {\bibfnamefont {M.~T.}\ \bibnamefont
  {Hansen}}\ and\ \bibinfo {author} {\bibfnamefont {S.~R.}\ \bibnamefont
  {Sharpe}},\ }\href {\doibase 10.1103/PhysRevD.86.016007} {\bibfield
  {journal} {\bibinfo  {journal} {Phys. Rev.}\ }\textbf {\bibinfo {volume}
  {D86}},\ \bibinfo {pages} {016007} (\bibinfo {year} {2012})},\ \Eprint
  {http://arxiv.org/abs/1204.0826} {arXiv:1204.0826 [hep-lat]} \BibitemShut
  {NoStop}%
\bibitem [{\citenamefont {Briceño}\ and\ \citenamefont
  {Davoudi}(2013{\natexlab{b}})}]{Briceno:2012rv}%
  \BibitemOpen
  \bibfield  {author} {\bibinfo {author} {\bibfnamefont {R.~A.}\ \bibnamefont
  {Briceño}}\ and\ \bibinfo {author} {\bibfnamefont {Z.}~\bibnamefont
  {Davoudi}},\ }\href {\doibase 10.1103/PhysRevD.87.094507} {\bibfield
  {journal} {\bibinfo  {journal} {Phys. Rev.}\ }\textbf {\bibinfo {volume}
  {D87}},\ \bibinfo {pages} {094507} (\bibinfo {year} {2013}{\natexlab{b}})},\
  \Eprint {http://arxiv.org/abs/1212.3398} {arXiv:1212.3398 [hep-lat]}
  \BibitemShut {NoStop}%
\bibitem [{\citenamefont {Hansen}\ and\ \citenamefont
  {Sharpe}(2014)}]{Hansen:2014eka}%
  \BibitemOpen
  \bibfield  {author} {\bibinfo {author} {\bibfnamefont {M.~T.}\ \bibnamefont
  {Hansen}}\ and\ \bibinfo {author} {\bibfnamefont {S.~R.}\ \bibnamefont
  {Sharpe}},\ }\href {\doibase 10.1103/PhysRevD.90.116003} {\bibfield
  {journal} {\bibinfo  {journal} {Phys. Rev.}\ }\textbf {\bibinfo {volume}
  {D90}},\ \bibinfo {pages} {116003} (\bibinfo {year} {2014})},\ \Eprint
  {http://arxiv.org/abs/1408.5933} {arXiv:1408.5933 [hep-lat]} \BibitemShut
  {NoStop}%
\bibitem [{\citenamefont {Hansen}\ and\ \citenamefont
  {Sharpe}(2015)}]{Hansen:2015zga}%
  \BibitemOpen
  \bibfield  {author} {\bibinfo {author} {\bibfnamefont {M.~T.}\ \bibnamefont
  {Hansen}}\ and\ \bibinfo {author} {\bibfnamefont {S.~R.}\ \bibnamefont
  {Sharpe}},\ }\href {\doibase 10.1103/PhysRevD.92.114509} {\bibfield
  {journal} {\bibinfo  {journal} {Phys. Rev.}\ }\textbf {\bibinfo {volume}
  {D92}},\ \bibinfo {pages} {114509} (\bibinfo {year} {2015})},\ \Eprint
  {http://arxiv.org/abs/1504.04248} {arXiv:1504.04248 [hep-lat]} \BibitemShut
  {NoStop}%
\bibitem [{\citenamefont {Polejaeva}\ and\ \citenamefont
  {Rusetsky}(2012)}]{Polejaeva:2012ut}%
  \BibitemOpen
  \bibfield  {author} {\bibinfo {author} {\bibfnamefont {K.}~\bibnamefont
  {Polejaeva}}\ and\ \bibinfo {author} {\bibfnamefont {A.}~\bibnamefont
  {Rusetsky}},\ }\href {\doibase 10.1140/epja/i2012-12067-8} {\bibfield
  {journal} {\bibinfo  {journal} {Eur. Phys. J.}\ }\textbf {\bibinfo {volume}
  {A48}},\ \bibinfo {pages} {67} (\bibinfo {year} {2012})},\ \Eprint
  {http://arxiv.org/abs/1203.1241} {arXiv:1203.1241 [hep-lat]} \BibitemShut
  {NoStop}%
\bibitem [{\citenamefont {Döring}\ \emph {et~al.}(2018)\citenamefont
  {Döring}, \citenamefont {Hammer}, \citenamefont {Mai}, \citenamefont {Pang},
  \citenamefont {Rusetsky},\ and\ \citenamefont {Wu}}]{Doring:2018xxx}%
  \BibitemOpen
  \bibfield  {author} {\bibinfo {author} {\bibfnamefont {M.}~\bibnamefont
  {Döring}}, \bibinfo {author} {\bibfnamefont {H.~W.}\ \bibnamefont {Hammer}},
  \bibinfo {author} {\bibfnamefont {M.}~\bibnamefont {Mai}}, \bibinfo {author}
  {\bibfnamefont {J.~Y.}\ \bibnamefont {Pang}}, \bibinfo {author}
  {\bibfnamefont {A.}~\bibnamefont {Rusetsky}}, \ and\ \bibinfo {author}
  {\bibfnamefont {J.}~\bibnamefont {Wu}},\ }\href@noop {} {\  (\bibinfo {year}
  {2018})},\ \Eprint {http://arxiv.org/abs/1802.03362} {arXiv:1802.03362
  [hep-lat]} \BibitemShut {NoStop}%
\bibitem [{\citenamefont {Chew}\ and\ \citenamefont
  {Mandelstam}(1960)}]{Chew:1960iv}%
  \BibitemOpen
  \bibfield  {author} {\bibinfo {author} {\bibfnamefont {G.~F.}\ \bibnamefont
  {Chew}}\ and\ \bibinfo {author} {\bibfnamefont {S.}~\bibnamefont
  {Mandelstam}},\ }\href {\doibase 10.1103/PhysRev.119.467} {\bibfield
  {journal} {\bibinfo  {journal} {Phys. Rev.}\ }\textbf {\bibinfo {volume}
  {119}},\ \bibinfo {pages} {467} (\bibinfo {year} {1960})}\BibitemShut
  {NoStop}%
\bibitem [{\citenamefont {Stapp}\ \emph {et~al.}(1957)\citenamefont {Stapp},
  \citenamefont {Ypsilantis},\ and\ \citenamefont {Metropolis}}]{Stapp:1956mz}%
  \BibitemOpen
  \bibfield  {author} {\bibinfo {author} {\bibfnamefont {H.~P.}\ \bibnamefont
  {Stapp}}, \bibinfo {author} {\bibfnamefont {T.~J.}\ \bibnamefont
  {Ypsilantis}}, \ and\ \bibinfo {author} {\bibfnamefont {N.}~\bibnamefont
  {Metropolis}},\ }\href {\doibase 10.1103/PhysRev.105.302} {\bibfield
  {journal} {\bibinfo  {journal} {Phys. Rev.}\ }\textbf {\bibinfo {volume}
  {105}},\ \bibinfo {pages} {302} (\bibinfo {year} {1957})}\BibitemShut
  {NoStop}%
\bibitem [{\citenamefont {Edwards}\ \emph {et~al.}(2011)\citenamefont
  {Edwards}, \citenamefont {Dudek}, \citenamefont {Richards},\ and\
  \citenamefont {Wallace}}]{Edwards:2011jj}%
  \BibitemOpen
  \bibfield  {author} {\bibinfo {author} {\bibfnamefont {R.~G.}\ \bibnamefont
  {Edwards}}, \bibinfo {author} {\bibfnamefont {J.~J.}\ \bibnamefont {Dudek}},
  \bibinfo {author} {\bibfnamefont {D.~G.}\ \bibnamefont {Richards}}, \ and\
  \bibinfo {author} {\bibfnamefont {S.~J.}\ \bibnamefont {Wallace}},\ }\href
  {\doibase 10.1103/PhysRevD.84.074508} {\bibfield  {journal} {\bibinfo
  {journal} {Phys. Rev.}\ }\textbf {\bibinfo {volume} {D84}},\ \bibinfo {pages}
  {074508} (\bibinfo {year} {2011})},\ \Eprint {http://arxiv.org/abs/1104.5152}
  {arXiv:1104.5152 [hep-ph]} \BibitemShut {NoStop}%
\bibitem [{\citenamefont {Andersen}\ \emph {et~al.}(2019)\citenamefont
  {Andersen}, \citenamefont {Bulava}, \citenamefont {Hörz},\ and\
  \citenamefont {Morningstar}}]{Andersen:2018mau}%
  \BibitemOpen
  \bibfield  {author} {\bibinfo {author} {\bibfnamefont {C.}~\bibnamefont
  {Andersen}}, \bibinfo {author} {\bibfnamefont {J.}~\bibnamefont {Bulava}},
  \bibinfo {author} {\bibfnamefont {B.}~\bibnamefont {Hörz}}, \ and\ \bibinfo
  {author} {\bibfnamefont {C.}~\bibnamefont {Morningstar}},\ }\href {\doibase
  10.1016/j.nuclphysb.2018.12.018} {\bibfield  {journal} {\bibinfo  {journal}
  {Nucl. Phys.}\ }\textbf {\bibinfo {volume} {B939}},\ \bibinfo {pages} {145}
  (\bibinfo {year} {2019})},\ \Eprint {http://arxiv.org/abs/1808.05007}
  {arXiv:1808.05007 [hep-lat]} \BibitemShut {NoStop}%
\bibitem [{\citenamefont {Leskovec}\ \emph {et~al.}(2018)\citenamefont
  {Leskovec}, \citenamefont {Alexandrou}, \citenamefont {Meinel}, \citenamefont
  {Negele}, \citenamefont {Paul}, \citenamefont {Petschlies}, \citenamefont
  {Pochinsky}, \citenamefont {Rendon},\ and\ \citenamefont
  {Syritsyn}}]{Leskovec:2018tju}%
  \BibitemOpen
  \bibfield  {author} {\bibinfo {author} {\bibfnamefont {L.}~\bibnamefont
  {Leskovec}}, \bibinfo {author} {\bibfnamefont {C.}~\bibnamefont
  {Alexandrou}}, \bibinfo {author} {\bibfnamefont {S.}~\bibnamefont {Meinel}},
  \bibinfo {author} {\bibfnamefont {J.~W.}\ \bibnamefont {Negele}}, \bibinfo
  {author} {\bibfnamefont {S.}~\bibnamefont {Paul}}, \bibinfo {author}
  {\bibfnamefont {M.}~\bibnamefont {Petschlies}}, \bibinfo {author}
  {\bibfnamefont {A.}~\bibnamefont {Pochinsky}}, \bibinfo {author}
  {\bibfnamefont {G.}~\bibnamefont {Rendon}}, \ and\ \bibinfo {author}
  {\bibfnamefont {S.}~\bibnamefont {Syritsyn}},\ }in\ \href@noop {} {\emph
  {\bibinfo {booktitle} {{13th Conference on the Intersections of Particle and
  Nuclear Physics (CIPANP 2018) Palm Springs, California, USA, May 29-June 3,
  2018}}}}\ (\bibinfo {year} {2018})\ \Eprint {http://arxiv.org/abs/1810.01927}
  {arXiv:1810.01927 [hep-lat]} \BibitemShut {NoStop}%
\bibitem [{\citenamefont {Alexandrou}\ \emph {et~al.}(2017)\citenamefont
  {Alexandrou}, \citenamefont {Leskovec}, \citenamefont {Meinel}, \citenamefont
  {Negele}, \citenamefont {Paul}, \citenamefont {Petschlies}, \citenamefont
  {Pochinsky}, \citenamefont {Rendon},\ and\ \citenamefont
  {Syritsyn}}]{Alexandrou:2017mpi}%
  \BibitemOpen
  \bibfield  {author} {\bibinfo {author} {\bibfnamefont {C.}~\bibnamefont
  {Alexandrou}}, \bibinfo {author} {\bibfnamefont {L.}~\bibnamefont
  {Leskovec}}, \bibinfo {author} {\bibfnamefont {S.}~\bibnamefont {Meinel}},
  \bibinfo {author} {\bibfnamefont {J.}~\bibnamefont {Negele}}, \bibinfo
  {author} {\bibfnamefont {S.}~\bibnamefont {Paul}}, \bibinfo {author}
  {\bibfnamefont {M.}~\bibnamefont {Petschlies}}, \bibinfo {author}
  {\bibfnamefont {A.}~\bibnamefont {Pochinsky}}, \bibinfo {author}
  {\bibfnamefont {G.}~\bibnamefont {Rendon}}, \ and\ \bibinfo {author}
  {\bibfnamefont {S.}~\bibnamefont {Syritsyn}},\ }\href {\doibase
  10.1103/PhysRevD.96.034525} {\bibfield  {journal} {\bibinfo  {journal} {Phys.
  Rev.}\ }\textbf {\bibinfo {volume} {D96}},\ \bibinfo {pages} {034525}
  (\bibinfo {year} {2017})},\ \Eprint {http://arxiv.org/abs/1704.05439}
  {arXiv:1704.05439 [hep-lat]} \BibitemShut {NoStop}%
\bibitem [{\citenamefont {Prelovsek}\ \emph {et~al.}(2013)\citenamefont
  {Prelovsek}, \citenamefont {Leskovec}, \citenamefont {Lang},\ and\
  \citenamefont {Mohler}}]{Prelovsek:2013ela}%
  \BibitemOpen
  \bibfield  {author} {\bibinfo {author} {\bibfnamefont {S.}~\bibnamefont
  {Prelovsek}}, \bibinfo {author} {\bibfnamefont {L.}~\bibnamefont {Leskovec}},
  \bibinfo {author} {\bibfnamefont {C.~B.}\ \bibnamefont {Lang}}, \ and\
  \bibinfo {author} {\bibfnamefont {D.}~\bibnamefont {Mohler}},\ }\href
  {\doibase 10.1103/PhysRevD.88.054508} {\bibfield  {journal} {\bibinfo
  {journal} {Phys. Rev.}\ }\textbf {\bibinfo {volume} {D88}},\ \bibinfo {pages}
  {054508} (\bibinfo {year} {2013})},\ \Eprint {http://arxiv.org/abs/1307.0736}
  {arXiv:1307.0736 [hep-lat]} \BibitemShut {NoStop}%
\bibitem [{\citenamefont {Bali}\ \emph {et~al.}(2016)\citenamefont {Bali},
  \citenamefont {Collins}, \citenamefont {Cox}, \citenamefont {Donald},
  \citenamefont {Göckeler}, \citenamefont {Lang},\ and\ \citenamefont
  {Schäfer}}]{Bali:2015gji}%
  \BibitemOpen
  \bibfield  {author} {\bibinfo {author} {\bibfnamefont {G.~S.}\ \bibnamefont
  {Bali}}, \bibinfo {author} {\bibfnamefont {S.}~\bibnamefont {Collins}},
  \bibinfo {author} {\bibfnamefont {A.}~\bibnamefont {Cox}}, \bibinfo {author}
  {\bibfnamefont {G.}~\bibnamefont {Donald}}, \bibinfo {author} {\bibfnamefont
  {M.}~\bibnamefont {Göckeler}}, \bibinfo {author} {\bibfnamefont {C.~B.}\
  \bibnamefont {Lang}}, \ and\ \bibinfo {author} {\bibfnamefont
  {A.}~\bibnamefont {Schäfer}} (\bibinfo {collaboration} {RQCD}),\ }\href
  {\doibase 10.1103/PhysRevD.93.054509} {\bibfield  {journal} {\bibinfo
  {journal} {Phys. Rev.}\ }\textbf {\bibinfo {volume} {D93}},\ \bibinfo {pages}
  {054509} (\bibinfo {year} {2016})},\ \Eprint
  {http://arxiv.org/abs/1512.08678} {arXiv:1512.08678 [hep-lat]} \BibitemShut
  {NoStop}%
\bibitem [{\citenamefont {Brett}\ \emph {et~al.}(2018)\citenamefont {Brett},
  \citenamefont {Bulava}, \citenamefont {Fallica}, \citenamefont {Hanlon},
  \citenamefont {Hörz},\ and\ \citenamefont {Morningstar}}]{Brett:2018jqw}%
  \BibitemOpen
  \bibfield  {author} {\bibinfo {author} {\bibfnamefont {R.}~\bibnamefont
  {Brett}}, \bibinfo {author} {\bibfnamefont {J.}~\bibnamefont {Bulava}},
  \bibinfo {author} {\bibfnamefont {J.}~\bibnamefont {Fallica}}, \bibinfo
  {author} {\bibfnamefont {A.}~\bibnamefont {Hanlon}}, \bibinfo {author}
  {\bibfnamefont {B.}~\bibnamefont {Hörz}}, \ and\ \bibinfo {author}
  {\bibfnamefont {C.}~\bibnamefont {Morningstar}},\ }\href {\doibase
  10.1016/j.nuclphysb.2018.05.008} {\bibfield  {journal} {\bibinfo  {journal}
  {Nucl. Phys.}\ }\textbf {\bibinfo {volume} {B932}},\ \bibinfo {pages} {29}
  (\bibinfo {year} {2018})},\ \Eprint {http://arxiv.org/abs/1802.03100}
  {arXiv:1802.03100 [hep-lat]} \BibitemShut {NoStop}%
\bibitem [{\citenamefont {Rendon}\ \emph {et~al.}(2018)\citenamefont {Rendon},
  \citenamefont {Leskovec}, \citenamefont {Meinel}, \citenamefont {Negele},
  \citenamefont {Paul}, \citenamefont {Petschlies}, \citenamefont {Pochinsky},
  \citenamefont {Silvi},\ and\ \citenamefont {Syritsyn}}]{Rendon:2018fem}%
  \BibitemOpen
  \bibfield  {author} {\bibinfo {author} {\bibfnamefont {G.}~\bibnamefont
  {Rendon}}, \bibinfo {author} {\bibfnamefont {L.}~\bibnamefont {Leskovec}},
  \bibinfo {author} {\bibfnamefont {S.}~\bibnamefont {Meinel}}, \bibinfo
  {author} {\bibfnamefont {J.}~\bibnamefont {Negele}}, \bibinfo {author}
  {\bibfnamefont {S.}~\bibnamefont {Paul}}, \bibinfo {author} {\bibfnamefont
  {M.}~\bibnamefont {Petschlies}}, \bibinfo {author} {\bibfnamefont
  {A.}~\bibnamefont {Pochinsky}}, \bibinfo {author} {\bibfnamefont
  {G.}~\bibnamefont {Silvi}}, \ and\ \bibinfo {author} {\bibfnamefont
  {S.}~\bibnamefont {Syritsyn}},\ }in\ \href@noop {} {\emph {\bibinfo
  {booktitle} {{36th International Symposium on Lattice Field Theory (Lattice
  2018) East Lansing, MI, United States, July 22-28, 2018}}}}\ (\bibinfo {year}
  {2018})\ \Eprint {http://arxiv.org/abs/1811.10750} {arXiv:1811.10750
  [hep-lat]} \BibitemShut {NoStop}%
\bibitem [{\citenamefont {Edwards}\ and\ \citenamefont
  {Joo}(2005)}]{Edwards:2004sx}%
  \BibitemOpen
  \bibfield  {author} {\bibinfo {author} {\bibfnamefont {R.~G.}\ \bibnamefont
  {Edwards}}\ and\ \bibinfo {author} {\bibfnamefont {B.}~\bibnamefont {Joo}}
  (\bibinfo {collaboration} {SciDAC, LHPC, UKQCD}),\ }\bibfield  {booktitle}
  {\emph {\bibinfo {booktitle} {{Lattice field theory. Proceedings, 22nd
  International Symposium, Lattice 2004, Batavia, USA, June 21-26, 2004}}},\
  }\href {\doibase 10.1016/j.nuclphysbps.2004.11.254} {\bibfield  {journal}
  {\bibinfo  {journal} {Nucl. Phys. Proc. Suppl.}\ }\textbf {\bibinfo {volume}
  {140}},\ \bibinfo {pages} {832} (\bibinfo {year} {2005})},\ \Eprint
  {http://arxiv.org/abs/hep-lat/0409003} {arXiv:hep-lat/0409003 [hep-lat]}
  \BibitemShut {NoStop}%
\bibitem [{\citenamefont {Clark}\ \emph {et~al.}(2010)\citenamefont {Clark},
  \citenamefont {Babich}, \citenamefont {Barros}, \citenamefont {Brower},\ and\
  \citenamefont {Rebbi}}]{Clark:2009wm}%
  \BibitemOpen
  \bibfield  {author} {\bibinfo {author} {\bibfnamefont {M.~A.}\ \bibnamefont
  {Clark}}, \bibinfo {author} {\bibfnamefont {R.}~\bibnamefont {Babich}},
  \bibinfo {author} {\bibfnamefont {K.}~\bibnamefont {Barros}}, \bibinfo
  {author} {\bibfnamefont {R.~C.}\ \bibnamefont {Brower}}, \ and\ \bibinfo
  {author} {\bibfnamefont {C.}~\bibnamefont {Rebbi}},\ }\href {\doibase
  10.1016/j.cpc.2010.05.002} {\bibfield  {journal} {\bibinfo  {journal}
  {Comput. Phys. Commun.}\ }\textbf {\bibinfo {volume} {181}},\ \bibinfo
  {pages} {1517} (\bibinfo {year} {2010})},\ \Eprint
  {http://arxiv.org/abs/0911.3191} {arXiv:0911.3191 [hep-lat]} \BibitemShut
  {NoStop}%
\bibitem [{\citenamefont {Babich}\ \emph {et~al.}(2010)\citenamefont {Babich},
  \citenamefont {Clark},\ and\ \citenamefont {Joo}}]{Babich:2010mu}%
  \BibitemOpen
  \bibfield  {author} {\bibinfo {author} {\bibfnamefont {R.}~\bibnamefont
  {Babich}}, \bibinfo {author} {\bibfnamefont {M.~A.}\ \bibnamefont {Clark}}, \
  and\ \bibinfo {author} {\bibfnamefont {B.}~\bibnamefont {Joo}},\ }in\ \href
  {http://www1.jlab.org/Ul/publications/view_pub.cfm?pub_id=10186} {\emph
  {\bibinfo {booktitle} {{SC 10 (Supercomputing 2010) New Orleans, Louisiana,
  November 13-19, 2010}}}}\ (\bibinfo {year} {2010})\ \Eprint
  {http://arxiv.org/abs/1011.0024} {arXiv:1011.0024 [hep-lat]} \BibitemShut
  {NoStop}%
\bibitem [{\citenamefont {Blatt}\ and\ \citenamefont
  {Biedenharn}(1952)}]{Blatt:1952zz}%
  \BibitemOpen
  \bibfield  {author} {\bibinfo {author} {\bibfnamefont {J.~M.}\ \bibnamefont
  {Blatt}}\ and\ \bibinfo {author} {\bibfnamefont {L.~C.}\ \bibnamefont
  {Biedenharn}},\ }\href {\doibase 10.1103/RevModPhys.24.258} {\bibfield
  {journal} {\bibinfo  {journal} {Rev. Mod. Phys.}\ }\textbf {\bibinfo {volume}
  {24}},\ \bibinfo {pages} {258} (\bibinfo {year} {1952})}\BibitemShut
  {NoStop}%
\end{thebibliography}%

\appendix

\section{Generalised $n$-channel Stapp-parameterization \label{App:Generalised-Stapp}}


In this appendix we present a construction for a parameterization that naturally extends the two-channel Stapp parameterization~\cite{Stapp:1956mz} to $n$-channels, preserving the notion of $n$ phase-shifts and $n(n-1)/2$ mixing-angles. We begin by defining the exponential map from the Lie Algebra $\text{LU}(n)$ to the Lie Group $\text{U}(n)$ as,
\begin{align}\label{Eq:exp_map}
 \text{Exp: }\text{LU}(n)&\rightarrow \text{U}(n) \nonumber \\
 X&\rightarrow \exp(iX).
\end{align}
With this definition, a basis for $\text{LU}(n)$ is given by the set of $n^2$, $n\times n$ Hermitian matrices. A convenient choice are the sets $\{\Delta_{i} | 1\leq i \leq n\}$, $\{\Theta_{ij} |\, 1\leq i < j \leq n \}$ and $\{\Psi_{ij} |\, 1\leq i < j \leq n \}$  where
\begin{align}
(\Delta_i)_{ab}=&\delta_{ia}\delta_{ib} \quad (\text{no sum on }i) \\
(\Theta_{ij})_{ab}=&\delta_{ia}\delta_{jb} + \delta_{ja}\delta_{ib}, \quad (i<j)\\
(\Psi_{ij})_{ab}=&i\delta_{ia}\delta_{jb} - i\delta_{ja}\delta_{ib} \quad (i<j). 
\end{align}
In order to construct a general $n\times n$ symmetric unitary matrix $S$, we exponentiate the subset of $n(n+1)/2$ symmetric matrices, $\{\Delta_i,\Theta_{jk}\}$, and take $S=BB^T$ where
\begin{align}\label{Eq:B}
B = & \exp(i\delta_1\Delta_1)\exp(i\delta_2\Delta_2)...\exp(i\delta_n\Delta_n)\nonumber \\
&\times \exp(i\bar{\epsilon}_{n-1 n}\Theta_{n-1 n})...\exp(i\bar{\epsilon}_{12}\Theta_{12}).
\end{align}
Here $B^T$ denotes the matrix transpose of $B$ and $\{\delta_i,\bar{\epsilon}_{jk}\}$ are a set of $n(n+1)/2$ real parameters. 

With this choice, for two channels, $\delta_1$, $\delta_2$ and $\bar{\epsilon}_{12}$ are exactly the Stapp phase-shifts and mixing-angle of Ref.~\cite{Stapp:1956mz}. If instead we take $S = \tilde{B} \tilde{B}^T $, where
\begin{align}\label{Eq:BB}
\tilde{B} = & \exp(i\theta_{n-1 n}\Psi_{n-1 n})...\exp(i\theta_{12}\Psi_{12})
 \nonumber \\
&\times\exp(i\tilde{\delta}_1\Delta_1) \exp(i\tilde{\delta}_2\Delta_2)...\exp(i\tilde{\delta}_n\Delta_n) \, ,
\end{align}
we obtain a parameterization similar to that of Blatt and Biedenharn~\cite{Blatt:1952zz} where $\tilde{\delta}$ are the eigen-phaseshifts and $\theta$ are some mixing-angles.

We use the indexing $\bar{\epsilon}_{ij}$ and $\Theta_{ij}$ to conveniently label the angle and matrix respectively that \emph{mix} channels $i$ and $j$. By construction, this parameterization gives a symmetric unitary matrix with $n(n+1)/2$ independent free parameters and provides a natural $n$-channel extension of the two-channel Stapp parameterization.

\subsection{$n=2$} 
For two-channels, the basis construction above gives the matrices
\begin{align*}
\Delta_{1} = 
\begin{pmatrix} 
1 & 0 \\
0 & 0 
\end{pmatrix} , \quad
& \Delta_{2} = 
\begin{pmatrix} 
0 & 0 \\
0 & 1 
\end{pmatrix} , \; \\
\Theta_{12} =
\begin{pmatrix} 
0 & 1 \\
1 & 0
\end{pmatrix} , \quad
& \Psi_{12} =
\begin{pmatrix} 
0 & i \\
-i & 0
\end{pmatrix} .
\end{align*}
It follows that setting $n=2$ in Eq.~\ref{Eq:B} gives,
\begin{equation}\label{Eq:Stapp}
S=\begin{pmatrix} 
\cos(2\bar{\epsilon}_{12})\, e^{2i\delta_1} & i\sin(2\bar{\epsilon}_{12})\, e^{i(\delta_1+ \delta_2)} \\
i\sin(2\bar{\epsilon}_{12})\, e^{i(\delta_1 + \delta_2)} & \cos(2\bar{\epsilon}_{12}) \, e^{2i\delta_2}
\end{pmatrix}
\end{equation}
which is precisely the Stapp-parameterization. 

\subsection{$n=3$}
The generalized three-channel Stapp-parameterization has $6$ free real-parameters (three phase-shifts and three mixing-angles) and is obtained by taking $n=3$ in Eq.~\ref{Eq:B}. Fixing $\bar{\epsilon}_{13}=0$ and $\bar{\epsilon}_{23}=0$ reduces to the two-channel Stapp-parameterization in channels $1$ and $2$, and leaves a single phase-shift in the channel $3$. An analogous reduction applies for other appropriate combinations of mixing-angles taken to be zero. Explicitly, the elements of the $S$-matrix are
\begin{align}\label{Eq:n3}
S_{11} =&\, \big( \chi_{12}\,c^2_{13}-s^2_{13}  \big) \, e^{2i\delta_1} \nonumber \\[0.6ex]
S_{12} =&\, c_{13} \big(i \sigma_{12}c_{23}-s_{13}s_{23}(1+\chi_{12}) \big) \, e^{i(\delta_1+ \delta_2)}\nonumber \\[0.6ex]
S_{13} =&\, c_{13} \big(ic_{23}s_{13}(1+ \chi_{12}) - \sigma_{12} s_{23}  \big) \, e^{i(\delta_1+\delta_3)} \nonumber\\[0.6ex]
S_{22} =&\, \big(\chi_{12} \,c^2_{23} + \chi_{12} \, s^2_{13}s^2_{23} - c^2_{13}s^2_{23} - 2 i \sigma_{12}s_{13}s_{23}c_{23} \big) \, e^{2i\delta_2}\nonumber\\[0.6ex]
S_{23} =&\, \big(\sigma_{12} s_{13}\, (s^2_{23}-c^2_{23}) + ic^2_{13}c_{23}s_{23}(1+ \chi_{12}) \big)e^{i(\delta_2+\delta_3)}\nonumber\\[0.6ex]
S_{33} =&\,  \big(c^2_{13} c^2_{23} - \chi_{12}s^2_{13}c^2_{23} - \chi_{12}s^2_{23} - 2i \sigma_{12}s_{13}s_{23}c_{23} \big) \, e^{2i\delta_3}
\end{align}
where
\begin{align*}
\chi_{12}&=\cos(2\bar{\epsilon}_{12}),\, c_{13}=\cos(\bar{\epsilon}_{13}),\, c_{23}=\cos(\bar{\epsilon}_{23}) \\
\sigma_{12}&=\sin(2\bar{\epsilon}_{12}),\, s_{13}=\sin(\bar{\epsilon}_{13}),\, s_{23}=\sin(\bar{\epsilon}_{23}) .
\end{align*}
These conventions mean that $\delta_1$ is equal to $\arg(S_{11})$, which is in agreement with the conventions in Refs.~\cite{Dudek:2016cru,Moir:2016srx,Briceno:2017qmb} where the phase-shift is defined as $\delta_i = \arg(S_{ii})$. However, we see for $\delta_2$ and $\delta_3$ there are corrections to the phase due to the imaginary components $\propto \sigma_{12}s_{13}s_{23}c_{23}$ in the expressions for $S_{22}$ and $S_{33}$, given in Eq.~\ref{Eq:n3}. For a very weakly mixed channel these corrections are very small and $\delta_i \approx \arg(S_{ii})$ for $i=2,3$.

\subsection{$n=5$}

\begin{figure}[tb]
	\centering
	\includegraphics[trim={0 0 0 0},clip,width=0.5\textwidth]{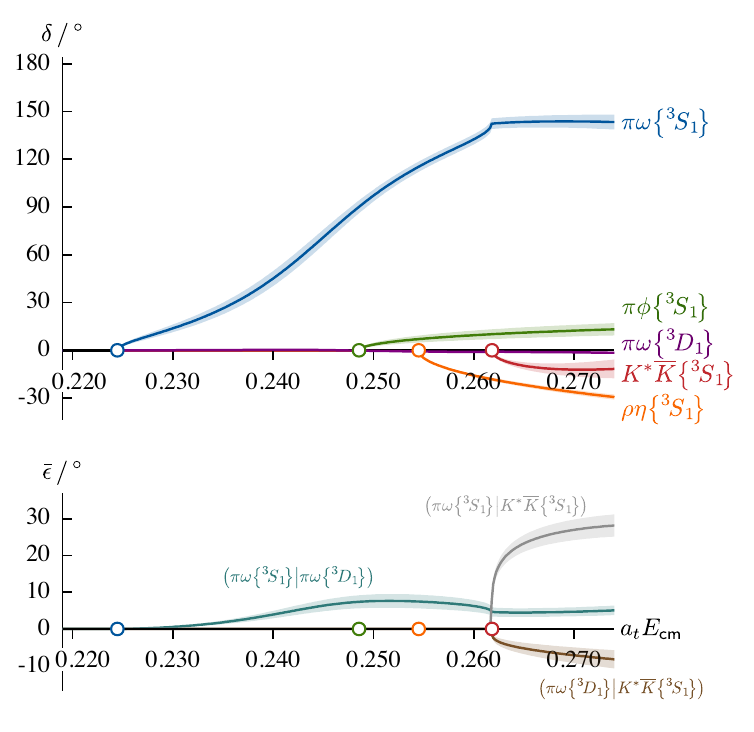}
	\caption{
	\textbf{Upper}: As in Figure~\ref{Fig:B_phase_shift} but for the $\piomegaS$ (blue), $\piomegaD$ (purple), $\piphiS$ (green), $\rho\eta\{\threeSone\}$ (orange) and $K^* \overline{K}\{\threeSone\}$ (red) phase-shifts for the reference amplitude in Eq.~\ref{Fit:E}. The faded error bands reflect the statistical uncertainty on the scattering parameters. The $\rho\eta$ and $K^* \overline{K}$ ``thresholds'' are calculated using the $\rho$ and $K^*$ masses given in Section~\ref{Subsec:5chan}.
	\textbf{Lower}: As upper but for the mixing-angles $\bar{\epsilon}(\piomegaS|\piomegaD)$ (blue), $\bar{\epsilon}(\piomegaS|K^* \overline{K}\{\threeSone\})$ (gray) and $\bar{\epsilon}(\piomegaD|K^* \overline{K}\{\threeSone\})$ (brown). All other mixing-angles are extremely small and consistent with zero as discussed in the text. 
	}
	\label{Fig:E_phase_shift_angles}
\end{figure}

For the limited five coupled-channel analysis of $\big( \piomegaS, \piomegaD, \piphiS, \rho\eta \big\{ \!\threeSone \big\}, K^* \overline{K}\big\{ \!\threeSone \big\} \big)$ given in Section~\ref{Subsec:5chan}, we calculate the \emph{five} phase-shifts and \emph{ten} mixing-angles. We find that seven of the mixing-angles, all featuring either $\piphiS$ and/or $\rho\eta\{\threeSone\}$, are extremely small and consistent with zero, in agreement with the observation that both are decoupled from the resonance as shown in Eq.~\ref{Eq:Couplings2}. This illustrates the natural reduction from the five-channel parameterization to the three-channel parameterization in the case that two channels decouple. The five phase-shifts and the remaining three non-zero mixing-angles are presented in Figure~\ref{Fig:E_phase_shift_angles}.

\section{Tables of Operators \label{App:Tables}}


We present here tables of operators as referred to in the text. \\[1.3ex]
\begin{table}[H]
\setlength{\tabcolsep}{2pt}
\small
{\renewcommand{\arraystretch}{1.2}
\begin{tabularx}{0.45\textwidth}{l @{\extracolsep{\fill}} ccc}
\toprule
$L/a_s$ & 16 & 20 & 24 \\
\cmidrule(lr{0em}){1-4}
\multirow{2}{*}{$\rho_{[000],\,T_1^-}$} & ${26}\times \bar{\psi}\bm{\Gamma}\psi$ & ${26}\times \bar{\psi}\bm{\Gamma}\psi$ & ${12}\times \bar{\psi}\bm{\Gamma}\psi$ \\[0.5ex]
& ${3}\times \pi\pi$ & ${2}\times \pi\pi$ & \\[0.5ex]
\cdashlinelr{1-4}
\multirow{2}{*}{$\rho_{[001],\,A_1}$} & ${8}\times \bar{\psi}\bm{\Gamma}\psi$ & ${18}\times \bar{\psi}\bm{\Gamma}\psi$ & ${18}\times \bar{\psi}\bm{\Gamma}\psi$ \\[0.5ex]
& ${4}\times \pi\pi$ & ${4}\times \pi\pi$ & ${4}\times \pi\pi$ \\[0.5ex]
\cdashlinelr{1-4}
\multirow{2}{*}{$\rho_{[011],\,A_1}$} & ${27}\times \bar{\psi}\bm{\Gamma}\psi$ & ${27}\times \bar{\psi}\bm{\Gamma}\psi$ & ${27}\times \bar{\psi}\bm{\Gamma}\psi$ \\[0.5ex]
& ${3}\times \pi\pi$ & ${3}\times \pi\pi$ & ${3}\times \pi\pi$ \\[0.5ex]
\cdashlinelr{1-4}
\multirow{2}{*}{$\rho_{[111],\,A_1}$} & ${8}\times \bar{\psi}\bm{\Gamma}\psi$ & ${21}\times \bar{\psi}\bm{\Gamma}\psi$ & ${21}\times \bar{\psi}\bm{\Gamma}\psi$ \\[0.5ex]
& ${3}\times \pi\pi$ & ${3}\times \pi\pi$ & ${3}\times \pi\pi$ \\[0.5ex]
\bottomrule
\end{tabularx}
}
\caption{Single-meson and two-meson operators used to compute optimised $\rho$ operators in the $[000]T_1^-$ irrep and $\vec{P}A_1$ irreps at various overall momenta on the three volumes. Momentum labels on the $\pi$'s that form the $\pi\pi$ operators are omitted for brevity.}
\label{Tab:ops_rho_proj}
\end{table} 
%
%
\begin{table}[H]
	\setlength{\tabcolsep}{2pt} 
\small
{\renewcommand{\arraystretch}{1.2}
\begin{tabularx}{0.45\textwidth}{l @{\extracolsep{\fill}} cc}
\toprule
$L/a_s$ & 16 & 20 \\
\cmidrule(lr{0em}){1-3}
\multirow{3}{*}{${a_0}_{[001],\,A_1}$} & ${14}\times \bar{\psi}\bm{\Gamma}\psi$ & ${14}\times \bar{\psi}\bm{\Gamma}\psi$ \\[0.5ex]
& ${4}\times \pi\eta$ & ${4}\times \pi\eta$\\[0.5ex]
& ${2}\times \bar{K}K$ & ${2}\times \bar{K}K$\\[0.5ex]
\cdashlinelr{1-3}
\multirow{3}{*}{${a_0}_{[011],\,A_1}$} & ${18}\times \bar{\psi}\bm{\Gamma}\psi$ & ${18}\times \bar{\psi}\bm{\Gamma}\psi$ \\[0.5ex]
& ${4}\times \pi\eta$ & ${4}\times \pi\eta$ \\[0.5ex]
& ${2}\times \bar{K}K$ & ${2}\times \bar{K}K$ \\[0.5ex]
\cdashlinelr{1-3}
\multirow{3}{*}{${a_0}_{[111],\,A_1}$} & & ${15}\times \bar{\psi}\bm{\Gamma}\psi$ \\[0.5ex]
& & ${4}\times \pi\eta$ \\[0.5ex]
& & ${2}\times \bar{K}K$ \\[0.5ex]
\bottomrule
\end{tabularx}
}
\caption{As Table~\ref{Tab:ops_rho_proj} but for optimised $a_0$ operators.}
\label{Tab:ops_a0_proj}
\end{table}
%
%
\begin{table}[H]
	\setlength{\tabcolsep}{3pt}
\small
{\renewcommand{\arraystretch}{1.2}
\begin{tabularx}{0.45\textwidth}{l @{\extracolsep{\fill}} ccc}
\toprule
$L/a_s$ & 16 & 20 & 24 \\
\cmidrule(lr{0em}){1-4}
\multirow{1}{*}{$K^*_{[000],\,T_1^-}$} & ${6}\times \bar{\psi}\bm{\Gamma}\psi$ & ${16}\times \bar{\psi}\bm{\Gamma}\psi$ & ${9}\times \bar{\psi}\bm{\Gamma}\psi$ \\[0.5ex]
\cdashlinelr{1-4}
\multirow{2}{*}{$K^*_{[001],\,A_1}$} & ${8}\times \bar{\psi}\bm{\Gamma}\psi$ & ${16}\times \bar{\psi}\bm{\Gamma}\psi$ & ${8}\times \bar{\psi}\bm{\Gamma}\psi$ \\[0.5ex]
& ${2}\times \pi K$ & ${6}\times \pi K$ &  \\[0.5ex]
\cdashlinelr{1-4}
\multirow{2}{*}{$K^*_{[011],\,A_1}$} & ${8}\times \bar{\psi}\bm{\Gamma}\psi$ & ${26}\times \bar{\psi}\bm{\Gamma}\psi$ &  \\[0.5ex]
& ${3}\times \pi K$ & ${6}\times \pi K$ & \\[0.5ex]
\cdashlinelr{1-4}
\multirow{2}{*}{$K^*_{[111],\,A_1}$} & ${8}\times \bar{\psi}\bm{\Gamma}\psi$ & ${9}\times \bar{\psi}\bm{\Gamma}\psi$ & ${9}\times \bar{\psi}\bm{\Gamma}\psi$ \\[0.5ex]
& ${4}\times \pi K$ & ${4}\times \pi K$ &  \\[0.5ex]
\bottomrule
\end{tabularx}
} 
\caption{As Table~\ref{Tab:ops_rho_proj} but for optimised $K^*$ operators.}
\label{Tab:ops_kstar_proj}
\end{table}
%
%
\begin{table}[H]
	\setlength{\tabcolsep}{2pt} 
	\small
	{\renewcommand{\arraystretch}{1.2}
		\begin{tabularx}{0.45\textwidth}{l @{\extracolsep{\fill}} ccc}
			\toprule
			$L/a_s$ & 16 & 20 & 24 \\
			\cmidrule(lr{.75em}){1-4}
			\multirow{13}{*}{$[001]A_2$} & ${12}\times \bar{\psi}\bm{\Gamma}\psi$ & ${12}\times \bar{\psi}\bm{\Gamma}\psi$ & ${12}\times \bar{\psi}\bm{\Gamma}\psi$ \\[0.5ex]
			\cdashlinelr{2-4}
			& $\pi_{[000]}\omega_{[001]}$ & $\pi_{[000]}\omega_{[001]}$ & $\pi_{[000]}\omega_{[001]}$ \\[0.5ex]
			& $\pi_{[000]}\phi_{[001]}$  & $\pi_{[001]}\omega_{[000]}$ & $\pi_{[001]}\omega_{[000]}$ \\[0.5ex]
			& $\rho_{[001]}\eta_{[000]}$  & $\pi_{[000]}\phi_{[001]}$ & $\pi_{[000]}\phi_{[001]}$  \\[0.5ex]
			& ${a_0}_{[001]}\pi_{[000]}$ & $\rho_{[001]}\eta_{[000]}$ & $\rho_{[001]}\eta_{[000]}$   \\[0.5ex]
			& $\pi_{[001]}\omega_{[000]}$ & ${a_0}_{[001]}\pi_{[000]}$ & ${\color{gray}K^*_{[001]}\bar{K}_{[000]}}$ \\[0.5ex]
			& $K^*_{[001]}\bar{K}_{[000]}$ & $K^*_{[001]}\bar{K}_{[000]}$ & ${\color{gray}{\rho^{\mathfrak{1}}_{[001]}\eta_{[000]}}}$ \\[0.5ex]
			& ${\color{gray}\rho_{[000]}\eta_{[001]}}$ &  &  ${\color{gray}\rho_{[000]}\eta_{[001]}}$\\[0.5ex]
			& ${\color{gray}\pi_{[001]}\phi_{[000]}}$ &  &  ${\color{gray}\pi_{[001]}\phi_{[000]}}$ \\[0.5ex]
			&  &  &   ${\color{gray}K^*_{[000]}\bar{K}_{[001]}}$ \\[0.5ex]
			&   &  & ${\color{gray}\{2\}\pi_{[001]}\omega_{[011]}}$  \\[0.5ex]
			& & & ${\color{gray}\{2\}\pi_{[011]}\omega_{[001]}}$  \\[0.5ex]
			\bottomrule
		\end{tabularx}
	} 
	\caption{As in Table~\ref{Tab:ops_b1_rest} but for irrep $[001]A_2$. For operators \smash{$\mathcal{O}^\dagger_{\mathbb{RM}}$}, the superscript $\mathfrak{n}$ on $\mathbb{R}^\mathfrak{n}$ denotes the $\mathfrak{n}^\text{th}$ excited state when $\mathfrak{n} \geq 1$. All $\rho$ and $K^*$ operators transform in $[000]T_1^-$ at $\vec{p}=\vec{0}$ and all $\rho$, $a_0$ and $K^*$ operators transform in $\vec{P}A_1$ for $\vec{p}\neq \vec{0}$. Operators shown in gray correspond to $E^{(2+1)}_{\text{n.i.}}$ greater than the $E^{(2+1)}_{\text{n.i.}}$ or $E^{(3)}_{\text{n.i.}}$ of operators that have not been included in the basis.}
	\label{Tab:ops_b1_p100}
\end{table}
%
%
\begin{table}[H]
	\setlength{\tabcolsep}{2pt} 
	\small
	{\renewcommand{\arraystretch}{1.2}
		\begin{tabularx}{0.45\textwidth}{l @{\extracolsep{\fill}} ccc}
			\toprule
			$L/a_s$ & 16 & 20 & 24 \\
			\cmidrule(lr{.75em}){1-4}
			\multirow{8}{*}{$[011]A_2$} & ${21}\times \bar{\psi}\bm{\Gamma}\psi$ & ${21}\times \bar{\psi}\bm{\Gamma}\psi$ & ${21}\times \bar{\psi}\bm{\Gamma}\psi$ \\[0.5ex]
			\cdashlinelr{2-4}
			& $\pi_{[000]}\omega_{[011]}$ & $\pi_{[000]}\omega_{[011]}$ & $\pi_{[000]}\omega_{[011]}$ \\[0.5ex]
			& $\pi_{[000]}\phi_{[011]}$  & $\pi_{[000]}\phi_{[011]}$  & $\{2\}\pi_{[001]}\omega_{[001]}$ \\[0.5ex]
			& $\rho_{[011]}\eta_{[000]}$  & $\{2\}\pi_{[001]}\omega_{[001]}$  & $\pi_{[000]}\phi_{[011]}$ \\[0.5ex]
			& $K^*_{[011]}\bar{K}_{[000]}$  & $\rho_{[011]}\eta_{[000]}$  & $\pi_{[011]}\omega_{[000]}$ \\[0.5ex]
			& $\{2\}\pi_{[001]}\omega_{[001]}$ &  ${a_0}_{[011]}\pi_{[000]}$  & $\rho_{[011]}\eta_{[000]}$ \\[0.5ex]
			& ${a_0}_{[011]}\pi_{[000]}$ & ${\color{gray}K^*_{[011]}\bar{K}_{[000]}}$  &  \\[0.5ex]
			& $\pi_{[011]}\omega_{[000]}$ & & \\[0.5ex]
			\bottomrule
		\end{tabularx}
	} 
	\caption{As in Table~\ref{Tab:ops_b1_p100} but for irrep $[011]A_2$.}
	\label{Tab:ops_b1_p110}
\end{table}
%
%
\begin{table}[H]
	\setlength{\tabcolsep}{2pt} 
	\small
	{\renewcommand{\arraystretch}{1.2}
		\begin{tabularx}{0.45\textwidth}{l @{\extracolsep{\fill}} ccc}
			\toprule
			$L/a_s$ & 16 & 20 & 24 \\
			\cmidrule(lr{.75em}){1-4}
			\multirow{8}{*}{$[111]A_2$} & ${15}\times \bar{\psi}\bm{\Gamma}\psi$ & ${15}\times \bar{\psi}\bm{\Gamma}\psi$ & ${15}\times \bar{\psi}\bm{\Gamma}\psi$ \\[0.5ex]
			\cdashlinelr{2-4}
			& $\pi_{[000]}\omega_{[111]}$ & $\pi_{[000]}\omega_{[111]}$ & $\pi_{[000]}\omega_{[111]}$ \\[0.5ex]
			& $\pi_{[000]}\phi_{[111]}$  & $\pi_{[000]}\phi_{[111]}$  & $\pi_{[000]}\phi_{[111]}$  \\[0.5ex]
			& $\rho_{[111]}\eta_{[000]}$  & $\{2\}\pi_{[001]}\omega_{[011]}$ & $\{2\}\pi_{[001]}\omega_{[011]}$ \\[0.5ex]
			& $K^*_{[111]}\bar{K}_{[000]}$  & $\rho_{[111]}\eta_{[000]}$  & $\{2\}\pi_{[011]}\omega_{[001]}$  \\[0.5ex]
			& ${\color{gray}\{2\}\pi_{[001]}\omega_{[011]}}$ & $K^*_{[111]}\bar{K}_{[000]}$ & $\rho_{[111]}\eta_{[000]}$  \\[0.5ex]
			& ${\color{gray}\pi_{[111]}\omega_{[000]}}$ & ${a_0}_{[111]}\pi_{[000]}$  & $\pi_{[111]}\omega_{[000]}$ \\[0.5ex]
			& ${\color{gray}\{2\}\pi_{[011]}\omega_{[001]}}$ &  & ${\color{gray}K^*_{[111]}\bar{K}_{[000]}}$ \\[0.5ex]
			\bottomrule
		\end{tabularx}
	} 
	\caption{As in Table~\ref{Tab:ops_b1_p100} but for irrep $[111]A_2$.}
	\label{Tab:ops_b1_p111}
\end{table}
%
%
\begin{table}[H]
	\setlength{\tabcolsep}{2pt} 
	\small
	{\renewcommand{\arraystretch}{1.2}
		\begin{tabularx}{0.45\textwidth}{l @{\extracolsep{\fill}} ccc}
			\toprule
			$L/a_s$ & 16 & 20 & 24 \\
			\cmidrule(lr{.75em}){1-4}
			\multirow{9}{*}{$[002]A_2$} & ${20}\times \bar{\psi}\bm{\Gamma}\psi$ & ${20}\times \bar{\psi}\bm{\Gamma}\psi$ & ${20}\times \bar{\psi}\bm{\Gamma}\psi$ \\[0.5ex]
			\cdashlinelr{2-4}
			& $\pi_{[001]}\omega_{[001]}$ & $\pi_{[001]}\omega_{[001]}$ & $\pi_{[001]}\omega_{[001]}$ \\[0.5ex]
			& $\rho_{[001]}\eta_{[001]}$ & $\rho_{[001]}\eta_{[001]}$ & $\pi_{[000]}\omega_{[002]}$ \\[0.5ex]
			& $K^*_{[001]}\bar{K}_{[001]}$  &  $\pi_{[000]}\omega_{[002]}$ & $\rho_{[001]}\eta_{[001]}$  \\[0.5ex]
			&  $\pi_{[000]}\omega_{[002]}$ & $\pi_{[001]}\phi_{[001]}$ & $\pi_{[001]}\phi_{[001]}$ \\[0.5ex]
			& $\pi_{[001]}\phi_{[001]}$  & $K^*_{[001]}\bar{K}_{[001]}$ & ${\color{gray}K^*_{[001]}\bar{K}_{[001]}}$\\[0.5ex]
			& ${\color{gray}\rho^{\mathfrak{1}}_{[001]}\eta_{[001]}}$ & ${a_0}_{[001]}\pi_{[001]}$ &  ${\color{gray}\pi_{[000]}\phi_{[002]}}$  \\[0.5ex]
			\bottomrule
		\end{tabularx}
	} 
	\caption{As in Table~\ref{Tab:ops_b1_p100} but for irrep $[002]A_2$.}
	\label{Tab:ops_b1_p200}
\end{table}
%
%
\begin{table}[H]
	\setlength{\tabcolsep}{2pt} 
	\small
	{\renewcommand{\arraystretch}{1.2}
		\begin{tabularx}{0.45\textwidth}{l @{\extracolsep{\fill}} ccc}
			\toprule
			$[000]T_2^+$ & $[000]E^-$ & $[001]B_1$ & $[001]B_2$ \\
			\cmidrule(lr{.75em}){1-4}
			${14}\times \bar{\psi}\bm{\Gamma}\psi$ & ${12}\times \bar{\psi}\bm{\Gamma}\psi$ & ${9}\times \bar{\psi}\bm{\Gamma}\psi$ & ${9}\times \bar{\psi}\bm{\Gamma}\psi$ \\[0.5ex]
			\cdashlinelr{1-4}
			$\pi_{[001]}\omega_{[001]}$ & $\pi_{[001]}\omega_{[001]}$ &  $\pi_{[011]}\pi_{[001]}$ & $\pi_{[111]}\pi_{[011]}$ \\[0.5ex]
			 & & $\bar{K}_{[011]}K_{[001]}$  & $\{2\}\pi_{[001]}\omega_{[011]}$ \\[0.5ex]
			 & & $\pi_{[001]}\omega_{[011]}$  & $\{2\}\pi_{[011]}\omega_{[001]}$ \\[0.5ex]
			 & & $\pi_{[011]}\omega_{[001]}$ & \\
			\bottomrule
		\end{tabularx}
	} 
	\caption{As Table~\ref{Tab:ops_b1_p100} for irreps $[000]T_2^+$, $[000]E^-$, $[001]B_1$ and $[001]B_2$ on the $(L/a_s)=24$ lattice.}
	\label{Tab:ops_extra}
\end{table}
%
%

\section{Tables of Scattering Parameterizations \label{App:Scattering}}


We present here tables of scattering parameterizations as referred to in Section~\ref{Sec:Scattering_Analysis}. \\[1.3ex]
\begin{table*}[tb]
\begin{tabular}{llcc}
\midrule
Parameterization & Further Restrictions & $N_\mathrm{pars}$ &  $\chi^2/N_\mathrm{dof}$  \\[1.3ex]
\midrule
 \multirow{1}{*}{Breit-Wigner} & -- & 2 & 0.84 \\[1.3ex]  
 \cdashlinelr{2-4}
 \multirow{2}{*}{\minitab[l]{Effective Range \\  $k_{cm}\cot(\delta) = a^{-1} + \frac{1}{2}r k_{cm}^2 $}} & \multirow{2}{*}{--} & \multirow{2}{*}{2} & \multirow{2}{*}{0.86}\\[1.3ex]  
 &&&\\
 \cdashlinelr{2-4}
 \multirow{4}{*}{\minitab[l]{$K=\frac{g^2}{m^2-s} + \gamma^{(0)} + \gamma^{(1)}s$ \\ $I(s)=-i\rho(s)$}} &--&4&0.80\\ 
 &$\gamma^{(1)}=0$&3 &0.76\\
 &$\gamma^{(0)}=0, \gamma^{(1)}=0$& 2 &0.84\\
 &$\gamma^{(0)}=0$&3 &0.75\\[1.3ex]
 \cdashlinelr{2-4}
 \multirow{4}{*}{\minitab[l]{$K=\frac{g^2}{m^2-s} + \gamma^{(0)}+ \gamma^{(1)}s$ \\ $\text{CM Re}\{I(s=s^{\text{thr}})=0\}$ }} &--&4&0.80\\
 &$\gamma^{(1)}=0$&3 &0.76\\
 &$\gamma^{(0)}=0, \gamma^{(1)}=0$& 2 &0.84\\
 &$\gamma^{(0)}=0$&3 &0.76\\[1.3ex]
 \cdashlinelr{2-4}
 \multirow{4}{*}{\minitab[l]{$K=\frac{g^2}{m^2-s} + \gamma^{(0)}+ \gamma^{(1)}s$ \\ $\text{CM Re}\{I(s=m^2)=0\}$ }} &--&4&0.80\\  
 &$\gamma^{(1)}=0$&3 &0.76\\
 &$\bm{\gamma^{(0)}=0, \gamma^{(1)}=0}$& \textbf{2} &\textbf{0.84}\\
 &$\gamma^{(0)}=0$&3 &0.76\\[1.3ex]
 \cdashlinelr{2-4}
 \multirow{3}{*}{\minitab[l]{$K^{-1}=c^{(0)} + c^{(1)}s$ \\ $I(s)=-i\rho(s)$}} & \multirow{3}{*}{--} & \multirow{3}{*}{2} & \multirow{3}{*}{0.84}\\ 
  &&&\\[3.3ex]
 \cdashlinelr{2-4}
 \multirow{2}{*}{\minitab[l]{$K^{-1}=c^{(0)} + c^{(1)}s$ \\ $\text{CM Re}\{I(s=s^{\text{thr}})=0\}$ }} &\multirow{2}{*}{--} & \multirow{2}{*}{2} & \multirow{2}{*}{0.84}\\
 &&&\\[1.3ex]
\bottomrule
\end{tabular}
\caption{Parameterizations of elastic $\pi\omega\{^3S_1\}$ scattering amplitudes with $N_\text{pars}$ free parameters. Fits used 20 energy levels below $\pi\phi$ threshold as described in the text. The reference amplitude, Eq.~\ref{Fit:A}, is in bold. `CM' denotes that the Chew-Mandelstam prescription was employed with subtraction at energy $m$ or at threshold $s^{\text{thr}} = (m_{\pi}+m_{\omega})^2$. Otherwise, we set $I(s)=-i\rho(s)$.}
\label{Tab:below_piphi_Swave}
\end{table*}
%
%
\begin{table*}
\begin{tabular}{lllcc}
\midrule
Parameterization & Further Restrictions & Phase-space & $N_\mathrm{pars}$ &  $\chi^2/N_\mathrm{dof}$  \\[1.3ex]
\midrule
  \multirow{5}{*}{\minitab[l]{ {\minitab[l]{ \\ $K_{\ell J,\ell' J}=\displaystyle\frac{g_{\ell J}g_{\ell' J}}{m^2-s} +\,\gamma^{(0)}_{\ell J,\ell' J} $ }} \\[1.3ex] \\ 
  where $\gamma^{(0)}_{\pi \omega\{^3D_1\},\,\pi \omega\{^3D_1\}} = 0$, \\[1.3ex] hence $6-1=5$ free real-parameters.}} 
  &\multirow{4}{*}{\minitab[l]{$\gamma^{(0)}_{\pi \omega\{^3S_1\},\,\pi \omega\{^3S_1\}} = 0$ \\  $\gamma^{(0)}_{\pi \omega\{^3S_1\},\,\pi \omega\{^3D_1\}} = 0$ \\ $\gamma^{(0)}_{\pi \omega\{^3D_1\},\,\pi \omega\{^3D_1\}} = 0$}}& \multirow{3}{*}{\minitab[l]{ \\[1.3ex] \textbf{CM} $\bm{\text{\textbf{Re}}\{I(s=m^2)=0\}}$ \\[1.3ex] }}&\multirow{3}{*}{\minitab[l]{ \\[1.3ex] \textbf{3} \\[1.3ex]}} &\multirow{3}{*}{\minitab[l]{ \\[1.3ex] \textbf{0.87} \\[1.3ex] }} \\[1.3ex]&&&\\[1.3ex]&&&\\[1.3ex]
   \cdashlinelr{2-5}
  &\multirow{3}{*}{\minitab[l]{$\gamma^{(0)}_{\pi \omega\{^3S_1\},\,\pi \omega\{^3D_1\}} = 0$ \\ $\gamma^{(0)}_{\pi \omega\{^3D_1\},\,\pi \omega\{^3D_1\}} = 0$}}& \multirow{3}{*}{ {CM} ${\text{Re}\{I(s=m^2)=0\}}$ }&\multirow{3}{*}{4} &\multirow{3}{*}{ 0.80} \\[1.3ex]&&&\\[1.3ex]
   \cdashlinelr{2-5}
  &\multirow{3}{*}{\minitab[l]{$\gamma^{(0)}_{\pi \omega\{^3S_1\},\,\pi \omega\{^3S_1\}} = 0$ \\ $\gamma^{(0)}_{\pi \omega\{^3D_1\},\,\pi \omega\{^3D_1\}} = 0$}}& \multirow{3}{*}{ {CM} ${\text{Re}\{I(s=m^2)=0\}}$ }&\multirow{3}{*}{4} &\multirow{3}{*}{ 0.93} \\[1.3ex]&&&\\[1.3ex]
   \cdashlinelr{2-5}
  &\multirow{4}{*}{\minitab[l]{ $g_{\pi \omega\{^3S_1\}}=0 $\\$\gamma^{(0)}_{\pi \omega\{^3S_1\},\,\pi \omega\{^3S_1\}} = 0$ \\ $\gamma^{(0)}_{\pi \omega\{^3D_1\},\,\pi \omega\{^3D_1\}} = 0$}}& \multirow{4}{*}{ {CM} ${\text{Re}\{I(s=m^2)=0\}}$ }&\multirow{4}{*}{3} &\multirow{4}{*}{ 0.89} \\[1.3ex]&&&\\[1.3ex]&&&\\[1.3ex]
%
\bottomrule
\end{tabular}
	\caption{Parameterizations of dynamically-coupled $\pi\omega\{^3S_1\}$ and $\pi\omega\{^3D_1\}$ scattering amplitudes. Fits were determined using 20 energy levels below $\pi\phi$ threshold as described in the text. Displayed in bold is the reference amplitude of Eq.~\ref{Fit:B}. `CM' denotes that the Chew-Mandelstam prescription was employed with subtraction at energy $m$, the `pole' parameter in the $K$-matrix.}
	\label{Tab:below_piphi_S+Dwave}
\end{table*}
%
%
\begin{table*}
\begin{tabular}{lllcc}
\midrule
Parameterization & Further Restrictions & Phase-space & $N_\mathrm{pars}$ &  $\chi^2/N_\mathrm{dof}$  \\[1.3ex]
\midrule
  \multirow{35}{*}{\rotatebox[origin=c]{90}{
  {\minitab[l]{ {\minitab[l]{$\qquad K_{\ell J a,\ell' J b}=\displaystyle\frac{\big(g^{(0)}_{\ell J a}+g^{(1)}_{\ell J a}s\big)\big(g^{(0)}_{\ell' J b}+g^{(1)}_{\ell' J b}s\big)}{m^2-s} +\,\gamma^{(0)}_{\ell J a,\ell' J b} + \gamma^{(1)}_{\ell J a,\ell' J b}s \quad$}} \\[1.3ex] \\ 
    where $\gamma^{(0,1)}_{\pi \omega\{^3D_1\},\,\pi \omega\{^3D_1\}} = 0$,
     $\quad\gamma^{(1)}_{\pi \phi\{^3S_1\},\,\pi \phi\{^3S_1\}} = 0$, 
    $\quad\gamma^{(1)}_{\pi \omega\{^3S_1\},\,\pi \omega\{^3D_1\}} = 0$, \\[1.3ex]
     $\quad\gamma^{(1)}_{\pi \omega\{^3S_1\},\,\pi \phi\{^3S_1\}} = 0$, 
     $\quad\gamma^{(0,1)}_{\pi \omega\{^3D_1\},\,\pi \phi\{^3S_1\}} = 0$,
    $\quad g^{(1)}_{\pi \omega\{^3D_1\}} = 0$,
    $\quad g^{(1)}_{\pi \phi\{^3S_1\}} = 0$, \\[1.3ex] hence $19-9=10$ free real-parameters. \\[2.9ex]}}
  }
  }
  &\multirow{4}{*}{\minitab[l]{$g^{(0)}_{\pi \phi\{^3S_1\}}=g^{(1)}_{\pi \omega\{^3S_1\}}=0\qquad$ \\  $\gamma^{(0)}_{\pi \omega\{^3S_1\},\,\pi \omega\{^3D_1\}} = 0$ \\  $\gamma^{(0)}_{\pi \omega\{^3S_1\},\,\pi \phi\{^3S_1\}} = 0$ \\ $\gamma^{(1)}_{\pi \omega\{^3S_1\},\,\pi \omega\{^3S_1\}} = 0 $}}& \multirow{3}{*}{\minitab[l]{ $I_a(s)=-i\rho_a(s)$ \\[1.3ex] \textbf{CM} $\bm{\text{\textbf{Re}}\{I_a(s=m^2)=0\}}$ \\[1.3ex] CM $\text{Re}\{I_a(s=s^{\text{thr}}_a)=0\}$ }}&\multirow{3}{*}{\minitab[l]{ \\[1.3ex] \textbf{5} \\[1.3ex]}} &\multirow{3}{*}{\minitab[l]{ 1.18\\[1.3ex] \textbf{1.19} \\[1.3ex] 1.19}} \\[1.3ex]&&&\\[1.3ex]&&&\\[1.3ex]&&&\\[1.3ex]
   \cdashlinelr{2-5}
  &\multirow{4}{*}{\minitab[l]{$g^{(0)}_{\pi \phi\{^3S_1\}}=g^{(1)}_{\pi \omega\{^3S_1\}}=0\qquad$ \\ $\gamma^{(0)}_{\pi \omega\{^3S_1\},\,\pi \phi\{^3S_1\}} = 0$ \\ $\gamma^{(1)}_{\pi \omega\{^3S_1\},\,\pi \omega\{^3S_1\}} = 0 $}}& \multirow{3}{*}{\minitab[l]{ $I_a(s)=-i\rho_a(s)$ \\[1.3ex] CM $\text{Re}\{I_a(s=m^2)=0\}$ \\[1.3ex] CM $\text{Re}\{I_a(s=s^{\text{thr}}_a)=0\}$ }}&\multirow{3}{*}{\minitab[l]{ \\[1.3ex] 6 \\[1.3ex]}} &\multirow{3}{*}{\minitab[l]{ 1.22\\[1.3ex] 1.22 \\[1.3ex] 1.22}} \\[1.3ex]&&&\\[1.3ex]&&&\\[1.3ex]
   \cdashlinelr{2-5}
  &\multirow{4}{*}{\minitab[l]{$g^{(1)}_{\pi \omega\{^3S_1\}}=0\qquad$ \\ $\gamma^{(0)}_{\pi \omega\{^3S_1\},\,\pi \phi\{^3S_1\}} = 0$ \\ $\gamma^{(1)}_{\pi \omega\{^3S_1\},\,\pi \omega\{^3S_1\}} = 0 $}}& \multirow{3}{*}{\minitab[l]{$I_a(s)=-i\rho_a(s)$ \\[1.3ex] CM $\text{Re}\{I_a(s=m^2)=0\}$ \\[1.3ex] CM $\text{Re}\{I_a(s=s^{\text{thr}}_a)=0\}$ }}&\multirow{3}{*}{\minitab[l]{ \\[1.3ex] 7 \\[1.3ex]}} &\multirow{3}{*}{\minitab[l]{ 1.27\\[1.3ex] 1.27 \\[1.3ex] 1.27}} \\[1.3ex]&&&\\[1.3ex]&&&\\[1.3ex]
   \cdashlinelr{2-5}
  &\multirow{4}{*}{\minitab[l]{$g^{(1)}_{\pi \omega\{^3S_1\}}=0\qquad$ \\ $\gamma^{(0)}_{\pi \omega\{^3S_1\},\,\pi \phi\{^3S_1\}} = 0$ \\ $\gamma^{(0)}_{\pi \omega\{^3S_1\},\,\pi \omega\{^3D_1\}} = 0 $}}& \multirow{3}{*}{\minitab[l]{ CM $\text{Re}\{I_a(s=m^2)=0\}$ \\[1.3ex] CM $\text{Re}\{I_a(s=s^{\text{thr}}_a)=0\}$ }}&\multirow{3}{*}{7} &\multirow{3}{*}{\minitab[l]{ 1.24 \\[1.3ex] 1.24}} \\[1.3ex]&&&\\[1.3ex]&&&\\[1.3ex]
   \cdashlinelr{2-5}
  &\multirow{4}{*}{\minitab[l]{$g^{(0)}_{\pi \phi\{^3S_1\}}=g^{(1)}_{\pi \omega\{^3S_1\}}=0\qquad$ \\ $\gamma^{(0)}_{\pi \omega\{^3S_1\},\,\pi \phi\{^3S_1\}} = 0$ \\ $\gamma^{(0)}_{\pi \omega\{^3S_1\},\,\pi \omega\{^3D_1\}} = 0 $}}& \multirow{3}{*}{\minitab[l]{ CM $\text{Re}\{I_a(s=m^2)=0\}$ \\[1.3ex] CM $\text{Re}\{I_a(s=s^{\text{thr}}_a)=0\}$ }}&\multirow{3}{*}{6} &\multirow{3}{*}{\minitab[l]{ 1.20 \\[1.3ex] 1.20}} \\[1.3ex]&&&\\[1.3ex]&&&\\[1.3ex]
   \cdashlinelr{2-5}
  &\multirow{4}{*}{\minitab[l]{$g^{(0)}_{\pi \omega\{^3D_1\}}=g^{(0)}_{\pi \phi\{^3S_1\}}=g^{(1)}_{\pi \omega\{^3S_1\}}=0\qquad$ \\ $\gamma^{(1)}_{\pi \omega\{^3S_1\},\,\pi \omega\{^3S_1\}} = 0 $}}& \multirow{3}{*}{\minitab[l]{ $I_a(s)=-i\rho_a(s)$ \\[1.3ex] CM $\text{Re}\{I_a(s=m^2)=0\}$ \\[1.3ex] CM $\text{Re}\{I_a(s=s^{\text{thr}}_a)=0\}$ }}&\multirow{3}{*}{\minitab[l]{ \\[1.3ex] 6 \\[1.3ex]}} &\multirow{3}{*}{\minitab[l]{ 1.35\\[1.3ex] 1.35 \\[1.3ex] 1.32}} \\[1.3ex]&&&\\[1.3ex]&&&\\[1.3ex]
   \cdashlinelr{2-5}  
  &\multirow{4}{*}{\minitab[l]{$g^{(0)}_{\pi \omega\{^3D_1\}}=g^{(0)}_{\pi \phi\{^3S_1\}}=0\qquad$ \\ $\gamma^{(0)}_{\pi \omega\{^3S_1\},\,\pi \phi\{^3S_1\}} = 0$ \\ $\gamma^{(1)}_{\pi \omega\{^3S_1\},\,\pi \omega\{^3S_1\}} = 0$}}& \multirow{3}{*}{\minitab[l]{$I_a(s)=-i\rho_a(s)$ \\[1.3ex] CM $\text{Re}\{I_a(s=m^2)=0\}$ }}&\multirow{3}{*}{6} &\multirow{3}{*}{\minitab[l]{ 1.35 \\[1.3ex] 1.35}} \\[1.3ex]&&&\\[1.3ex]&&&\\[1.3ex]
  \cdashlinelr{2-5}
    &\multirow{4}{*}{\minitab[l]{$g^{(0)}_{\pi \omega\{^3D_1\}}=g^{(0)}_{\pi \phi\{^3S_1\}}=g^{(1)}_{\pi \omega\{^3S_1\}}=0\qquad$ \\ $\gamma^{(1)}_{\pi \omega\{^3S_1\},\,\pi \omega\{^3S_1\}} = 0 $ \\ $\gamma^{(0)}_{\pi \omega\{^3S_1\},\,\pi \phi\{^3S_1\}} = 0$ }}& \multirow{3}{*}{\minitab[l]{ $I_a(s)=-i\rho_a(s)$ \\[1.3ex] CM $\text{Re}\{I_a(s=m^2)=0\}$ \\[1.3ex] CM $\text{Re}\{I_a(s=s^{\text{thr}}_a)=0\}$ }}&\multirow{3}{*}{\minitab[l]{ \\[1.3ex] 5 \\[1.3ex]}} &\multirow{3}{*}{\minitab[l]{ 1.31\\[1.3ex] 1.31 \\[1.3ex] 1.28}} \\[1.3ex]&&&\\[1.3ex]&&&\\[1.3ex]
\bottomrule
\end{tabular}
\caption{Parameterizations of coupled $\pi\omega\{^3S_1\}$, $\pi\omega\{^3D_1\}$ and $\pi\phi\{^3S_1\}$ scattering amplitudes. Fits used 36 energy levels below $\pi\pi\pi\pi$ threshold as described in the text. Displayed in bold is the reference amplitude of Eq.~\ref{Fit:C}. `CM' denotes that the Chew-Mandelstam prescription was employed with subtraction at energy $m$ or at threshold $s^{\text{thr}}_a$ where $s^{\text{thr}}_a = (m^{(a)}_{1}+m^{(a)}_{2})^2$. Otherwise, we set $I_a(s)=-i\rho_a(s)$. Results of fits to these parameterizations can be found in the Supplemental Material.}
\label{Tab:below_isobars_params}
\end{table*}
%
%

\end{document}